\documentclass[a4paper,12pt]{article}
\pdfoutput=1
\usepackage{graphicx}

\usepackage{amssymb,amsmath}

\usepackage[colon,authoryear]{natbib}

\begin{document}

\title{2/1 resonant periodic orbits in three dimensional planetary systems}

\author{K. I. Antoniadou, G. Voyatzis\\
Department of Physics, Aristotle University of Thessaloniki, \\54124, Thessaloniki, Greece 
\\kyant@auth.gr, voyatzis@auth.gr}
\maketitle

The final publication is available at springerlink.com\\
http://www.springerlink.com/openurl.asp?genre=article\&id=doi:10.1007/s10569-012-9457-4

\begin{abstract}
We consider the general spatial three body problem and study the dynamics of planetary systems consisting of a star and two planets which evolve into 2/1 mean motion resonance and into inclined orbits. Our study is focused on the periodic orbits of the system given in a suitable rotating frame. The stability of periodic orbits characterize the evolution of any planetary system with initial conditions in their vicinity. Stable periodic orbits are associated with long term regular evolution, while unstable periodic orbits are surrounded by regions of chaotic motion. We compute many families of symmetric periodic orbits by applying two schemes of analytical continuation. In the first scheme, we start from the 2/1 (or 1/2) resonant periodic orbits of the restricted problem and in the second scheme, we start from vertical critical periodic orbits of the general planar problem. Most of the periodic orbits are unstable, but many stable periodic orbits have been, also, found with mutual inclination up to $50^\circ$ - $60^\circ$, which may be related with the existence of real planetary systems.
\end{abstract}
{\bf keywords} 2/1 resonance, 3D general three body problem, periodic orbits, vertical stability, planetary systems.

\section{Introduction}
The dynamics of planetary systems consisting of two massive planets can be studied by considering the general three body problem (GTBP). Many of such systems seem to be locked in mean motion resonances and particularly in 2/1 resonance (e.g. GJ 876, HD 82943, HD 73526 and 47 Uma). Such resonances are related with periodic orbits of the three body problem in a rotating frame and are very important, since they are associated with regions of stability and instability in phase space (\citealt{psyhadj05,gozd05,voyatzis08}). Particularly, the phase space structure near the 2/1 resonance and its connection with periodic evolution is studied in \citet{mbf08a,mbf08b} and \citet{mfpro}. Also, it has been shown that stable periodic orbits can drive the migration process of planets (\citealt{leepeal02,mebeaumich03,hadjvoy10}). 

In the previous years, most of the studies of planetary evolution in mean motion resonances were restricted in planar motion. In such cases, resonant periodic solutions can be found in various ways, see e.g. (\citealt{varadi99,haghi03,voyhadj05,mbf06}). It is shown that periodic orbits are classified into various configurations, where some of them are stable and others unstable. Setting initial conditions close to a stable periodic orbit the planetary system evolves regularly, since in phase space and around stable periodic orbits invariant tori exist. Along this motion the resonant angles show librations around the values that correspond to the particular periodic orbit (see e.g. \citet{voyatzis08} and \citet{mfpro}).  

Concerning the dynamics of multiple planetary systems in space, mutual inclinations are a very interesting feature. The dynamics of such systems can be studied, for instance, by examining the phase space structure (\citealt{mfb06}) or the Hill's stability  (\citealt{verasa04}).  Numerical simulations show that inclined planetary systems can be stable even for relatively high mutual inclinations at systems as 47 Uma (\citealt{lauchfi02}) and  $\upsilon$ And (\citealt{bgqmb11}). Stability of inclined systems may be associated with the Kozai resonance (\citealt{litsi09}) or the Lagrangian equilibrium points (\citealt{l4}). Some possible mechanisms that lead to the excitation of planetary inclinations are the planetary scattering (\citealt{mawei02,chaford08}), the differential migration (\citealt{thommes03,litsi09b,leetho09}) and the tidal evolution (\citealt{cor11}). In these mechanisms, resonance trapping is possible to occur and the evolution of the system may be associated with particular inclined periodic orbits.

In this paper, we study resonant periodic orbits of the spatial three body problem that model a possible planetary system in 2/1 resonance. We attempt to determine all possible periodic configurations and especially the stable ones. We apply the method of analytic continuation described and proved in (\citealt{ichmich80}). Based on this method, families of periodic orbits of the general three body problem in space have been computed  by \citet{mich79b}, \citet{mich80}, \citet{mich81} and \citet{katichmich}. However, these families are not associated with planetary resonant dynamics and stability aspects are not sufficiently discussed.            
   
In Section 2, we present the particular TBP model and refer to periodic conditions, stability and resonances. In Section 3, we demonstrate typical examples of periodic orbits, we present the evolution of orbits starting from the vicinity of these periodic orbits and discuss their long-term stability. In section 4, we compute and present families of resonant periodic orbits and indicate their stability. Some miscellaneous cases are discussed in section 5 and we conclude in section 6.

\section{The general TBP in space and periodic orbits}

\subsection{The model}
Assume a planetary system consisting of a star $S$ and two planets $P_1$ and $P_2$ with masses $m_0$, $m_1$ and $m_2$, respectively, moving in the space under their mutual gravitational attraction. We consider these bodies as point masses and let $O$ be their center of mass. The three bodies are isolated, thus their center of mass is fixed with respect to any  inertial system of reference and its angular momentum vector, ${\bf L}$ , is constant. We now introduce a particular inertial coordinate system, $OXYZ$, such that its origin coincides with their center of mass $O$ and its $Z$-axis is parallel to  ${\bf L}$. The system is described by six degrees of freedom defined for instance, by the position vectors of the two planets.

We can reduce the number of degrees of freedom by introducing a suitable rotating frame of reference. Following \citet{mich79}, we introduce the rotating frame $Gxyz$, such that:
\begin{enumerate} 
	\item Its origin coincides with the center of mass $G$ of the bodies $S$ and $P_1$.
	\item Its $z$-axis is always parallel to the $Z$-axis.
	\item $S$ and $P_1$ move always on $xz$-plane.
\end{enumerate}
We define $x_i$, $y_i$, $z_i$ $(i=1, 2, 3)$ as the Cartesian coordinates of the bodies with respect to $Gxyz$ and an angle $\theta$ as the angle between $X$ and $x$ axes assuming always that $\theta(0)=0$. According to the definition of $Gxyz$, it always holds that $y_1=0$. 
Considering the normalized gravitational constant $G=1$ and setting $m = m_0 + m_1 + m_2$ (thus $2\pi$ time units (t.u.) correspond to one year), the Lagrangian of the system in the rotating frame of reference is given by \citet{mich79}:
\begin{equation}
\begin{array}{l}
\mathfrak{L}=\frac{\displaystyle 1}{\displaystyle 2} (m_0+m_1)[a(\dot x_1^2+\dot z_1^2+x_1^2\dot \theta^2)+\displaystyle b [(\dot x_2^2+\dot y_2^2+\dot z_2^2)\\[0.2cm] 
\quad +\dot\theta^2(x_2^2+y_2^2)+2\dot\theta(x_2\dot y_2-\dot x_2y_2)]]-V,
\label{Lagrangian}
\end{array}
\end{equation}
where $V=-\frac{\displaystyle m_0 m_1}{\displaystyle r_{01}}-\frac{\displaystyle m_0 m_2}{\displaystyle r_{02}}-\frac{\displaystyle m_1 m_2}{\displaystyle r_{12}}$, $a=m_1/m_0$, $b=m_2/m$, 
\begin{equation}
\begin{array}{l}
\displaystyle r_{01}^2=(\displaystyle 1+\displaystyle a)^2(\displaystyle x_1^2+z_1^2) , \\[0.3cm] \nonumber
\displaystyle r_{02}^2=(\displaystyle a x_1 +\displaystyle x_2)^2+y_2^2+(\displaystyle a  z_1+z_2)^2, \\[0.3cm] 
\displaystyle r_{12}^2=(\displaystyle x_1 -\displaystyle x_2)^2+y_2^2+ (z_1-z_2)^2.
\label{Lr}
\end{array}
\end{equation}
The angle $\theta$ is ignorable and, consequently, the angular momentum $p_\theta=\partial{\mathfrak{L}}/\partial{\dot\theta}$ is constant and given by
\begin{equation}
p_\theta=(m_0+m_1)[ax_1^2\dot \theta+\displaystyle b[\dot \theta (x_2^2+y_2^2)+(x_2\dot y_2-\dot x_2y_2)]]=\textnormal{const.}
\label{Lz}
\end{equation}
Solving the Eq. \eqref{Lz} with respect to $\dot \theta\ $  we find
\begin{equation}
\dot \theta=\frac{\frac{p_\theta}{m_0+m_1}-\displaystyle b (x_2\dot y_2-\dot x_2 y_2)}{\displaystyle a x_1^2+\displaystyle b (x_2^2+y_2^2)}.
\label{dth}
\end{equation}
As far as the components of the angular momentum vector are concerned, due to the choice of the inertial system it always holds:
\begin{equation}
\begin{array}{l}
L_X=(m_0+m_1)[b(y_2 \dot z_2 -\dot y_2 z_2)-\dot \theta (a x_1 z_1 +b x_2 z_2)]=0,\\[0.3cm] L_Y=(m_0 +m_1)[a (\dot x_1 z_1 -x_1 \dot z_1)+b(\dot x_2 z_2-x_2\dot z_2)-b \dot \theta y_2 z_2]=0
\label{Lxyz}
\end{array}
\end{equation}
and
$$
L_Z=(m_0 + m_1)[b(x_2 \dot y_2 -\dot x_2 y_2)+\dot \theta[a x_1^2+b(x_2^2 +y_2^2)]],
$$
which coincides with integral \eqref{Lz}. The constrains defined by Eqs. \eqref{Lxyz} give:
\begin{equation}
\begin{array}{l}
 z_1=\frac{m_0 m_2}{m_1 m} \left ( \frac{y_2\dot z_2-\dot y_2 z_2}{x_1 \dot \theta}-\frac{x_2z_2}{x_1} \right ), \\[10pt] 
\dot z_1=\frac{m_0 m_2}{m_1 m}\frac{1}{x_1}(\dot x_2 z_2-x_2\dot z_2 -\dot{\theta}y_2z_2)+\frac{\dot x_1}{x_1} z_1. 
\label{Lxy}
\end{array}
\end{equation}
Thus, for a particular value of the angular momentum, $p_\theta$, the position of the system is defined by the coordinate $x_1$ of the planet $P_1$ and the position $(x_2,y_2,z_2)$ of the planet $P_2$. Namely, the system has been reduced to four degrees of freedom.    
We can derive the equations of motion from the Lagrangian \eqref{Lagrangian}:
\begin{equation}
\begin{array}{l}
\ddot x_1=-\frac{m_0 m_2 (x_1-x_2)}{(m_0+m_1)[(x_1-x_2)^2+y_2^2+(z_1-z_2)^2]^{3/2}}-\frac{m_0 m_2 ( a x_1+x_2)}{(m_0+m_1)[(a x_1+x_2)^2+y_2^2+(a z_1+z_2)^2]^{3/2}} \\[0.3cm] 
\quad -\frac{(m_0+m_1) x_1}{[(1+a)^2(x_1^2+z_1^2)]^{3/2}}+x_1\dot{\theta}^2
\label{x1}
\end{array}
\end{equation}
\begin{equation}
\begin{array}{l}
\ddot x_2=\frac{m m_1 (x_1-x_2)}{( m_0+m_1)[(x_1-x_2)^2+y_2^2+(z_1-z_2)^2]^{3/2}}-\frac{m m_0 (a x_1+x_2)}{(m_0+m_1)[( a x_1+x_2)^2+y_2^2+( a z_1+z_2)^2]^{3/2}}\\[0.3cm] 
\quad +x_2\dot{\theta}^2+2 \dot y_2 \dot{\theta}+y_2\ddot{\theta}
\label{x2}
\end{array}
\end{equation}
\begin{equation}
\begin{array}{l}
\ddot y_2=-\frac{m m_1 y_2}{( m_0+m_1)[(x_1-x_2)^2+y_2^2+(z_1-z_2)^2]^{3/2}}-\frac{m m_0 y_2}{(m_0+m_1)[( a x_1+x_2)^2+y_2^2+( a z_1+z_2)^2]^{3/2}}\\[0.3cm] 
\quad +y_2\dot{\theta}^2-2 \dot x_2 \dot{\theta}-x_2\ddot{\theta}
\label{y2}
\end{array}
\end{equation}
\begin{equation}
\begin{array}{l}
\ddot z_2=\frac{m m_1 (z_1-z_2)}{( m_0+m_1)[(x_1-x_2)^2+y_2^2+(z_1-z_2)^2]^{3/2}}-\frac{m m_0 (a x_1+x_2)}{(m_0+m_1)[( a x_1+x_2)^2+y_2^2+( a z_1+z_2)^2]^{3/2}}
\label{z2}
\end{array}
\end{equation}
We should remark some essential differences between the restricted and the general 3D-TBP described in the rotating frame. In the general problem, the primaries do not remain fixed on the rotating axis $Gx$, but move on the rotating plane $xz$. Also, the rotating frame $Gxyz$ does not revolve with a constant angular velocity, see Eq. \eqref{dth}, and its origin $G$ does not remain fixed with respect to the inertial system $OXYZ$.

In our numerical computations we should always take that $P_1$ is the inner planet, $P_2$ is the outer one and the total mass of the system is $m=1$. Also, we should always take that one of the planets (either the inner or the outer) has the mass of Jupiter, namely $m_i=10^{-3}$, with either $i=1$ or $i=2$.

\subsection{Periodic orbits}

Considering the rotating frame and the Poincar\'e map $y_2=0$, periodic orbits are defined as fixed or periodic points of the Poincar\'e map satisfying the conditions
\begin{equation} \label{apocon}
\begin{array}{l}
x_1(0)=x_1(T),\; x_2(0)=x_2(T), \; z_2(0)=z_2(T),\\
\dot x_1(0)=\dot x_1(T),\; \dot x_2(0)=\dot x_2(T), \; \dot z_2(0)=\dot z_2(T), \dot y_2(0)=\dot y_2(T),
\end{array}
\end{equation}
provided that $y_2(0)=y_2(T)$ and $T$ is the period. Due to the existence of the Jacobi integral, one of the above conditions is always fulfilled, when the rest ones are fulfilled.   

There exist four transformations under which the Lagrangian \eqref{Lagrangian} remains invariant and, subsequently, there exist four symmetries for the orbits of the 3D-GTBP in the rotating frame (\citealt{mich79}). A periodic orbit defined by \eqref{apocon} is called {\em symmetric}, if it is invariant under one of the symmetries of the system. In the planetary TBP, there exist periodic orbits which are symmetric with respect either to the $xz$-plane or to the $x$-axis. 

A periodic orbit obeying the symmetry with respect to $xz$-plane has two perpendicular crossings with the $xz$-plane. Planet, $P_1$, moves on that plane and, without loss of generality, we may consider that planet, $P_2$, is also on the $xz$-plane at $t=0$. Thus, the initial conditions of such a periodic orbit should be 
\begin{equation}
\begin{array}{llll}
x_1(0)=x_{10}, &\quad x_2(0)=x_{20}, &\quad y_2(0)=0 & \quad z_2(0)=z_{20},  \\
\dot x_1(0)=0, &\quad \dot x_2(0)=0, &\quad \dot y_2(0)=\dot y_{20}, &\quad \dot z_2(0)=0.
\label{xzsym}
\end{array}
\end{equation}
Subsequently, a $xz$-{\em symmetric} periodic orbit can be represented by a point in the four-dimensional space of initial conditions 
$$
\Pi_4=\{(x_{10},x_{20},z_{20},\dot{y}_{20})\}.
$$ 

If an orbit has two perpendicular crossings with the $x$-axis, then it is symmetric with respect to that axis. We can assume that at $t=0$ both planets start perpendicularly from the $x$-axis and the periodic orbit has initial conditions
\begin{equation}
\begin{array}{llll}
x_1(0)=x_{10}, &\quad x_2(0)=x_{20}, &\quad y_2(0)=0, &\quad z_2(0)=0,\\
\dot x_1(0)=0, &\quad \dot x_2(0)=0, &\quad \dot y_2(0)=\dot y_{20}, &\quad \dot z_2(0)=\dot z_{20}.
\label{xsym}
\end{array}
\end{equation}
Thus, a $x$-{\em symmetric} periodic orbit can be represented by a point in the four-dimensional space of initial conditions 
$$
\Pi'_4=\{(x_{10},x_{20},\dot{y}_{20},\dot{z}_{20})\}.
$$ 
By changing the value of $z_2$ (for the $xz$-symmetry) or $\dot z_2$ (for the $x$-symmetry) a monoparametric family of periodic orbits is formed. Also, a monoparametric family can be formed by changing the mass of a planet, $P_1$ or $P_2$, but keeping the value $z_2$ (or $\dot z_2$) constant. 

\subsection{Linear stability of periodic orbits}
The linear stability of the periodic orbits can be found by computing the monodromy matrix $\mathbf{M}$ of the variational equations of the system Eqs. \eqref{x1}-\eqref{z2}. Since the system is of four degrees of freedom, $\mathbf{M}$ is an $8 \times 8$ symplectic matrix and has four pairs of conjugate eigenvalues. One pair is always equal to unity, because of the existence of the energy integral.  In order for the periodic orbit to be linearly stable, all of the eigenvalues have to be on the unit circle. Single instability is considered if there exists one pair of real eigenvalues $\{\mu,\mu^{-1}\}$. If an eigenvalue $\mu$ is complex, then there exist also the eigenvalues $\bar\mu$, $\mu^{-1}$ and $\bar\mu^{-1}$ and the stability is characterized as complex instability, while the remaining eigenvalue pair may either lie or not on the unit circle (\citealt{marchal90,hadjbook06}). Nevertheless, apart from the pair $(1,1)$, $\mathbf{M}$ may have two or three pairs of real eigenvalues.

\subsection{Periodic orbits and resonances}
When the planetary masses are very small compared to the mass of the Star, periodic orbits in the rotating frame correspond to almost Keplerian planetary orbits in space with orbital elements $a_i$ (semimajor axis), $e_i$ (eccentricity), $i_i$ (inclination), $\omega_i$ (argument of pericenter), $\Omega_i$ (longitude of ascending node) and $\lambda_i$ (mean longitude) or $M_i$ (mean anomaly), where $i=1,2$ indicates the particular planet. 

A periodic orbit defines an {\em exact resonance} for the underlying dynamical system and is locked to a mean motion resonance $n_P/n_J=(p+q)/p$, where $n_J$ is the mean motion of the Jupiter and $n_P$ is the mean motion of the other planet. In this paper, we study the $1/2$ (Jupiter is the inner planet $P_1$) or the $2/1$ (Jupiter is the outer planet $P_2$) resonant cases. We distinguish 2/1 from 1/2 resonance because different dynamics is obtained for the restricted problem. However, this distinction is not essential for the general problem. We can define the resonant angles $\sigma_1$ and $\sigma_2$ of the system as:
\begin{equation}
\begin{array}{c}
q\sigma _{1}=(p+q)\lambda _{1}-p\lambda _{2}-q\varpi _{1}\\ 
q\sigma _{2}=(p+q)\lambda _{1}-p\lambda _{2}-q\varpi _{2}\nonumber
\end{array}
\end{equation}
In the following, we refer to the resonant angles $\sigma _1$, $\Delta \varpi$=$\sigma _1-\sigma _2$ and  $\Delta \Omega=\Omega_2-\Omega_1$. For the initial conditions (\ref{xzsym}) the orbital elements $\omega_i$ and $\Omega_i$ can take the  values $\frac{\pi}{2}$ or $\frac{3\pi}{2}$.  For the initial conditions (\ref{xsym}) the orbital elements $\omega_i$ and $\Omega_i$ can take the  values $0$ or $\pi$. Also, for both symmetries the planets at $t=0$ are at their periastron or apoastron. Subsequently, the resonant angles that correspond to a periodic orbit can take values $0$ or $\pi$.

\section{Resonant Periodic Orbits and planetary evolution in their vicinity}
The computation of periodic orbits is based on the satisfaction of the periodicity conditions
$$
\dot x_1(T/2)= 0,\quad \dot x_2(T/2)= 0,\quad \dot{z_2}(T/2)= 0  \quad (xz-\textnormal{symmetry})
$$
or
$$
\dot x_1(T/2)= 0,\quad \dot x_2(T/2)= 0,\quad z_2(T/2)= 0  \quad (x-\textnormal{symmetry})
$$
where $T/2$ is the time of the first crossing of the section plane $y_2=0$. Keeping $x_{10}$ fixed in the set of initial conditions, $\Pi_4$ or $\Pi'_4$, we differentially correct the rest initial conditions, until they satisfy the periodicity conditions with accuracy of 11-12 digits.  
 
As already stated, the stability of a periodic orbit is defined by the 3 pairs of non unit eigenvalues of the monodromy matrix. However, due to the limited accuracy, we cannot be sure, whether the eigenvalues lie on the unit circle, or not. An estimation of the accuracy of computations can be retrieved by the accuracy of computation of the unit eigenvalues which is about 10 digits. Nevertheless, apart from the linear stability, we estimate the evolution stability by using as index the Fast Lyapunov Indicator (FLI), (\citealt{froe97}) and, particularly, we compute the de-trended FLI, (\citealt{voyatzis08}), defined as 
$$
DFLI(t)=log \left ( \frac{1}{t}\max\{|\mathbf{\xi_1}(t)|,|\mathbf{\xi_2}(t)|\} \right ),
$$
where $\mathbf{\xi_i}$ are deviation vectors (initially orthogonal) computed after numerical integration of the variational equations.

\subsection{Examples of periodic orbits}
\begin{figure}
\begin{center}
$\begin{array}{ccc}
\includegraphics[width=6cm]{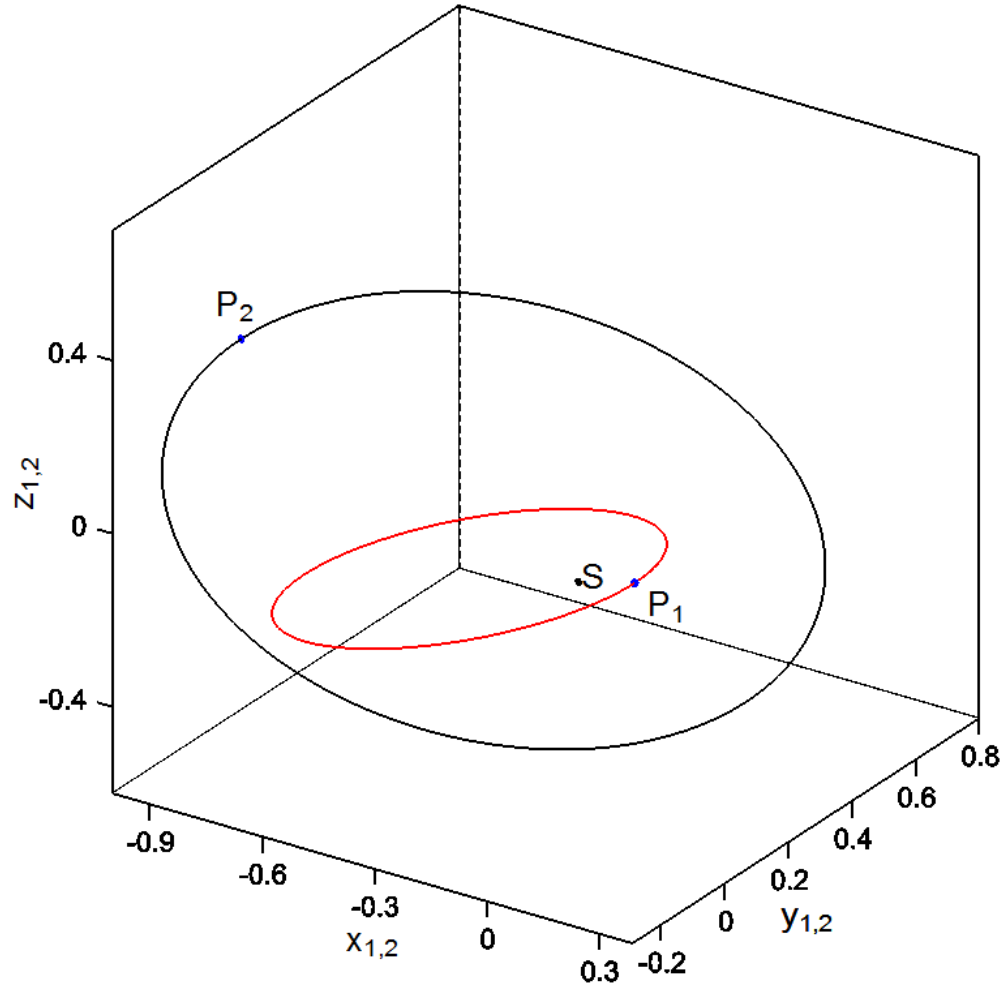}  & \quad&
\includegraphics[width=6cm]{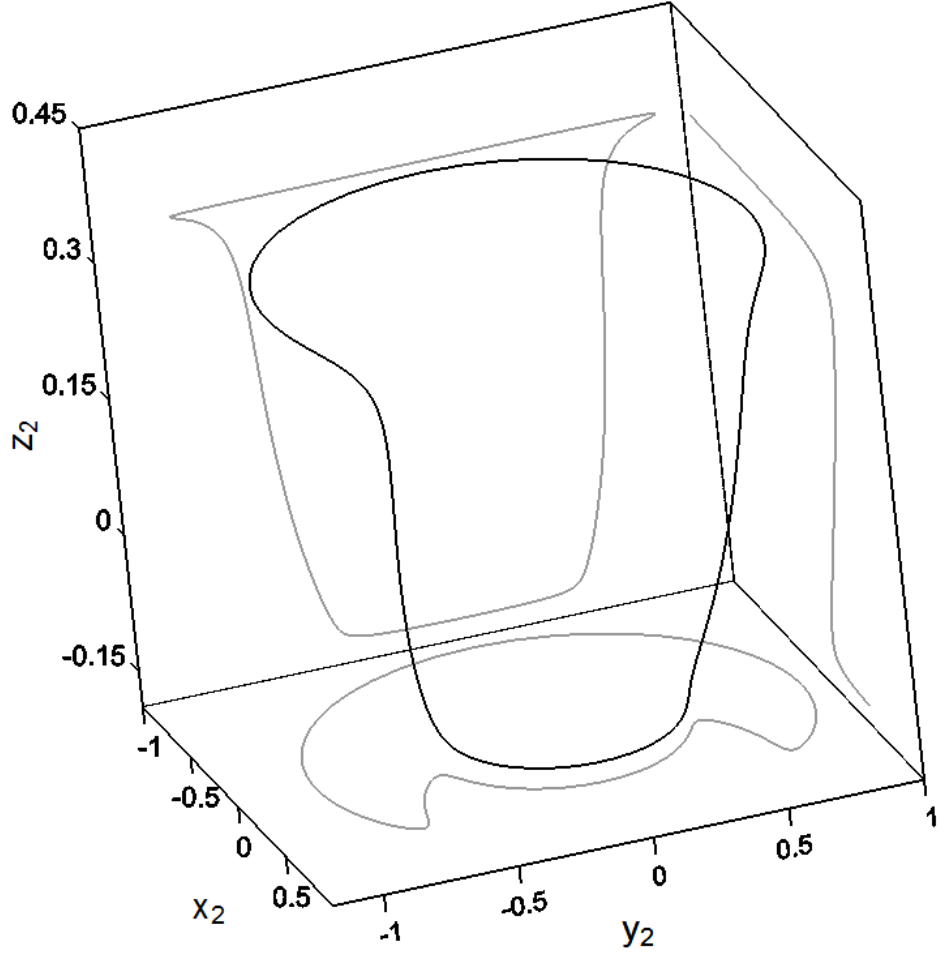} \\
\textnormal{(a)} & \quad & \textnormal{(b)} 
\end{array} $
\end{center}
\caption{A symmetric periodic orbit with respect to the $xz$-plane. The planetary mass ratio $\rho=m_2/m_1$ equals to $0.4$. {\bf a} The planetary orbits in the inertial frame (the mutual inclination is $\Delta i\approx 33^\circ$). {\bf b} The periodic orbit in the rotating frame together with its projection to the planes of the frame.}
\label{XZx}
\end{figure}

In Fig. \ref{XZx}, we illustrate an 1/2 resonant stable $xz$-symmetric periodic orbit\footnote{This orbit belongs to the family $F^{1/2}_{g1,i}$, see Section \ref{SII}.} in the inertial frame and in the projection space $x_2y_2z_2$ of the rotating frame and for planetary masses $m_1=10^{-3}$ and $m_2=4\times 10^{-4}$. The initial conditions correspond to the planetary orbital elements
\begin{equation}
\begin{array}{cccccc}
a_1=0.45, & e_1=0.66,&i_1=12.79^{\circ}, &\Omega_1=270^{\circ},&\omega_1=90^{\circ},& M_1=0^{\circ}\\
a_2=0.71, & e_2=0.34,&i_2=20.46^{\circ}, &\Omega_2=90^{\circ},&\omega_2=270^{\circ},& M_2=180^{\circ}.\nonumber
\label{XZoeincon}
\end{array}
\end{equation} 
Its $xz$ symmetry is revealed by the projections of the orbit in the planes of the space $x_2y_2z_2$ (see Fig. \ref{XZx}b). The corresponding resonant angles take the values $\sigma _1=0^{\circ}$, $\Delta \varpi=0^{\circ}$ and $\Delta\Omega=180^{\circ}$. 

\begin{figure}
\begin{center}
$\begin{array}{ccc}
\includegraphics[width=6cm]{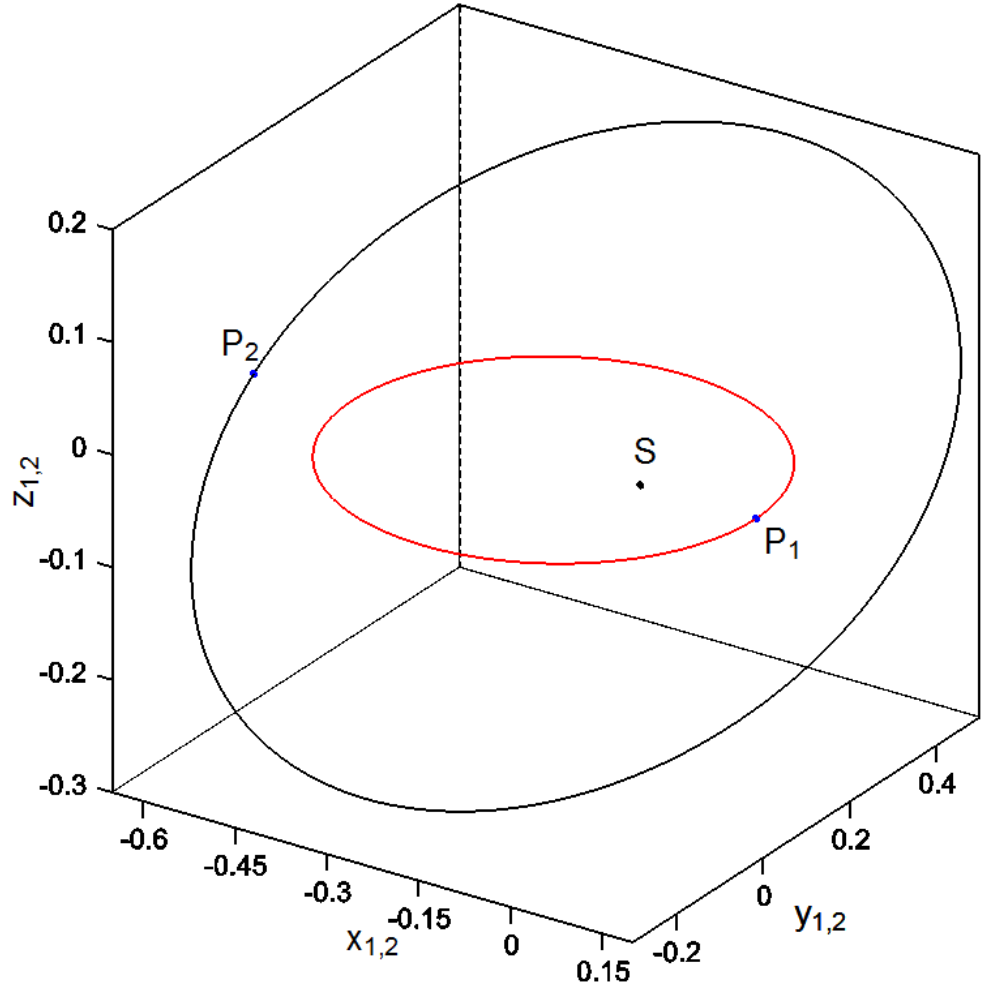}  & \quad&
\includegraphics[width=6cm]{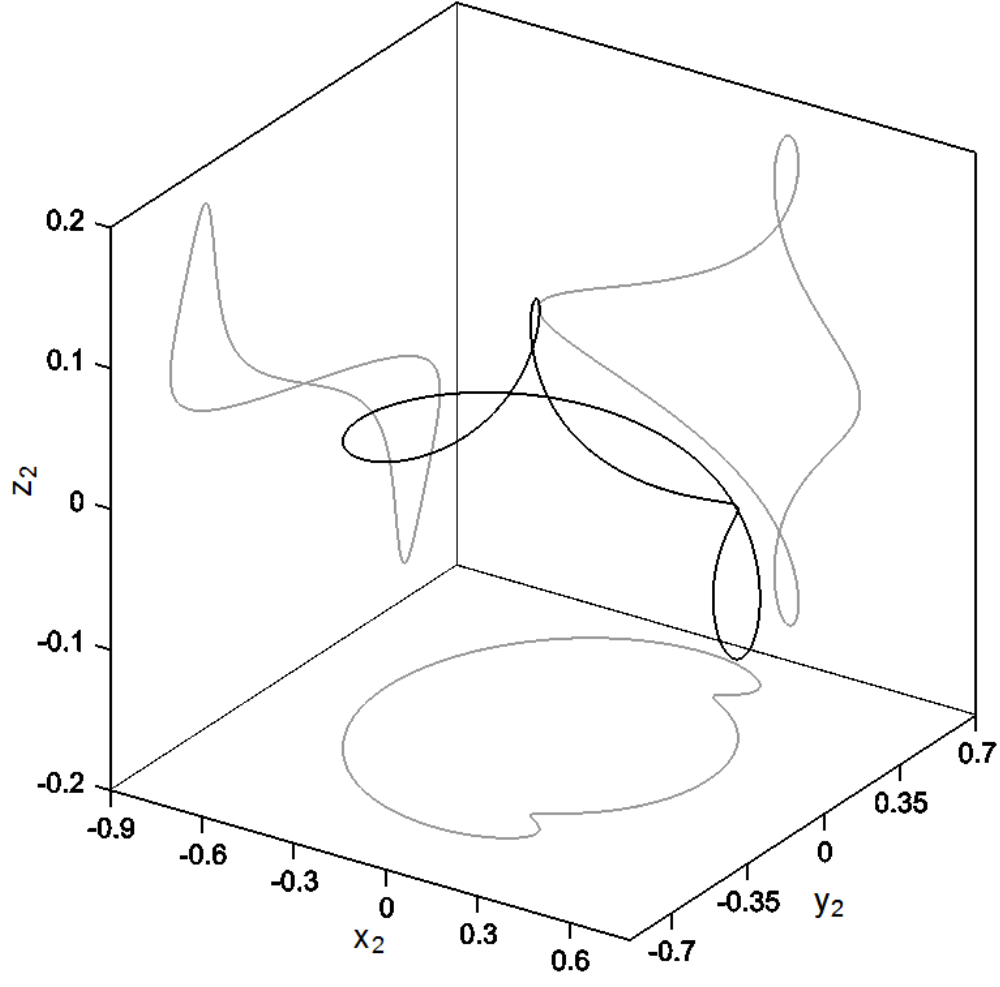} \\
\textnormal{(a)}  & \quad & \textnormal{(b)} 
\end{array} $
\end{center}
\caption{A symmetric periodic orbit with respect to the $x$-axis. The planetary mass ratio $\rho=m_2/m_1$ equals to $0.01$. {\bf a} The periodic orbits of the planets in the inertial frame (the mutual inclination is $\Delta i\approx 20^\circ$). {\bf b} The periodic orbit of the third body together with its projection to the planes in the rotating frame.}
\label{Xx}
\end{figure}

In Fig. \ref{Xx}, we show an 1/2 resonant $x$-symmetric unstable periodic orbit\footnote{This orbit belongs to the family $G^{1/2}_{g1,i}$, see Section \ref{SII}.}. The planetary masses are $m_1=10^{-3}$ and $m_2=10^{-5}$ and the initial conditions correspond to the planetary orbital elements
\begin{equation}
\begin{array}{cccccc}
a_1=0.33, & e_1=0.42,&i_1=0.26^{\circ}, &\Omega_1=0^{\circ},&\omega_1=0^{\circ},& M_1=0^{\circ}\\
a_2=0.52, & e_2=0.20,&i_2=19.79^{\circ}, &\Omega_2=0^{\circ},&\omega_2=0^{\circ},& M_2=180^{\circ}.\nonumber
\label{Xoeincon}
\end{array}
\end{equation}
Its $x$ symmetry is revealed by the projections of the orbit in the planes of the space $x_2y_2z_2$ (see Fig. \ref{Xx}b). The corresponding resonant angles take the values $\sigma _1=0^{\circ}$, $\Delta \varpi=0^{\circ}$ and $\Delta\Omega=0^{\circ}$.

\begin{figure}
$\begin{array}[t]{cc}
\includegraphics[width=5cm]{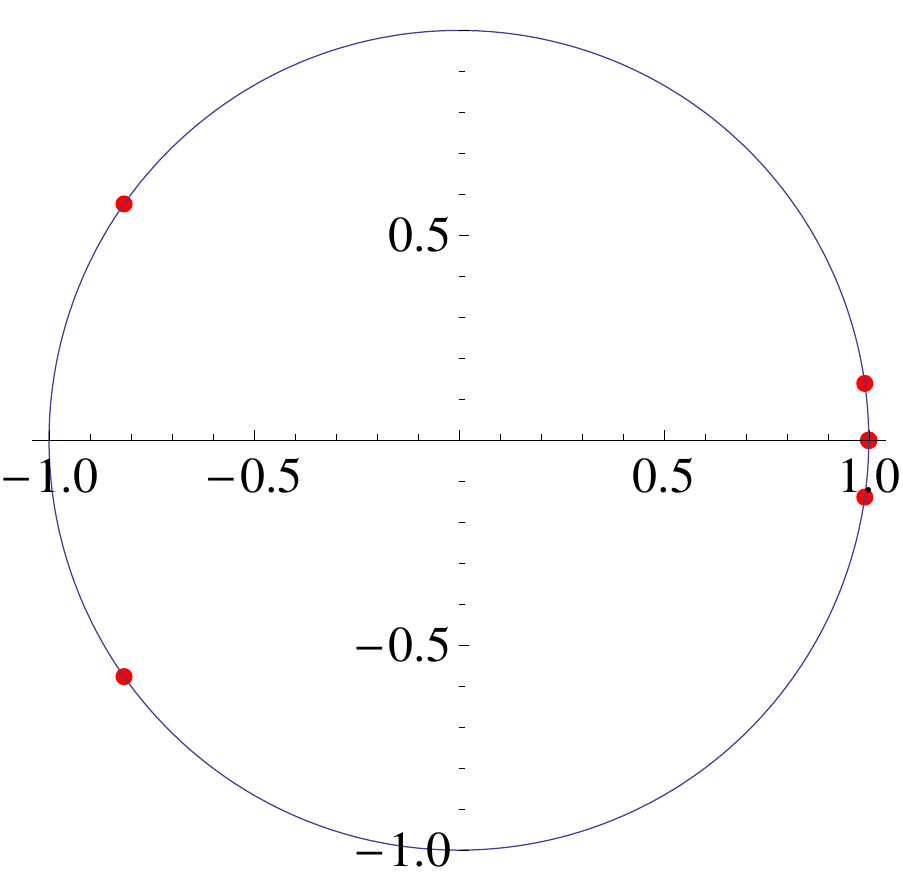}  &
\qquad \includegraphics[width=5cm]{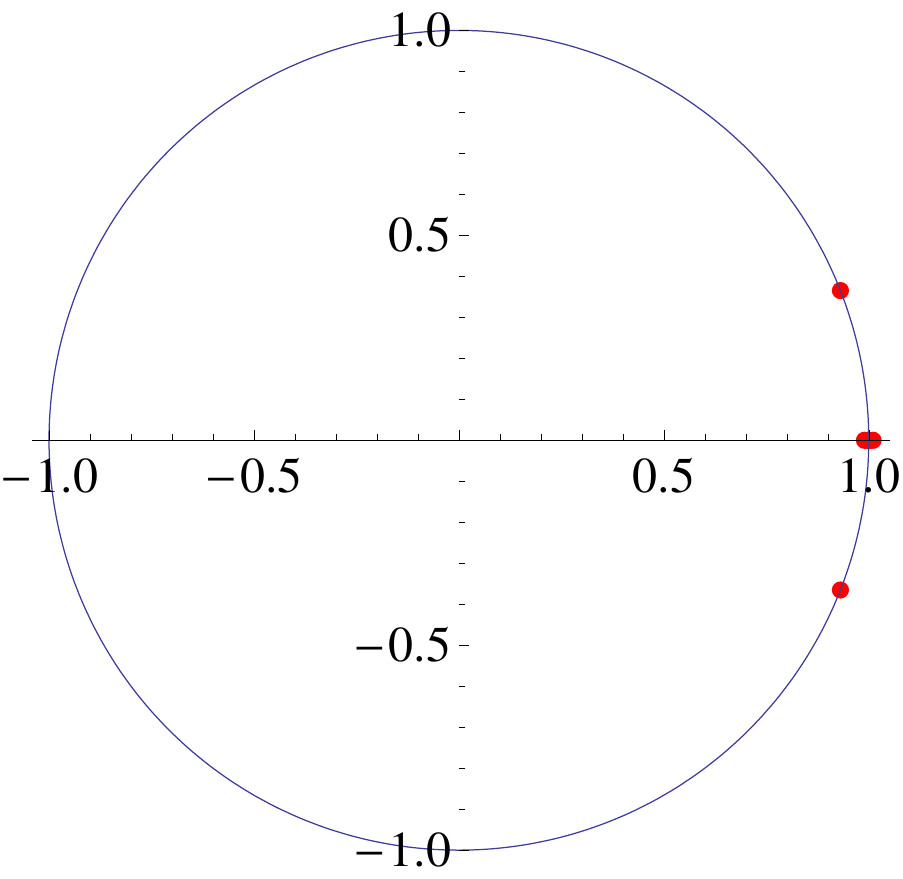}  \\
\textnormal{(a)} &  \textnormal{(b)} \\ 
\includegraphics[width=5cm]{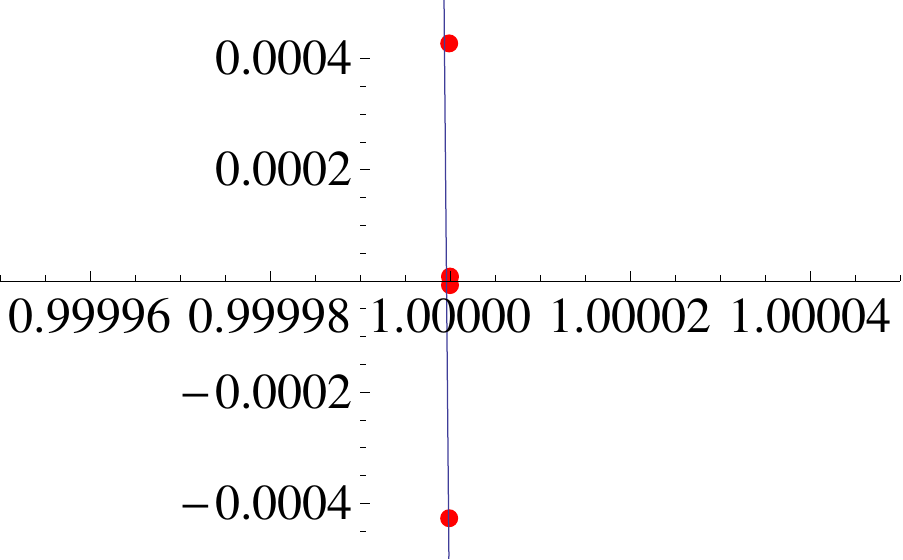} &
\qquad \includegraphics[width=5cm]{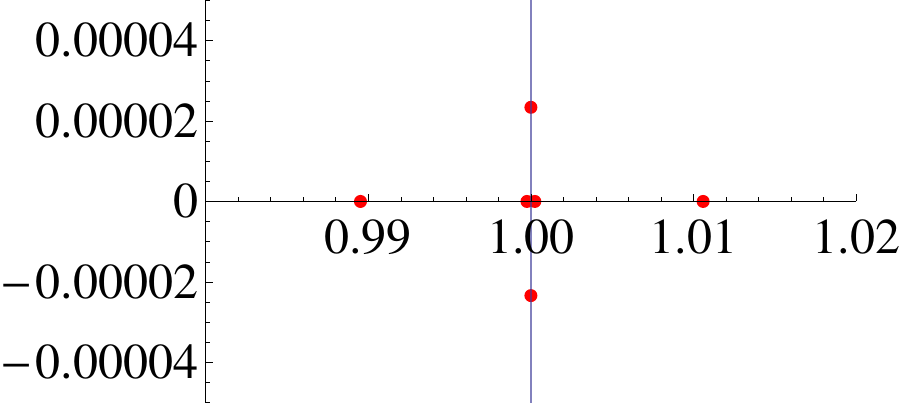} \\
\textnormal{(c)} & \textnormal{(d)}\\
\end{array} $
\caption{Presentation of the linear stability. {\bf a} The eigenvalues of the stable periodic orbit of Fig. \ref{XZx}. {\bf b} The eigenvalues of the unstable periodic orbit of Fig. \ref{Xx}. {\bf c} and {\bf d} Magnifications of cases (a) and (b), respectively.}
\label{eigfliXZX}
\end{figure}

The distribution of eigenvalues of the linear stability analysis is presented in Fig. {\ref{eigfliXZX}}. Regarding the stable $xz$-symmetric periodic orbit, all eigenvalues lie on the unit circle (Fig. {\ref{eigfliXZX}}a). One pair (of non unit eigenvalues) is very close to 1 and this is shown in the corresponding magnification (Fig. {\ref{eigfliXZX}}c). Concerning the unstable $x$-symmetric orbit there exists one pair of real eigenvalues indicating the instability (Fig. {\ref{eigfliXZX}}b). Also, one pair of eigenvalues is very close to 1 (see the magnification in Fig. {\ref{eigfliXZX}}d).  

\begin{figure}
\begin{center}
$\begin{array}{l}
\includegraphics[width=10cm]{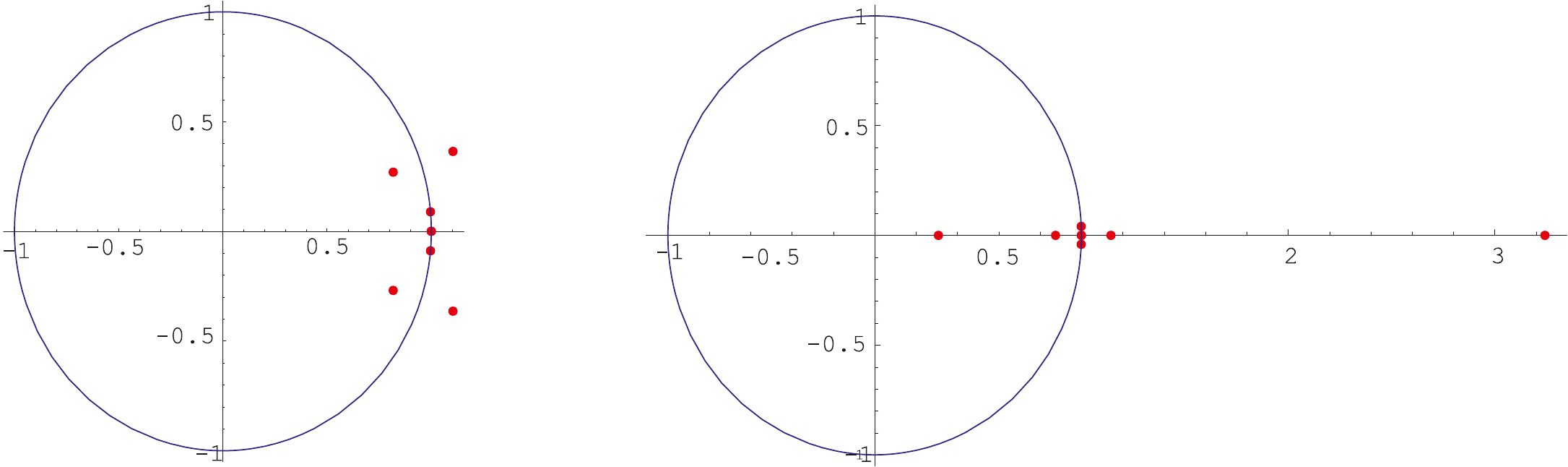}\\
\textnormal{\hspace{1cm} (a) \hspace{4cm} (b)}  \\
\end{array} $
\end{center}
\caption{An example of transition from {\bf a} complex instability to {\bf b} real instability along the family $F^{1/2}_{g2,i}$ for $\rho=0.4$ (see Section \ref{SII}).}
\label{eig}
\end{figure}

In Fig.{\ref{eig}}, we present two different distributions of eigenvalues obtained for two orbits that belong to the family $F^{1/2}_{g2,i}$ (see Section \ref{SII}). In the first case, we have the presence of complex instability, while in the second one, two pairs of real eigenvalues exist. Along a family of periodic orbits the eigenvalues move smoothly either on the unit circle, or the real axis and can either enter the real axis, or the unit circle. Therefore, the stability type can change along the families. In the rest of the paper, we will indicate whether an orbit is stable or unstable without declaring the particular type of instability.

\subsection{Stability of evolution in the vicinity of periodic orbits}
It is known, that in the phase space, a stable periodic orbit is surrounded by invariant tori, where the motion is regular and stays in the neighbourhood of the periodic orbit. In the case of an unstable periodic orbit, chaotic domains exist at least close to the periodic orbit. In general, starting from the neigbourhood of an unstable periodic orbit, the evolution of planetary system shows instabilities and chaotic behaviour.  Numerical integrations show that such instabilities appear stronger as far as highly eccentric orbits are concerned (\citealt{voyatzis08}). Hereafter, we present some numerical results that show the above mentioned behaviour of orbits. 

\begin{figure}
\begin{center}
$\begin{array}{cc}
\includegraphics[width=6cm]{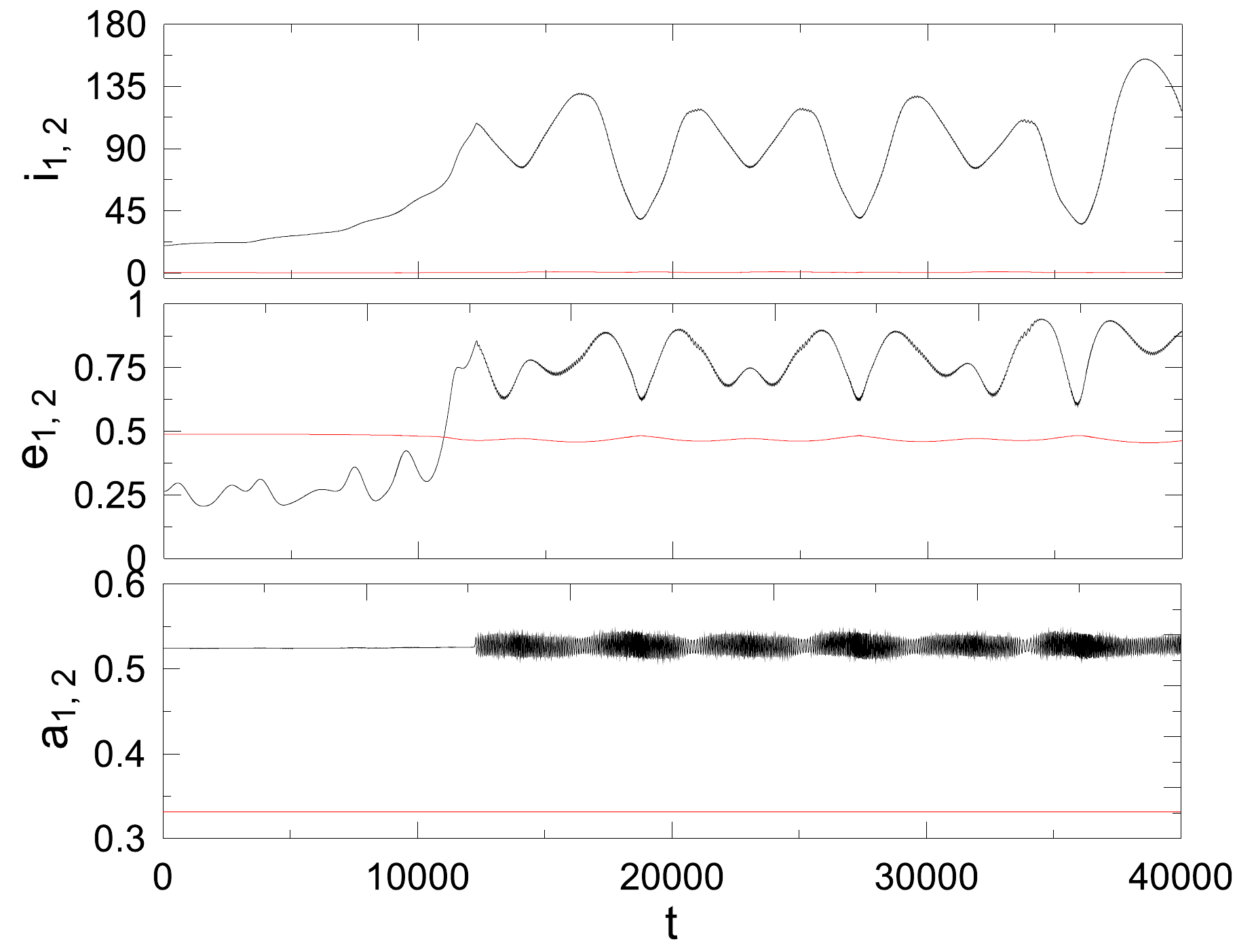} &
\includegraphics[width=6cm]{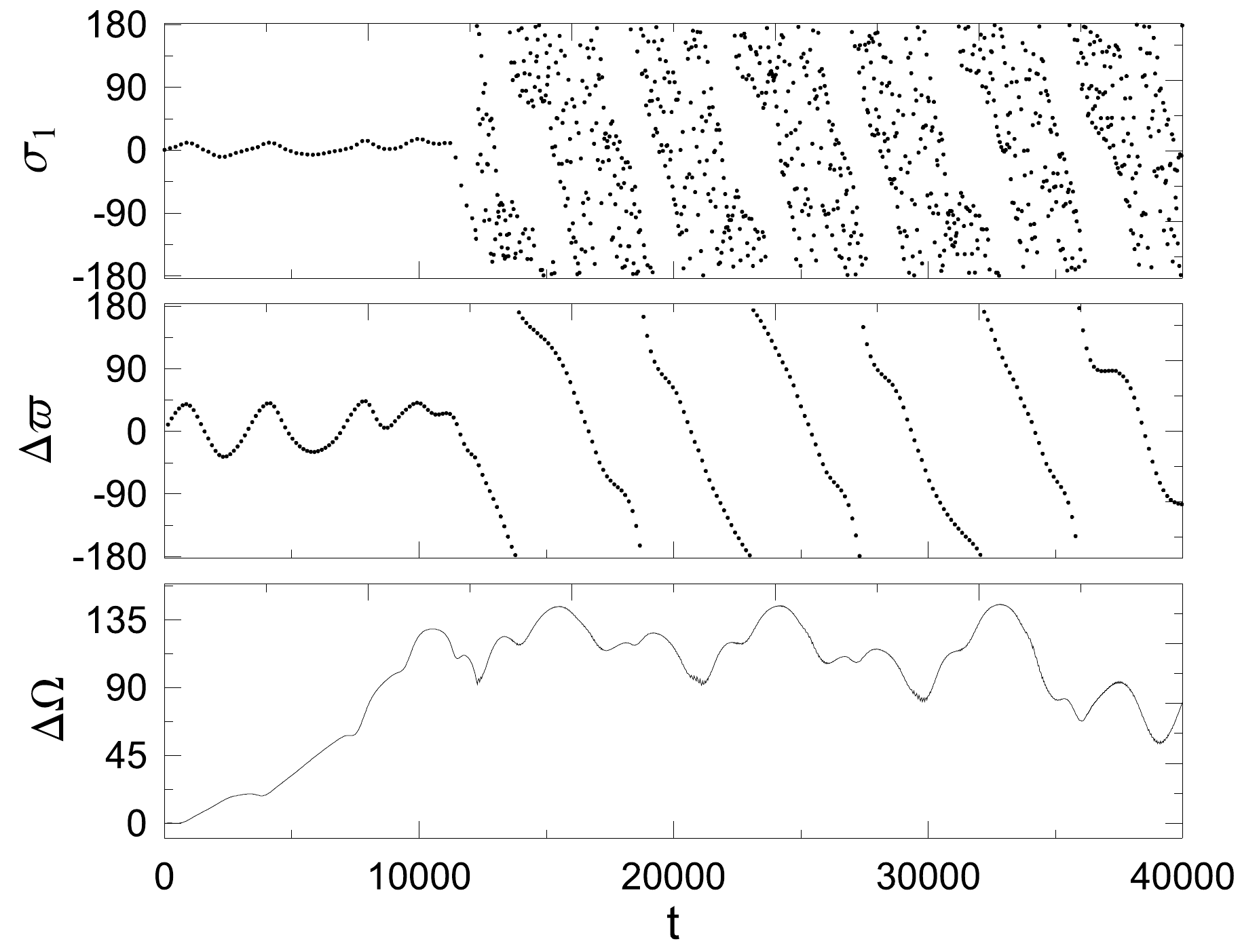}\\ 
\textnormal{(a)} & \textnormal{(b)}
\end{array} $
\end{center}
\caption{Evolution of orbital elements and resonant angles starting with initial conditions (of limited accuracy) of the unstable periodic orbit of Fig. \ref{Xx}. Red and black line stands for  Jupiter ($P_1$) and planet $P_2$, respectively.}
\label{001u}
\end{figure}

We consider the initial conditions of the unstable $x$-symmetric periodic orbit given in the previous section (Fig. \ref{Xx}). Since the initial conditions of the periodic orbit are not exact, the instability leads the evolution far from the periodic orbit. In Fig. \ref{001u}, we present the evolution of the orbital elements $a_i$, $e_i$ and $i_i$ and the resonant angles $\sigma _1$, $\Delta \varpi$ and $\Delta \Omega$. We observe that the eccentricity and the inclination of the heavier planet $P_1$ remain almost constant during the whole time of integration. However, the planet $P_2$ after the critical time $12\times 10^3$ t.u. destabilizes, its eccentricity increases and then oscillates around a high value. Also, its inclination takes large values and shows large oscillation around $90^\circ$, i.e. its motion turns from prograde to retrograde and vice versa. A significant oscillation of the semimajor axis of $P_2$ after $12\times 10^3$ t.u. is also, apparent, though the system remains in the mean motion resonance. The resonant angles $\sigma _1$ and $\Delta \varpi$ initially librate arround $0^{\circ}$, while $\Delta \Omega$ increases. After the critical time, $\sigma _1$ and $\Delta \varpi$ start to rotate, while $\Delta \Omega$ librates irregularly in the domain $[52^\circ,145^\circ]$ . The computation of the DFLI (see Fig. \ref{dfli414}, case II) shows clearly the chaotic nature of the evolution. 

\begin{figure}
\begin{center}
$\begin{array}{cc}
\includegraphics[width=6cm]{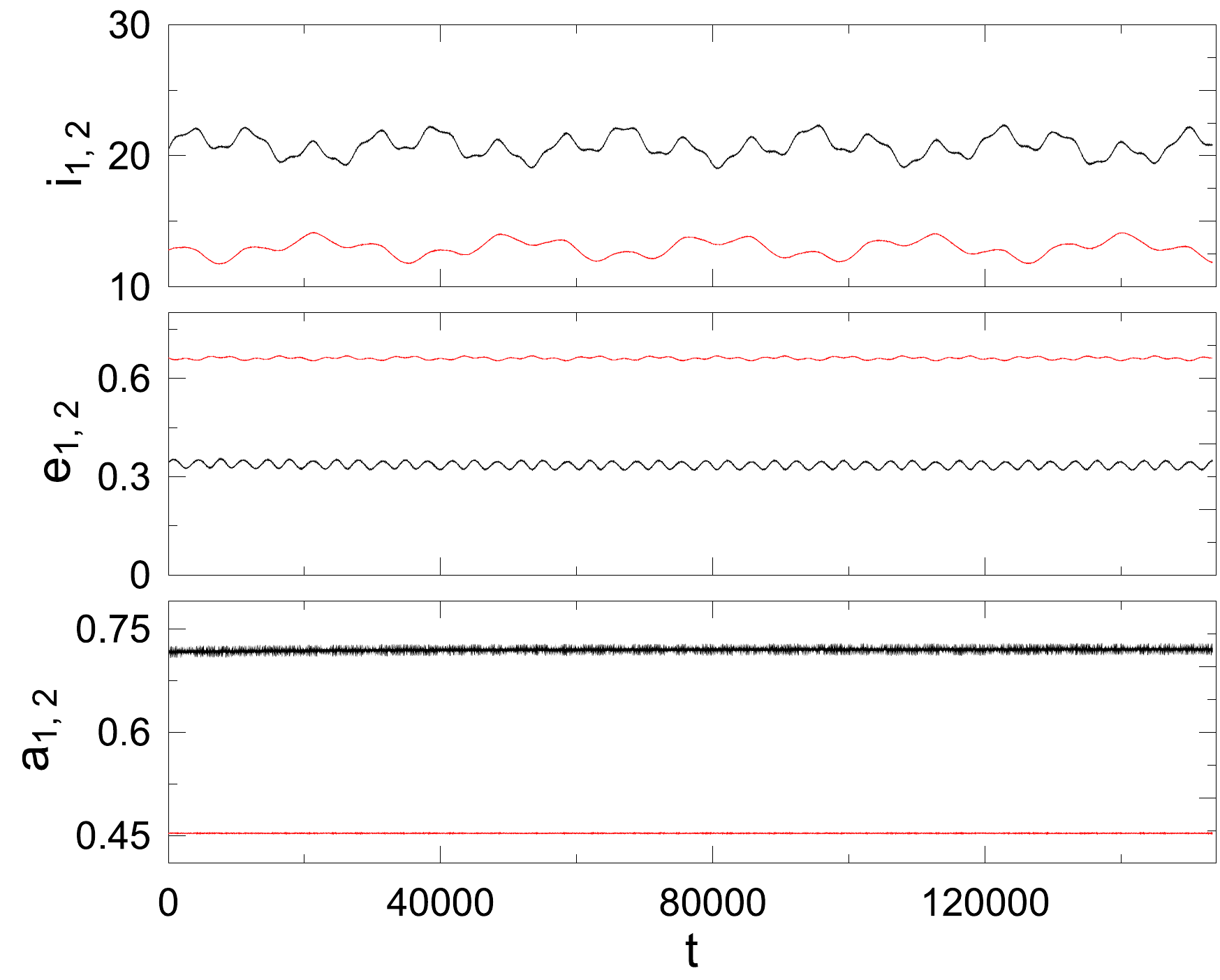} &
\includegraphics[width=6cm]{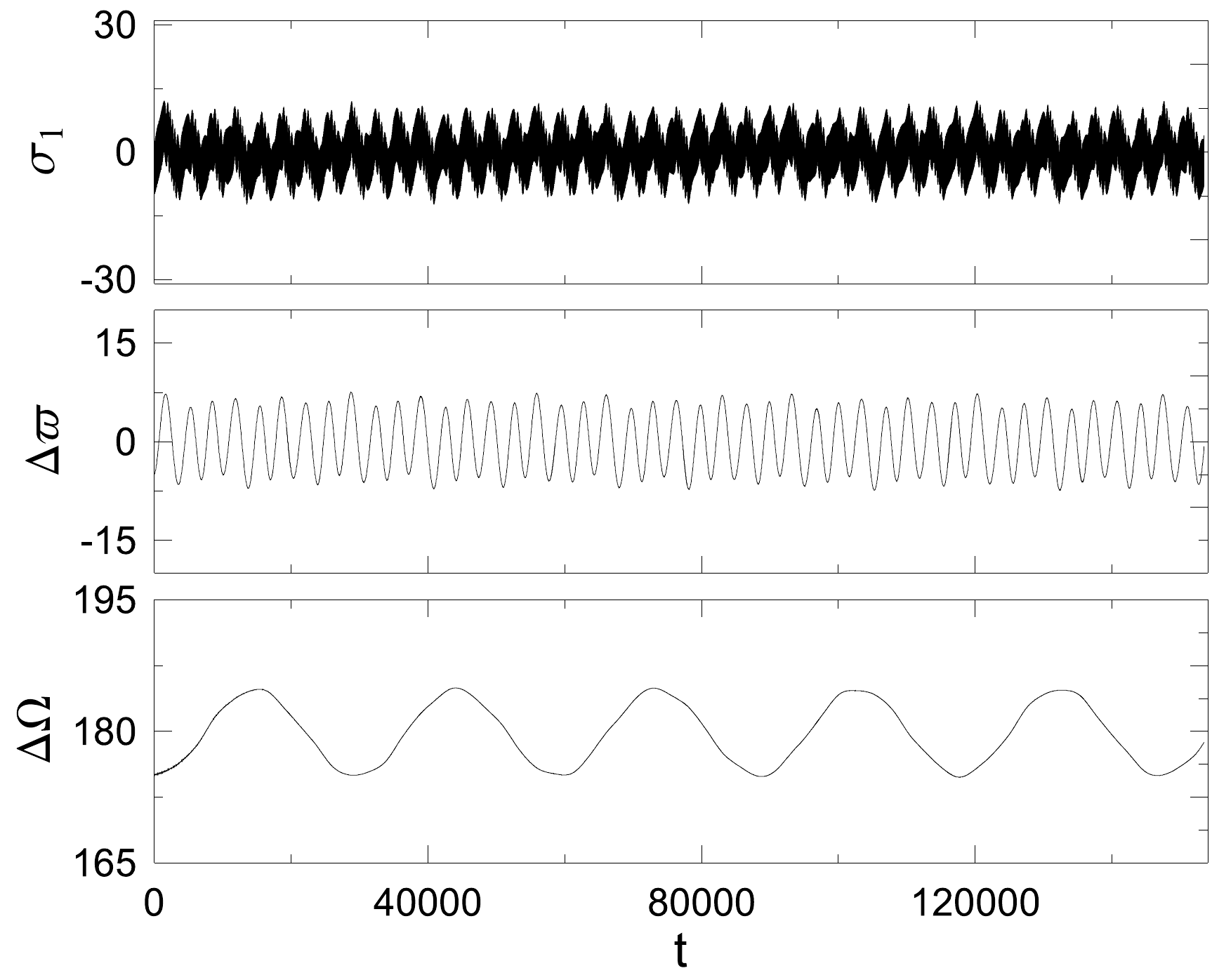}\\ 
\textnormal{(a)} &\textnormal{(b)}  
\end{array} $
\end{center}
\caption{Evolution of orbital elements and resonant angles of an orbit which starts in the vicinity of the stable periodic orbit of Fig. \ref{XZx}. Red and black line stands for Jupiter ($P_1$) and planet $P_2$, respectively.}
\label{04s}
\end{figure}

If we start with the initial conditions given for the stable $xz$-symmetric periodic orbit (Fig. \ref{XZx}), we will obtain that the orbital elements $a_i$, $e_i$,  $i_i$ and the resonant angles remain constant for ever (assuming sufficient integration accuracy). Starting near the periodic orbit, and particularly considering a deviation of $5^\circ$ in $\Omega_1$ (i.e. we set now $\Omega_1=275^\circ$), we obtain a regular evolution as it is shown in Fig. \ref{04s}.
The inclination and the eccentricity values of both planets oscillate arround their initial values,  while the resonant angles $\sigma_1$, $\Delta \varpi$ and $\Delta\Omega$ show librations around the values which correspond to the periodic orbit, namely $0^\circ$, $0^\circ$ and $180^\circ$, respectively. Now, the DFLI does not increase (Fig. \ref{dfli414}, case I)

\begin{figure}
\begin{center}
$\begin{array}{cc}
\includegraphics[width=6cm]{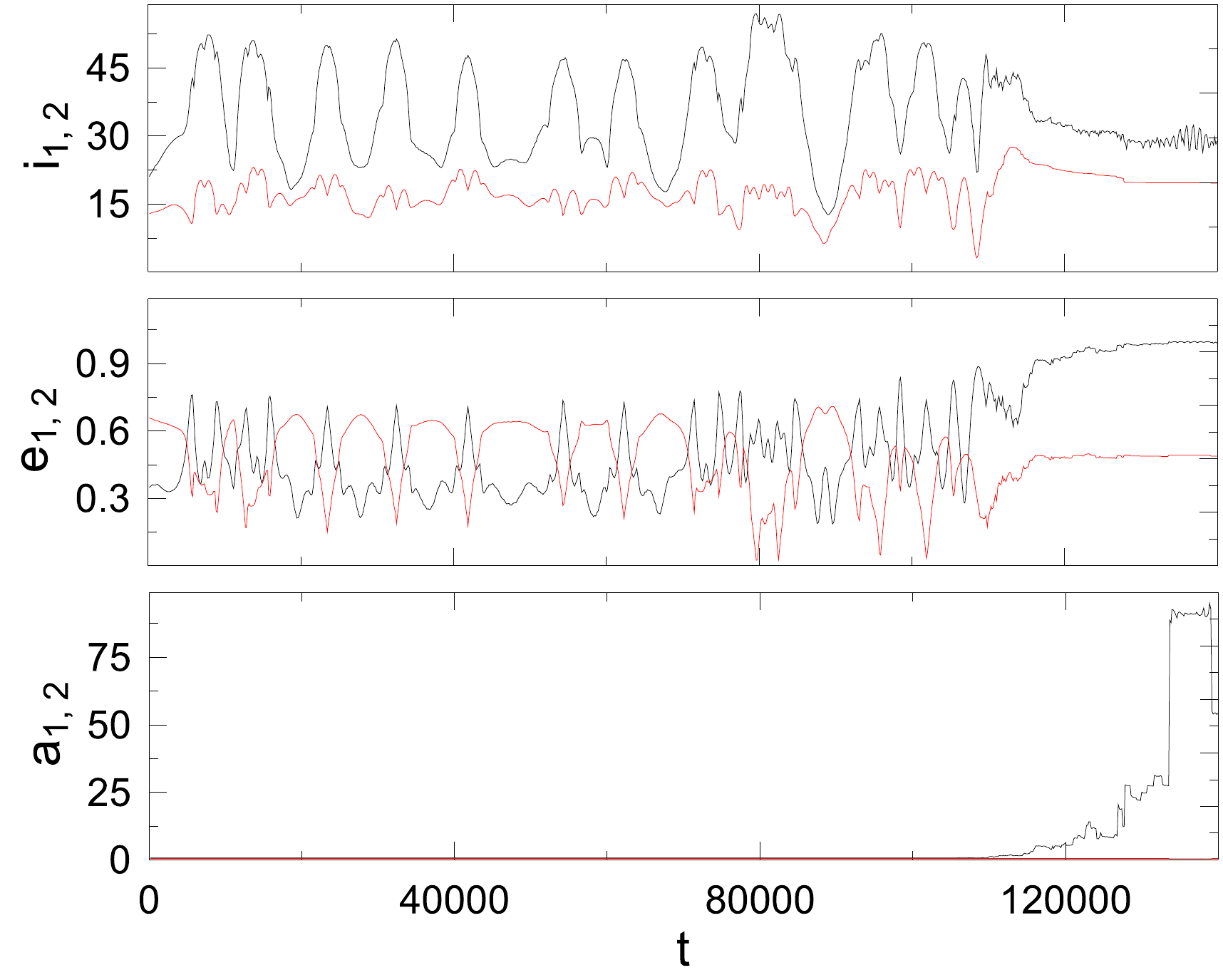} &
\includegraphics[width=6cm]{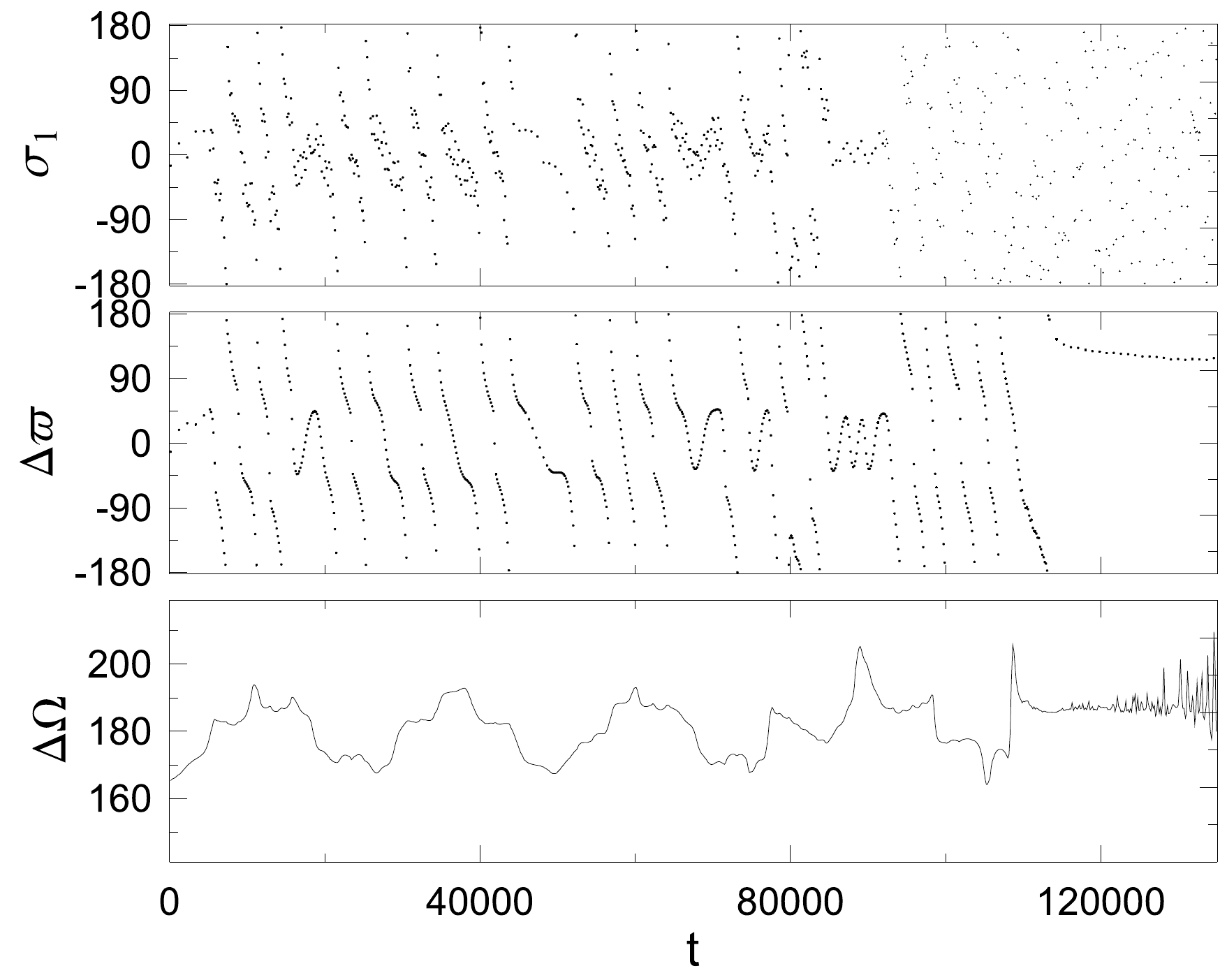}\\ 
\textnormal{(a)} &\textnormal{(b)}
\end{array} $
\end{center}
\caption{Evolution of orbital elements and resonant angles of an orbit that starts relatively far from the stable periodic orbit of Fig. \ref{XZx}. The orbit is chaotic and after a time span close encounters between planets cause the scattering of the outer planet $P_2$. Red and black line stands for  Jupiter ($P_1$) and planet, $P_2$, respectively.}
\label{04u}
\end{figure}

\begin{figure}
\begin{center}
\includegraphics[width=6cm]{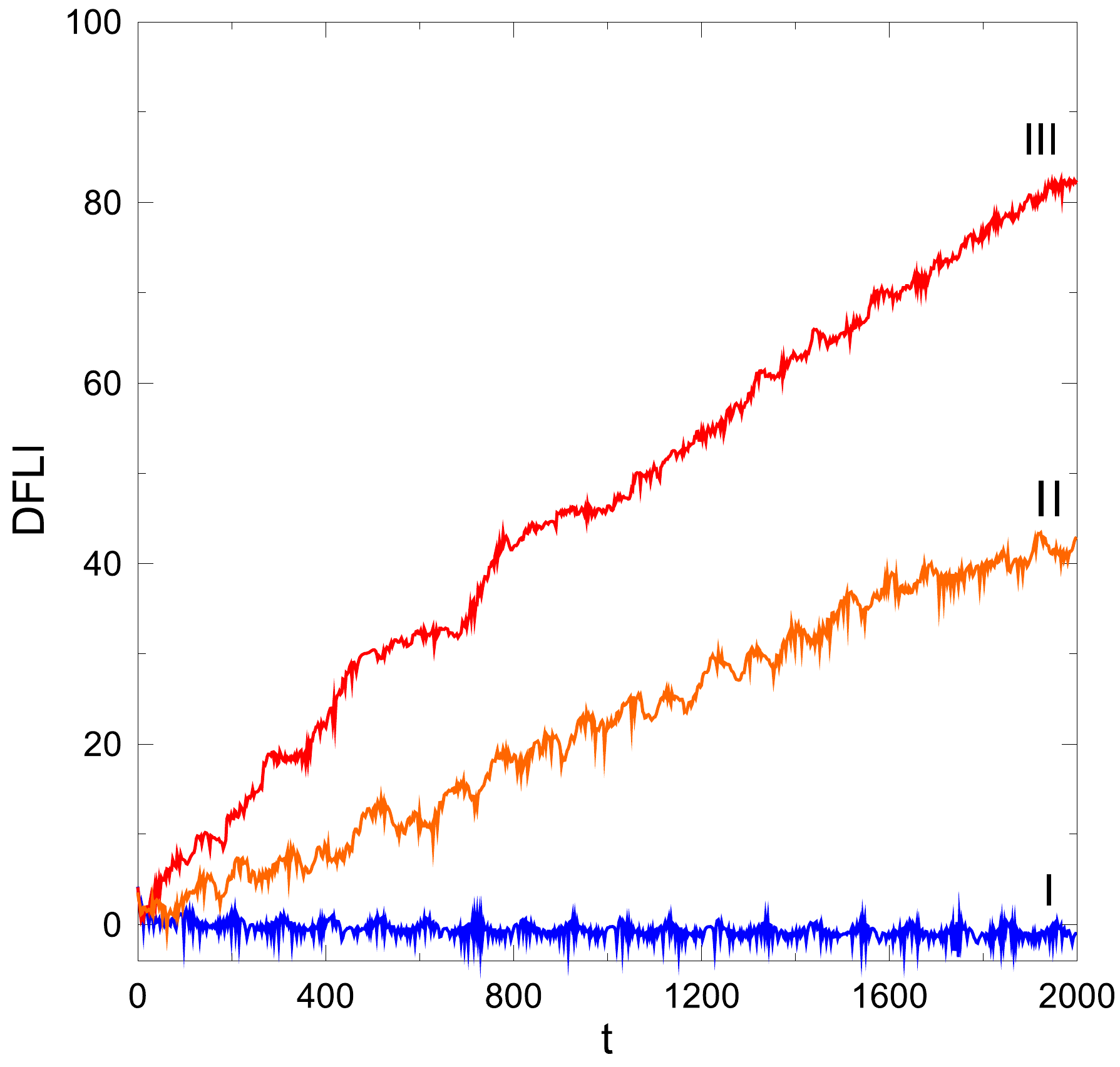} 
\caption{Computation of the DFLI. Case I corresponds to the regular evolution of  Fig. \ref{04s}, case II and case III to the chaotic evolution presented in Figs. \ref{001u} and \ref{04u}, respectively.}
\label{dfli414}
\end{center}
\end{figure}

If we start not sufficiently close to the stable periodic orbit, the stability of evolution is not guaranteed. We consider a larger deviation, compared to the deviation used in the previous case. Particularly, we consider a deviation of  $15^\circ$ in $\Omega_1$ and we integrate the orbit. The results are shown in Fig. \ref{04u}. The chaotic behaviour of the evolution is obvious and finally the planet $P_2$ is scattered. The resonant angles $\sigma_1$ and $\Delta \varpi$ rotate, but it is remarkable that $\Delta\Omega$ librates irregularly around $180^\circ$.  The DFLI clearly shows the chaotic evolution (Fig. \ref{dfli414}, case III).                   

\section{Families of periodic orbits}
We compute families of periodic orbits for the general spatial problem (3D-GTBP) by analytical continuation of periodic orbits either from the spatial circular restricted problem (3D-CRTBP), called {\em Scheme I}, or from the planar general problem (2D-GTBP), called {\em Scheme II}. The presentation of such families will be given in projection planes defined by the planetary eccentricity and inclination values.

\begin{figure}
\begin{center}
$\begin{array}{cc}
\includegraphics[width=6cm]{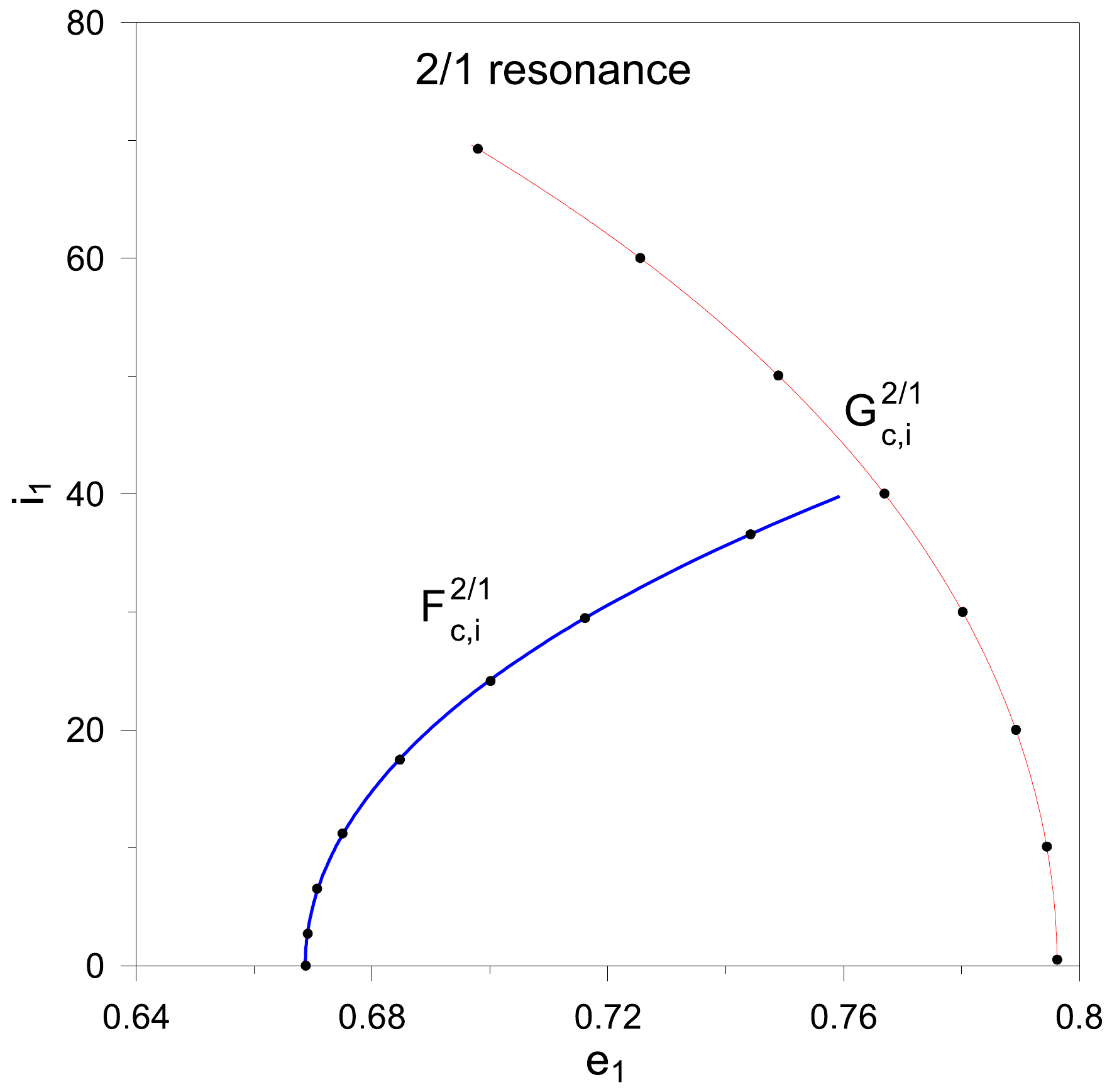} &
\includegraphics[width=6cm]{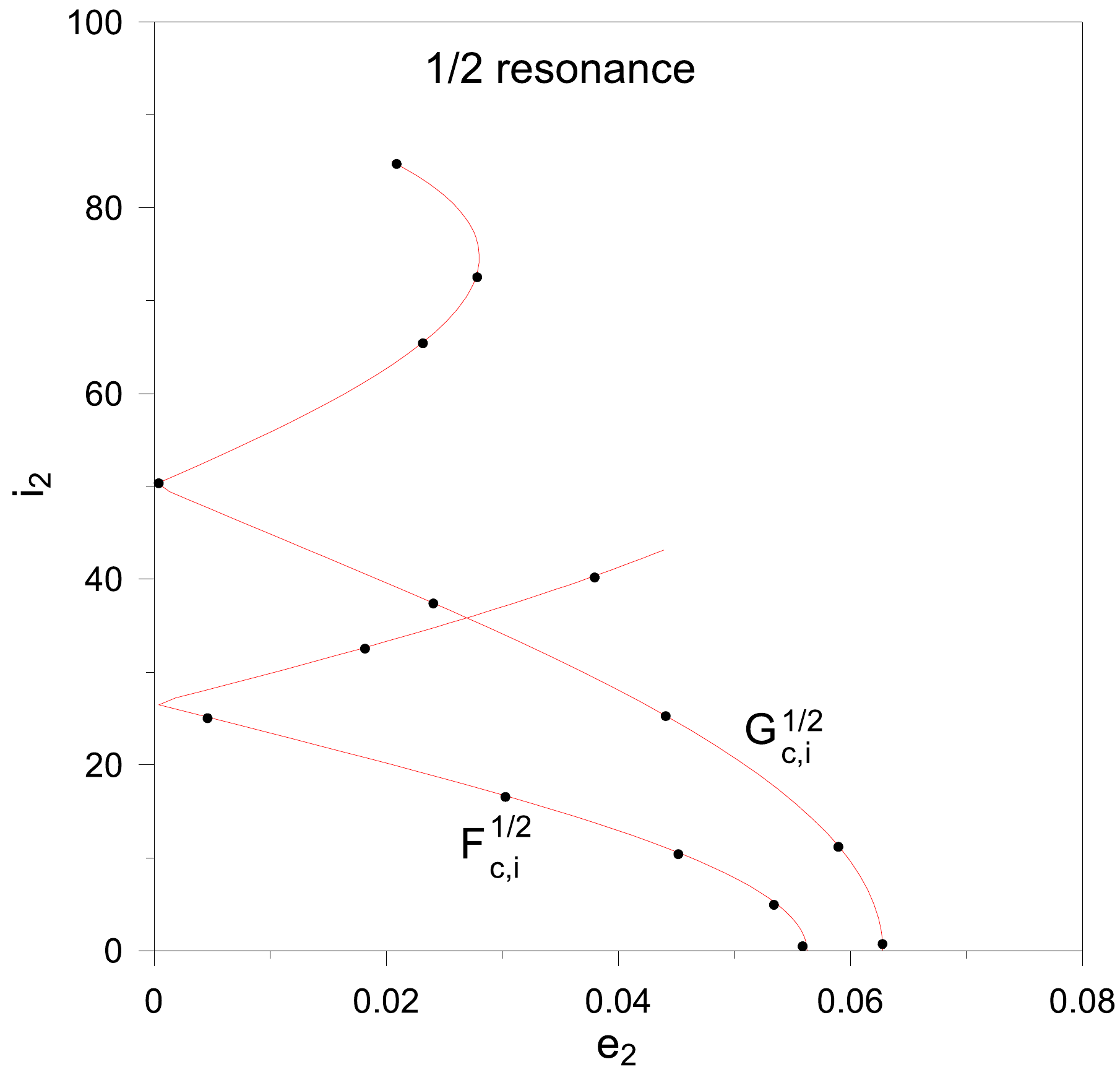}\\ 
\textnormal{(a)} & \textnormal{(b)}
\end{array} $
\end{center}
\caption{Symmetric periodic orbits of the 3D-CRTBP in {\bf a} 2:1 resonance and {\bf b} 1:2 resonance. Blue color corresponds to stable periodic orbits and the red one to unstable orbits. The initial conditions of orbits presented by dots used as starting points for the continuation in the general problem.}
\label{3DC}
\end{figure}

\subsection{Scheme I: Continuation from 3D-CRTBP to 3D-GTBP}

We consider the restricted problem with Jupiter moving in a circular planar motion with period $2\pi$ and a massless planet which moves in the space. In this problem, we can obtain families of periodic orbits which bifurcate from vertical critical periodic orbits of the planar restricted problem (\citealt{hen}). These families exist either when Jupiter is the inner, or the outer planet (resonance either 1/2, or 2/1, respectively). 

Following the notation of \citet{kotvoy05}, families of periodic orbits which are symmetric with respect to the $xz$-plane are denoted by $F$ and families of periodic orbits symmetric with respect to the $x$-axis are denoted by $G$. The subscript $c$ or $e$ indicates the spatial restricted problem (circular or elliptic, respectively) from which these families bifurcate. Moreover, $m$ indicates the continuation with respect to the mass of the initially massless body. Also, the corresponding resonance is indicated by a superscript. 

We computed and present the families of 2/1 and 1/2 resonant symmetric periodic orbits of the 3D-RTBP in Fig. \ref{3DC} (see also \citet{hadjvoy00} and \citet{kot05}, respectively). We denote these families by $F_{c,i}^{2/1}$, where the subscript $i$ indicates the continuation from plane to space (see Section \ref{SII}). For the 2/1 resonance such families correspond to relatively high eccentricity values of the massless planet $P_1$. Conversely, in the 1/2 resonance all periodic orbits correspond to almost circular orbits of the massless planet $P_2$ and are unstable.      

It is known that the periodic orbits of the 3D-RCTBP of period $T$ can be analytically continued, by increasing the planetary mass of the initially massless body to a periodic orbit of the 3D-GTBP of the same symmetry, provided that $T\neq 2k\pi$, where $k$ is an integer (\citealt{ichmich80}). This is the case for all orbits of the families given in Fig. \ref{3DC}. For the particular computations presented hereafter, we continue analytically the periodic orbits indicated by the dots in Fig. \ref{3DC} with respect to the mass $m_1$ (inner planet) or $m_2$ (outer planet) according to the resonance 2/1 or 1/2, respectively. These periodic orbits  correspond to some initial inclination value of the massless body, $i_{10}$ or $i_{20}$, and identify the generated families. 

\begin{figure}
\begin{center}
$\begin{array}{cc}
\includegraphics[width=6cm]{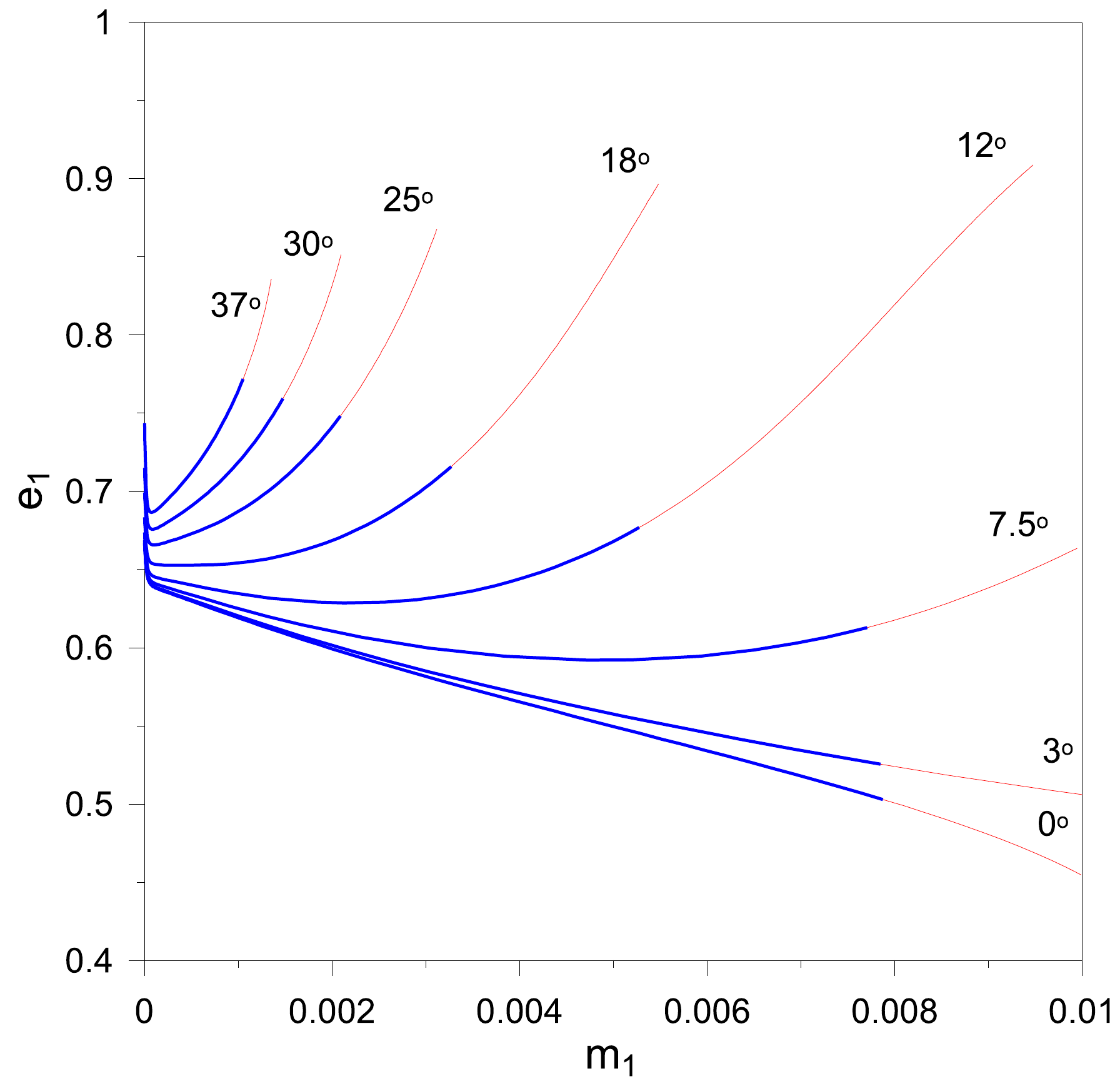} &\includegraphics[width=6cm]{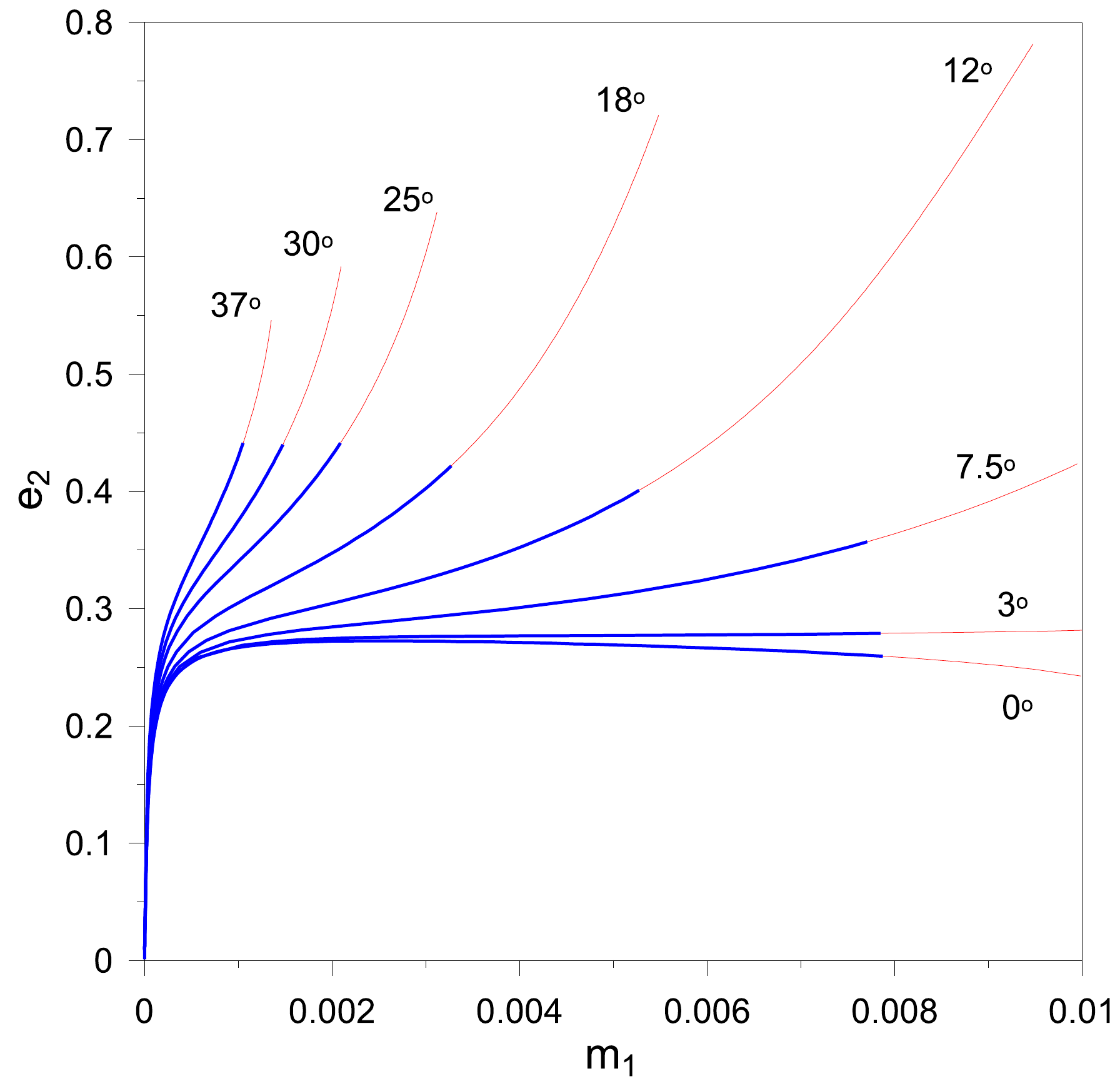}\\
\textnormal{(a)} &\textnormal{(b)}  
\end{array} $
\end{center}
\begin{center}
$\begin{array}{cc}
\includegraphics[width=6cm]{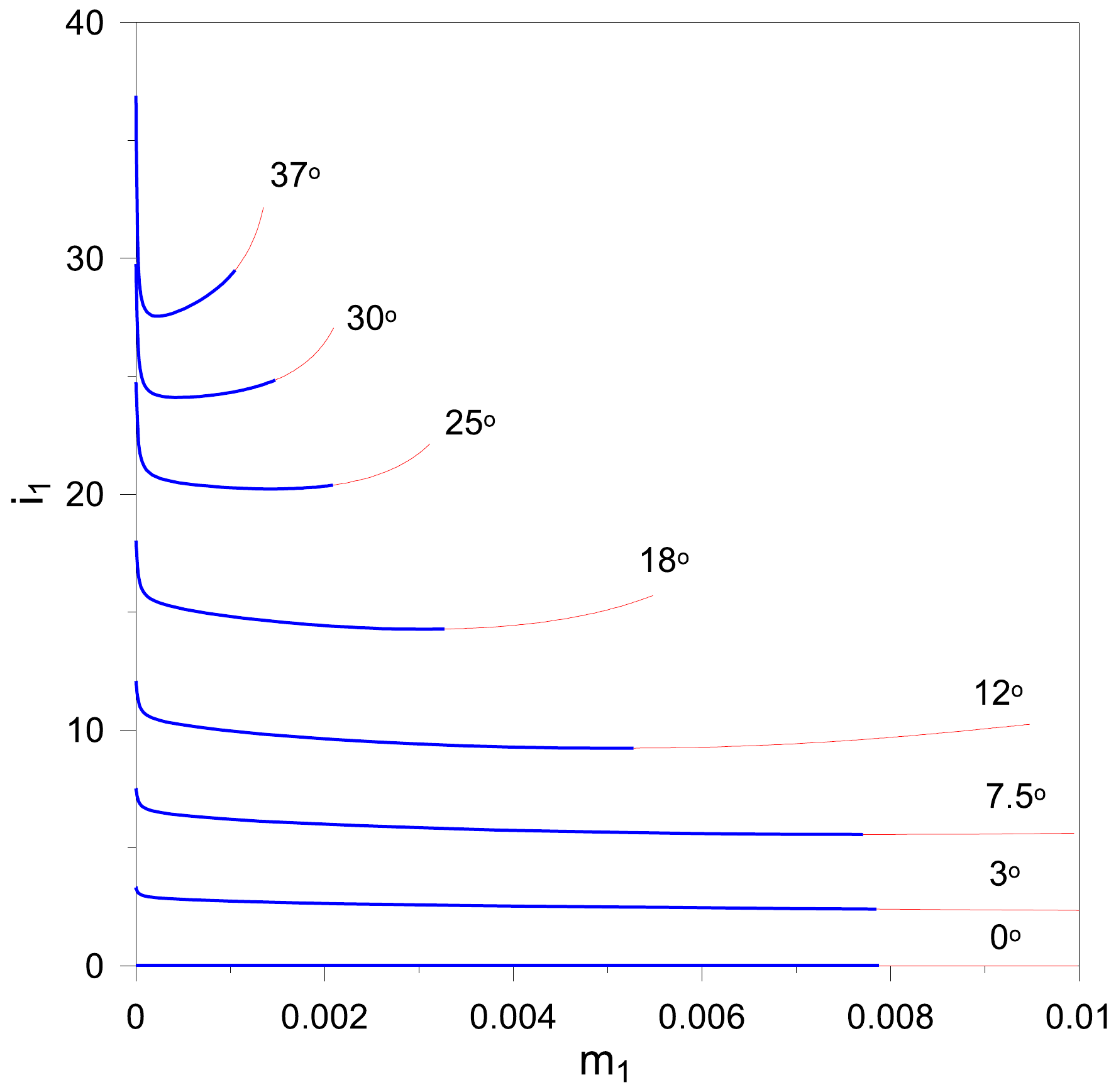} &\includegraphics[width=6cm]{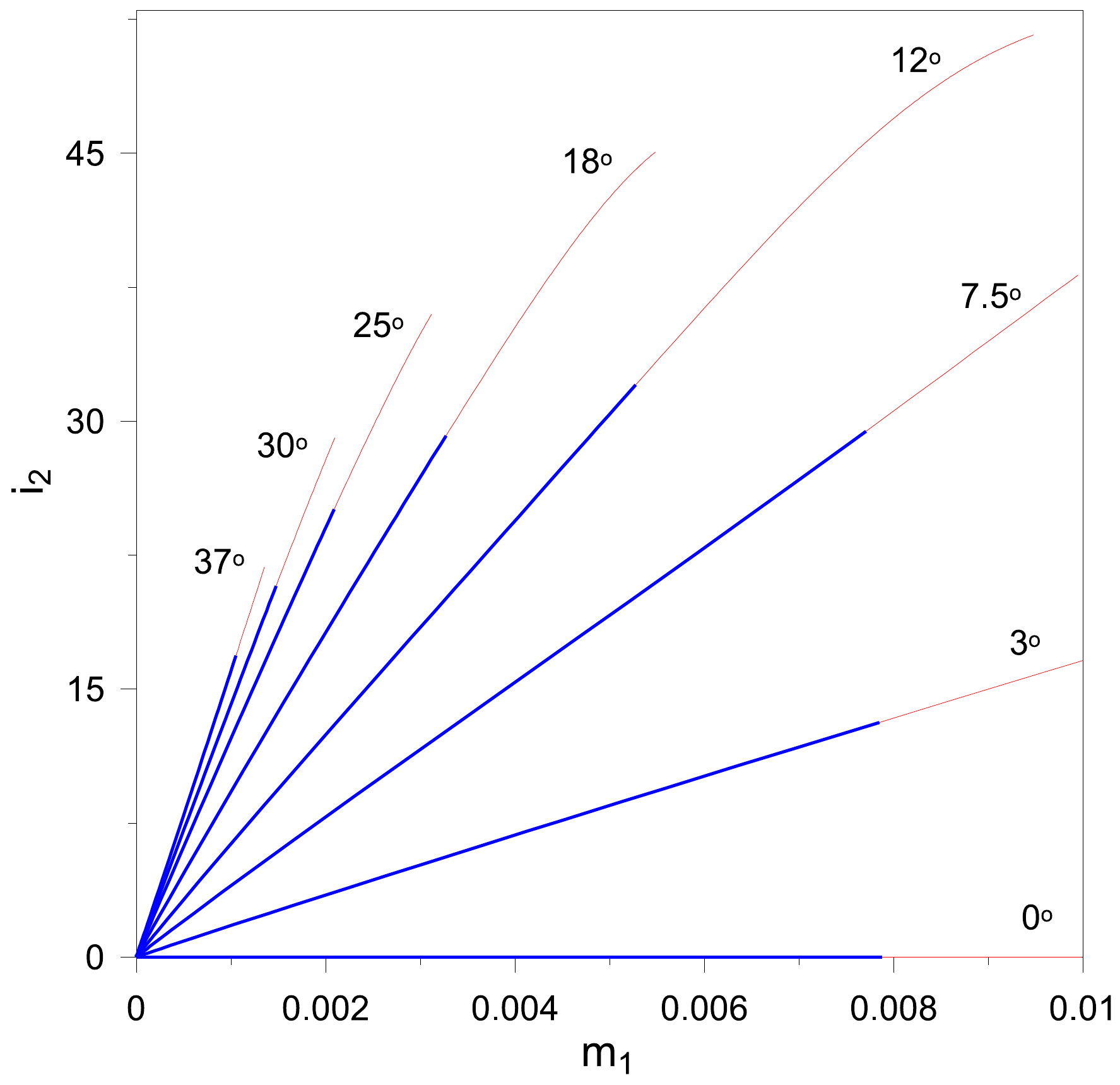}\\
\textnormal{(c)} &\textnormal{(d)}  
\end{array} $
\end{center}
\caption{Families $F_{c,m}^{2/1}$ of periodic orbits which are continued from the restricted problem (3D-CRTBP) and are symmetric with respect to $xz$-plane. The families are presented by characteristic curves in different projection planes. For each family the inclination value, $i_{10}$, of the starting orbit is indicated. Bold blue and red coloured segments represent stable or unstable orbits, respectively.}
\label{21XZm2}
\end{figure}

\begin{figure}
\begin{center}
\includegraphics[width=6cm]{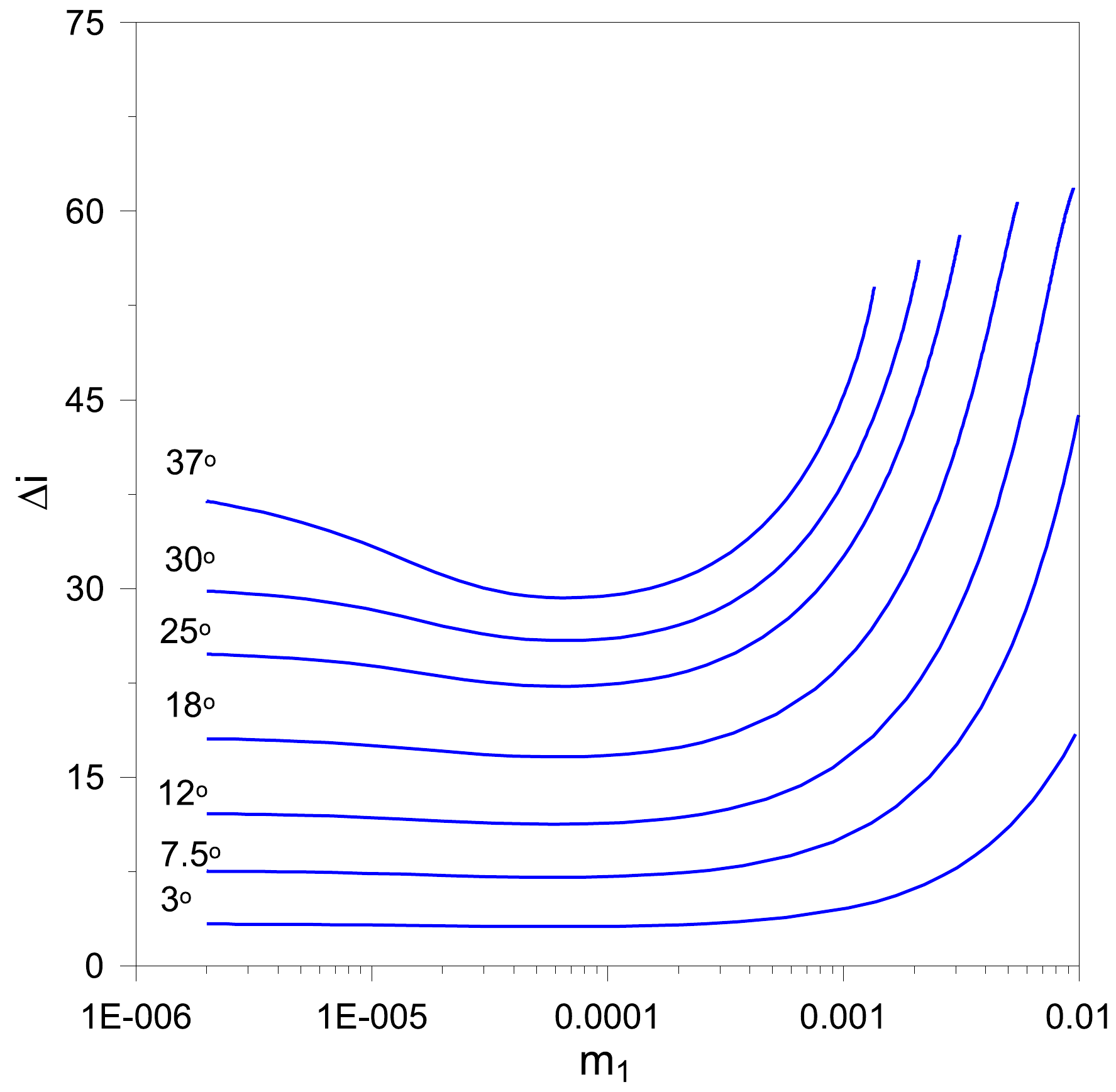} 
\end{center}
\caption{The variation of mutual inclination $\Delta i$ of planets along the stable segments of families $F_{c,m}^{2/1}$. The parameter value $i_{10}$ is shown for each family.}
\label{21XZm2i}
\end{figure}

\begin{figure}
\begin{center}
$\begin{array}{cc}
\includegraphics[width=6cm]{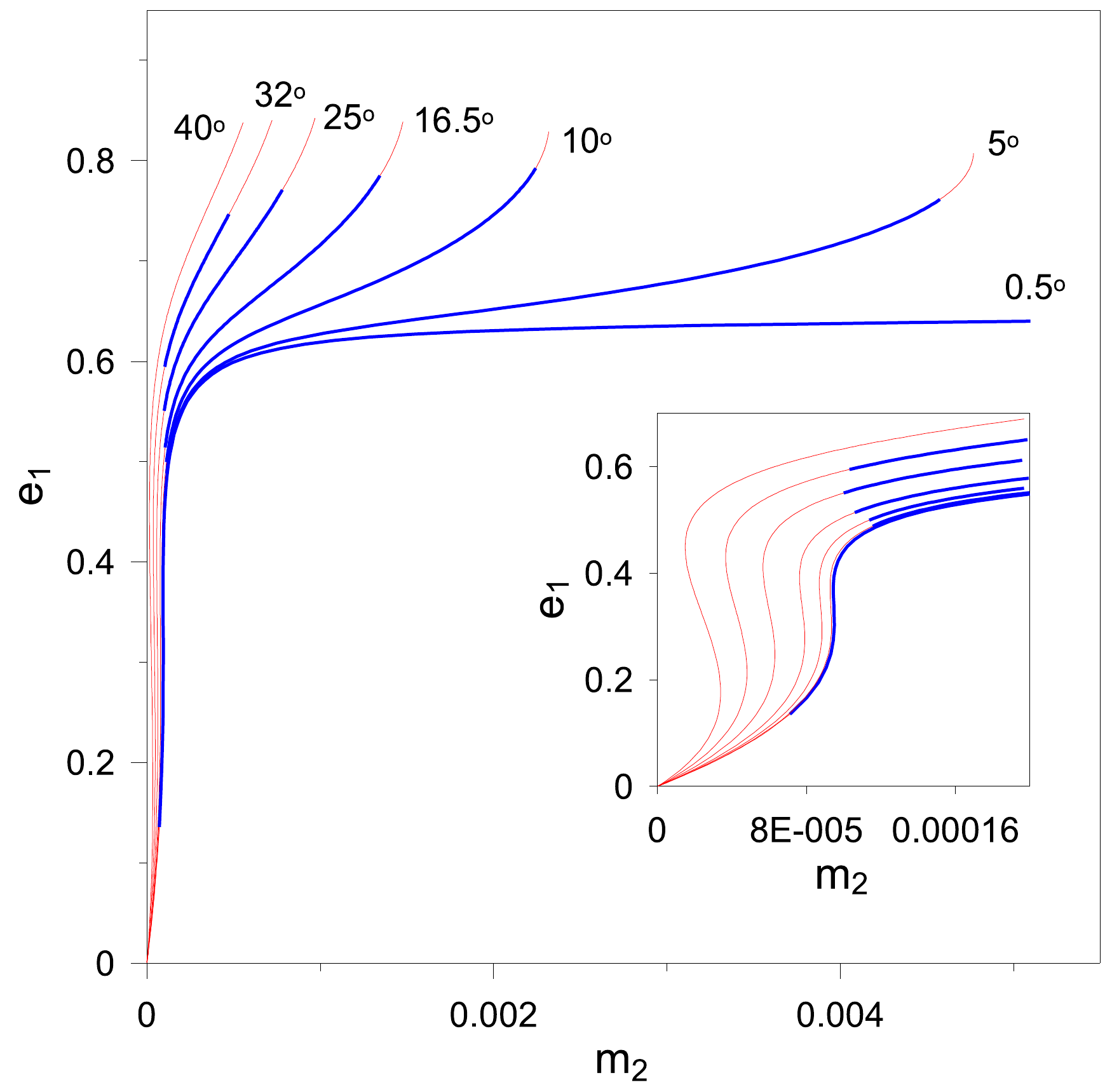} &\includegraphics[width=6cm]{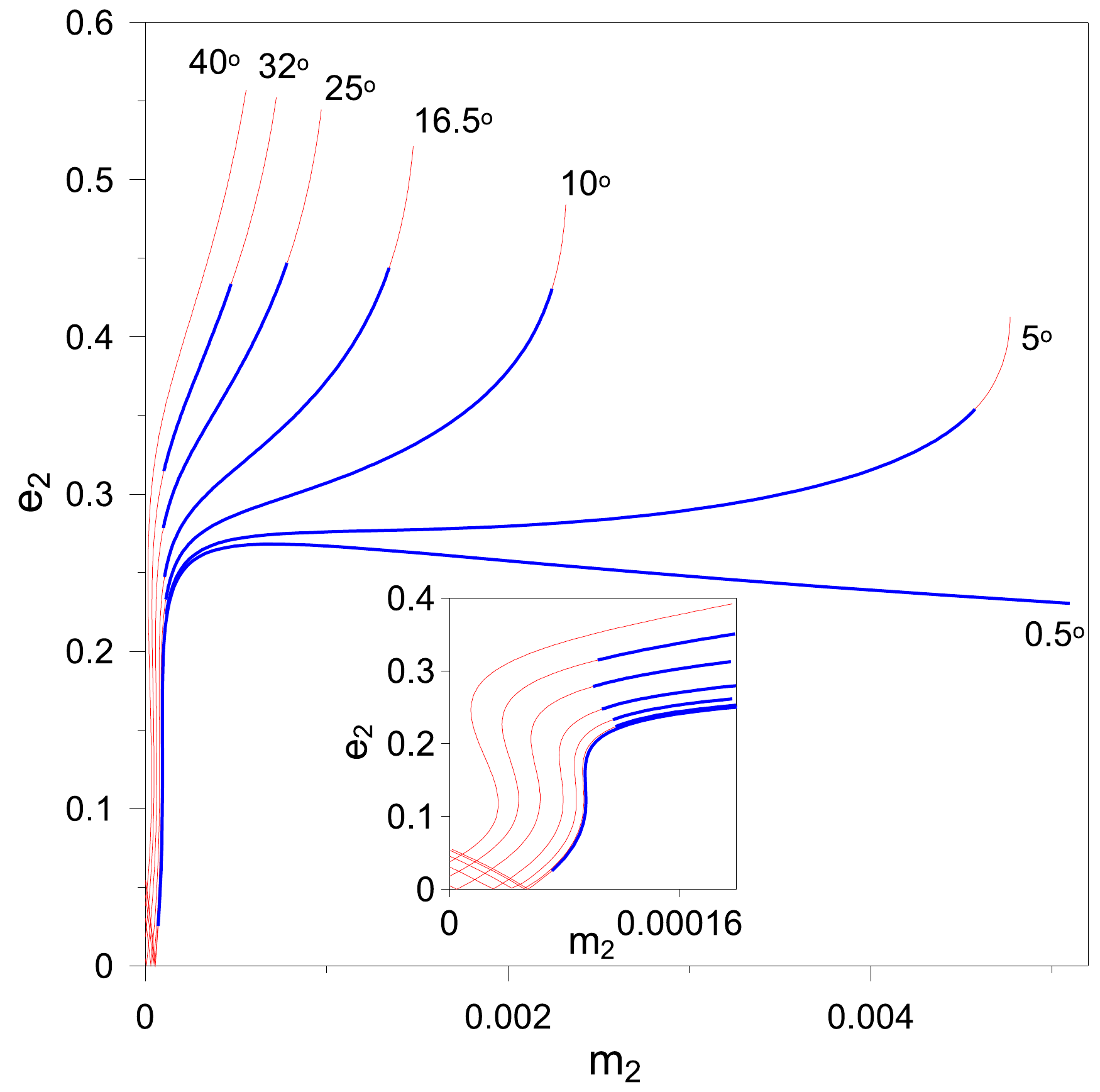}\\
\textnormal{(a)} &\textnormal{(b)}  
\end{array} $
\end{center}
\begin{center}
$\begin{array}{cc}
\includegraphics[width=6cm]{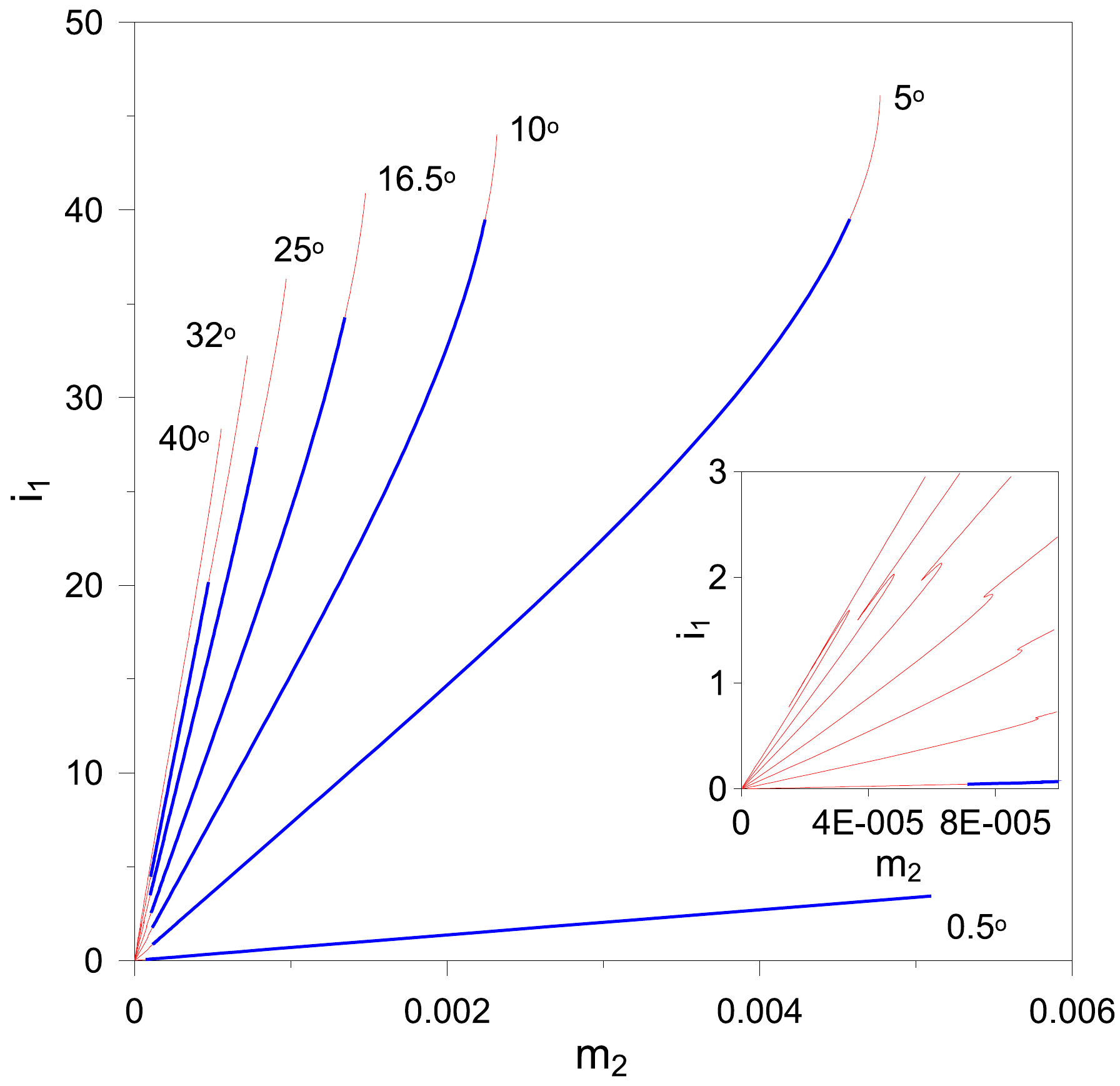} &\includegraphics[width=6cm]{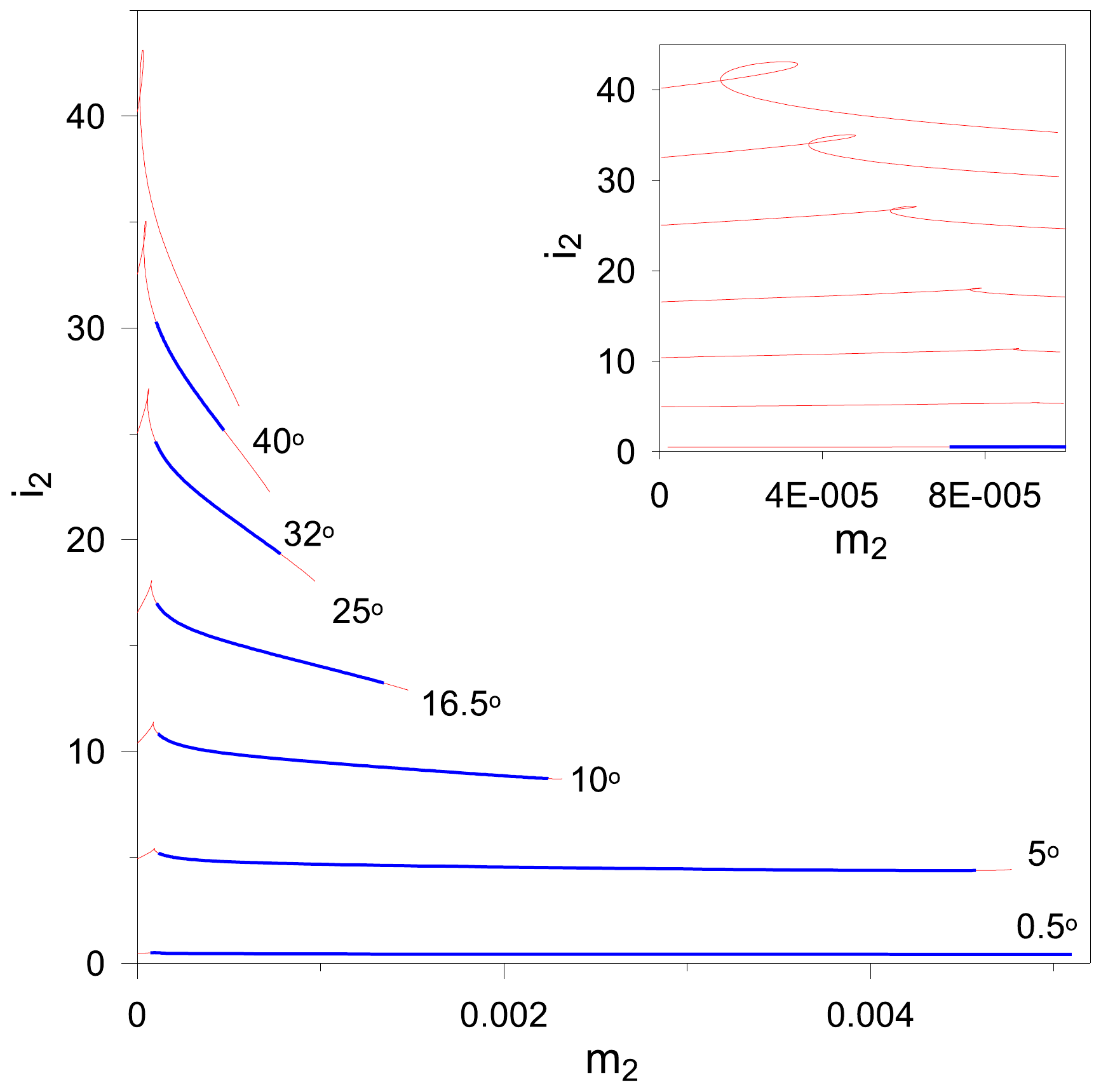}\\
\textnormal{(c)} &\textnormal{(d)}  
\end{array} $
\end{center}
\caption{Families $F_{c,m}^{1/2}$ of periodic orbits presented in a similar manner as in Fig. \ref{21XZm2}. Each family is identified by the inclination value, $i_{20}$, of the starting orbit.}
\label{12XZm2}
\end{figure}

\begin{figure}
\begin{center}
\includegraphics[width=6cm]{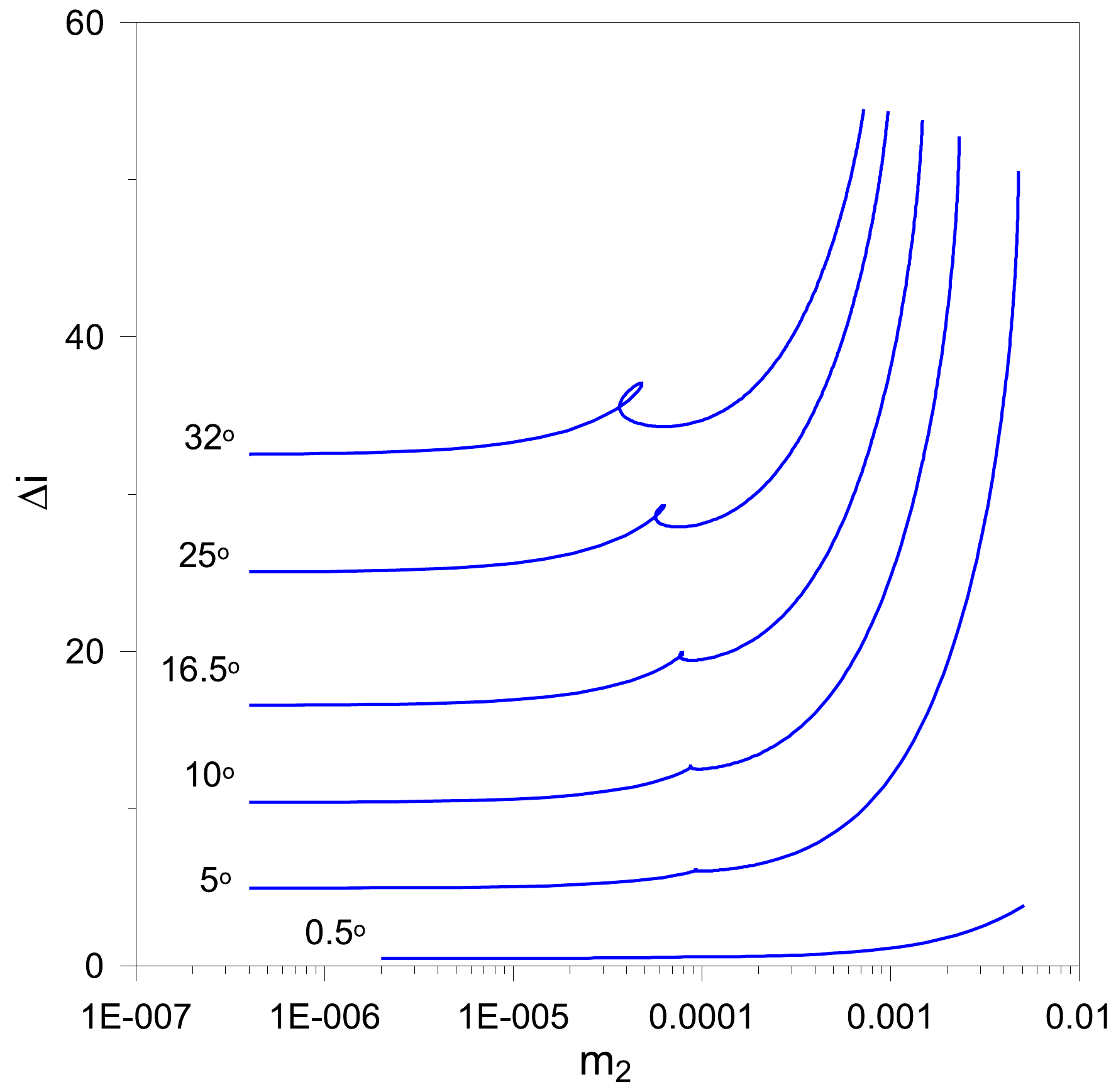} 
\end{center}
\caption{The variation of mutual inclination $\Delta i$ of planets along the stable segments of families $F_{c,m}^{1/2}$. The parameter value $i_{20}$ is shown for each family.}
\label{12XZm2i}
\end{figure}

\subsubsection{Families  $F_{c,m}^{2/1}$: Continuation from orbits of family $F_{c,i}^{2/1}$ with respect to $m_1$}
In Fig. \ref{21XZm2}, we present the continuation of the $xz$-symmetric periodic orbits of family $F_{c,i}^{2/1}$. The initial inclination values, $i_{10}$, are indicated. The formed families are denoted by $F_{c,m}^{2/1}$ and their periodic orbits correspond to the resonant angles 
$$
\Delta \varpi=0^{\circ}, \Delta \Omega=180^{\circ}, \sigma _1=0^{\circ}.
$$
The variation of the eccentricity of the planet $P_1$ along the families with respect to the mass $m_1$ is shown in Fig. \ref{21XZm2}a.
As $m_1$ increases from 0 to $10^{-4}$ (approximately), the eccentricity, $e_2$, of Jupiter increases rapidly from 0 to 0.25 for all computed families. Then, for small initial inclinations, $i_{10}$, of $P_1$, $e_2$ remains almost constant along the families, but continues to increase for higher initial inclinations, $i_{10}$ (see panel b). The diagram $m_1$-$i_1$ (panel c) shows a plateau for the inclination of the inner planet, whose range extends up to almost $m_1=0.01$ for the families of small inclination values of $P_1$. In contrary, Jupiter ($P_2$) shows an almost linear increment of its inclination, as $m_1$ increases (panel d). 

Families $F_{c,m}^{2/1}$ start having stable periodic orbits. The stability extends up to critical value of $m_1$, which depends on the particular initial value $i_{10}$. In these stable segments the mutual planetary inclination varies, as it is shown in Fig. \ref{21XZm2i}. We observe that we can have stable planetary orbits with mutual inclination, $\Delta i$, up to approximately $50^\circ$ - $60^\circ$ when the inner planet becomes very massive with respect to Jupiter.                     
 
\subsubsection{Families  $F_{c,m}^{1/2}$ : Continuation from orbits of family $F_{c,i}^{1/2}$ with respect to $m_2$}
The continuation of the 1/2 resonant $xz$-symmetric periodic orbits that belong to the family $F_{c,i}^{1/2}$ of the restricted problem is presented (in a similar way as above) in Fig. \ref{12XZm2}. Since the family $F_{c,i}^{1/2}$ is unstable, the families $F_{c,m}^{1/2}$,  which are obtained by the analytical continuation with respect to $m_2$, start having unstable periodic orbits, too. However, as $m_2$ increases the orbits with $i_{20}<35^\circ$ become stable. Instability becomes again apparent after a larger value of $m_2$, which depends on the initial inclination, $i_{20}$. It is interesting to note that the characteristic curves, which represent the families in the projection planes, show a {\em folding}. Namely, for a particular value of $m_2$ we may obtain more than one periodic orbits that belong in the same family. This folding becomes clear in the magnified plot in each panel of Fig. \ref{12XZm2}. The mutual planetary inclination along the stable segments is shown in Fig. \ref{12XZm2i}. As in the previous case, we obtain stable planetary orbits up to relatively high mutual inclination values (approximately $50^\circ$). 

As we have mentioned, the families $F_{c,m}^{1/2}$ start with orbits where the eccentricity of the planet $P_2$ is very small. For $i_{20}<26^\circ$ the corresponding resonant angles of the family $F_{c,i}^{1/2}$ of the restricted problem are 
$$
\textnormal{(i)}\;\; \Delta \varpi=180^{\circ}, \;\;\Delta \Omega=180^{\circ}, \;\; \sigma _1=0^{\circ}.
$$ 
For $i_{20}>26^\circ$ the orbit of $P_2$ changes apsidal orientation and the resonant angles take the values
$$
\textnormal{(ii)}\;\; \Delta \varpi=0^{\circ}, \;\;\Delta \Omega=180^{\circ}, \;\;\sigma _1=180^{\circ}.
$$ 
The orbits of the generated families $F_{c,m}^{1/2}$ start with the above resonant angle values, (i) or (ii), accordingly. However, the families that start with the values (i), show a decrement of the eccentricity $e_2$, as $m_2$ increases. At $e_2=0$ we obtain a change of the apsidal orientation of $P_2$ and the rest of the family consists of orbits with the resonant angle values (ii). 

\begin{figure}
\begin{center}
$\begin{array}{cc}
\includegraphics[width=6cm]{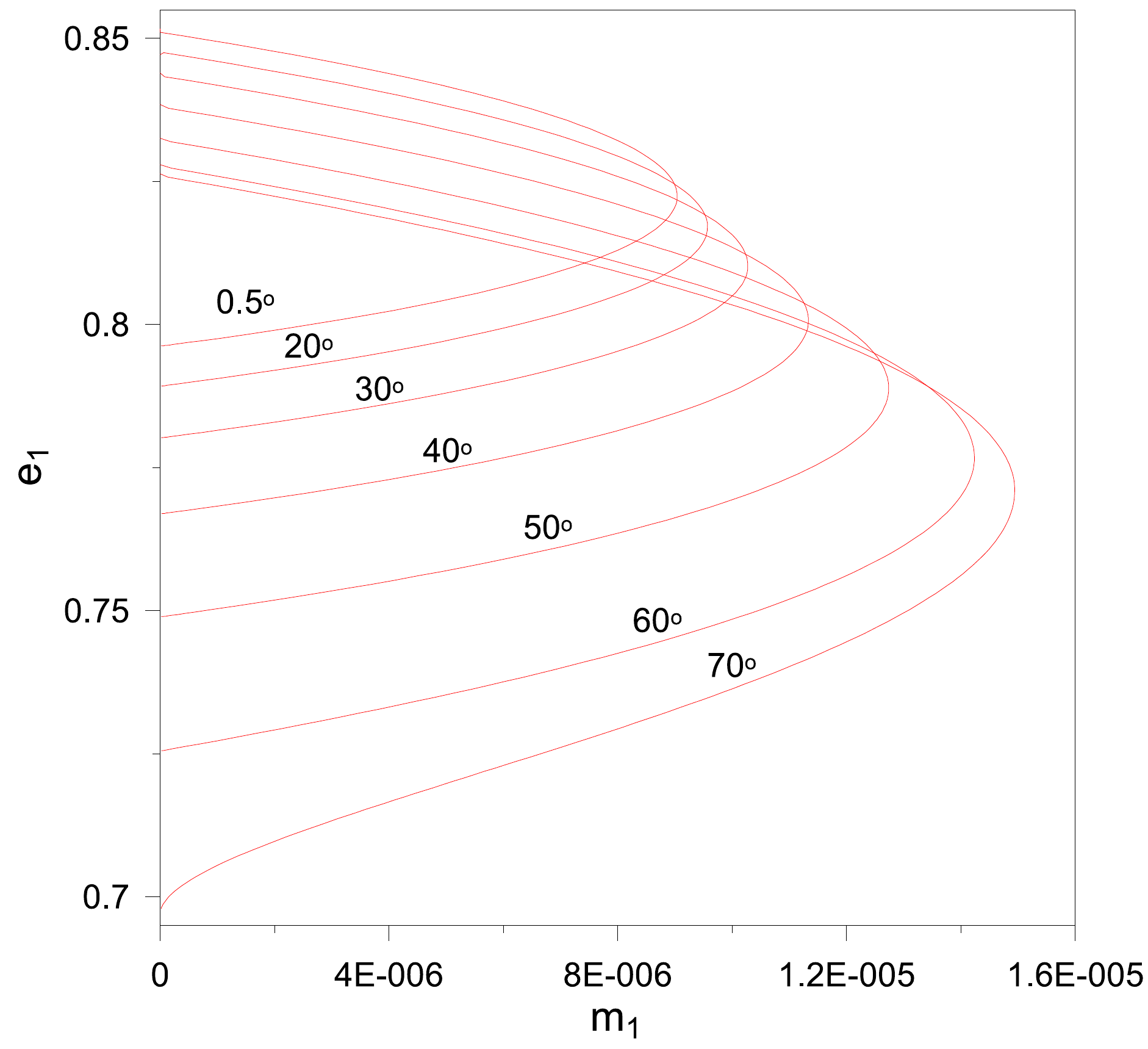} &\includegraphics[width=6cm]{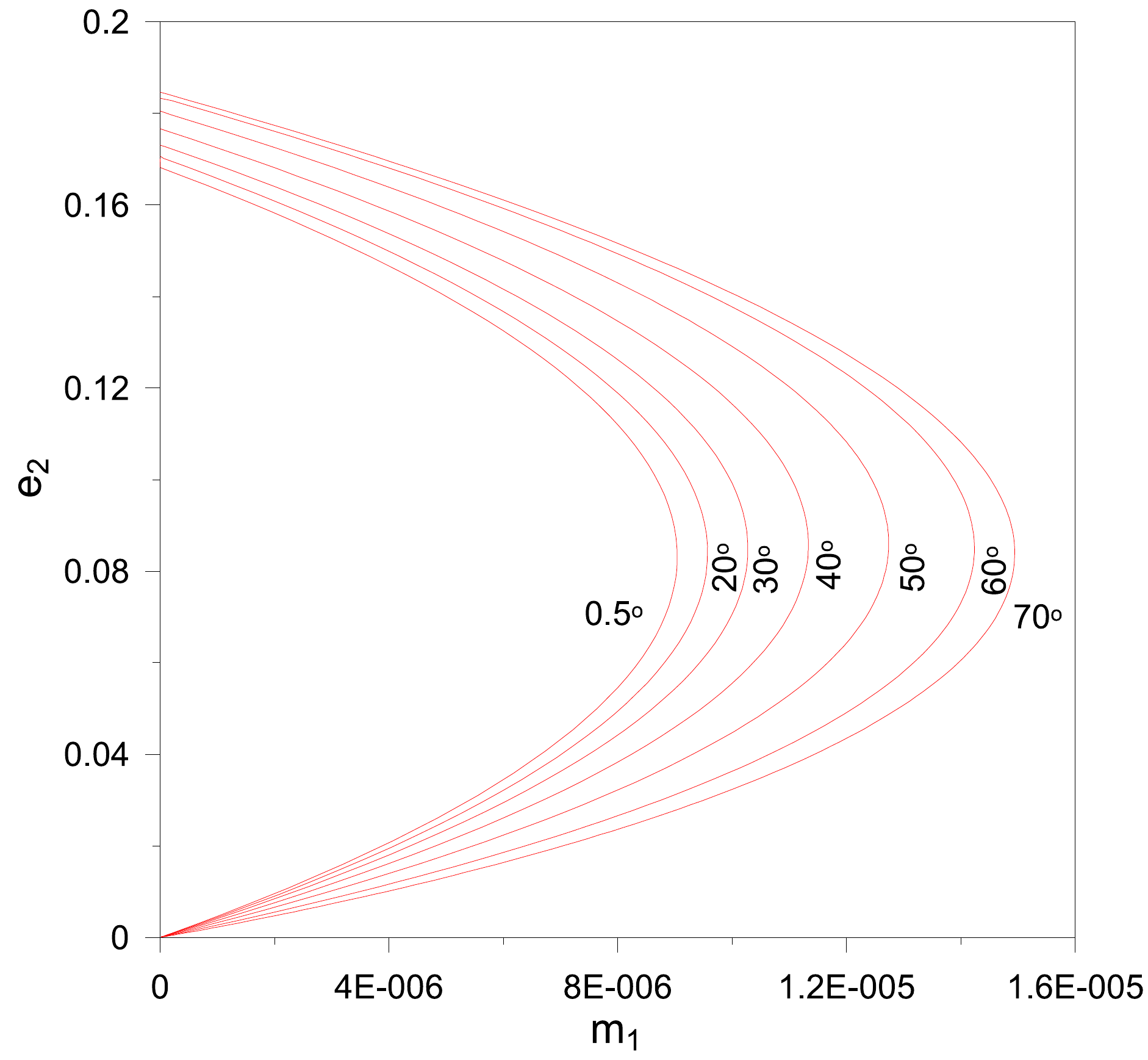}\\
\textnormal{(a)} &\textnormal{(b)}  
\end{array} $
\end{center}
\begin{center}
$\begin{array}{cc}
\includegraphics[width=6cm]{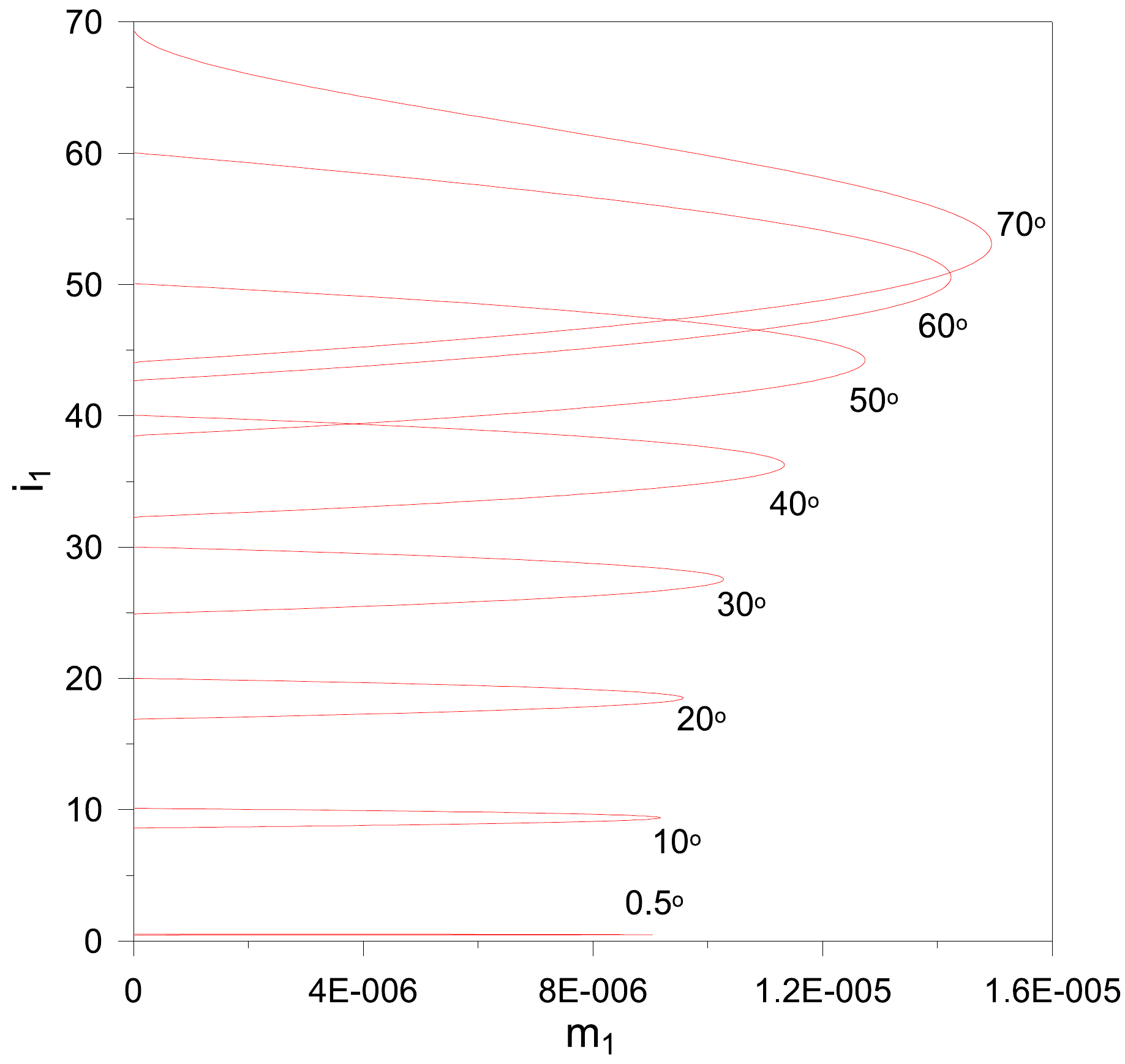} &\includegraphics[width=6cm]{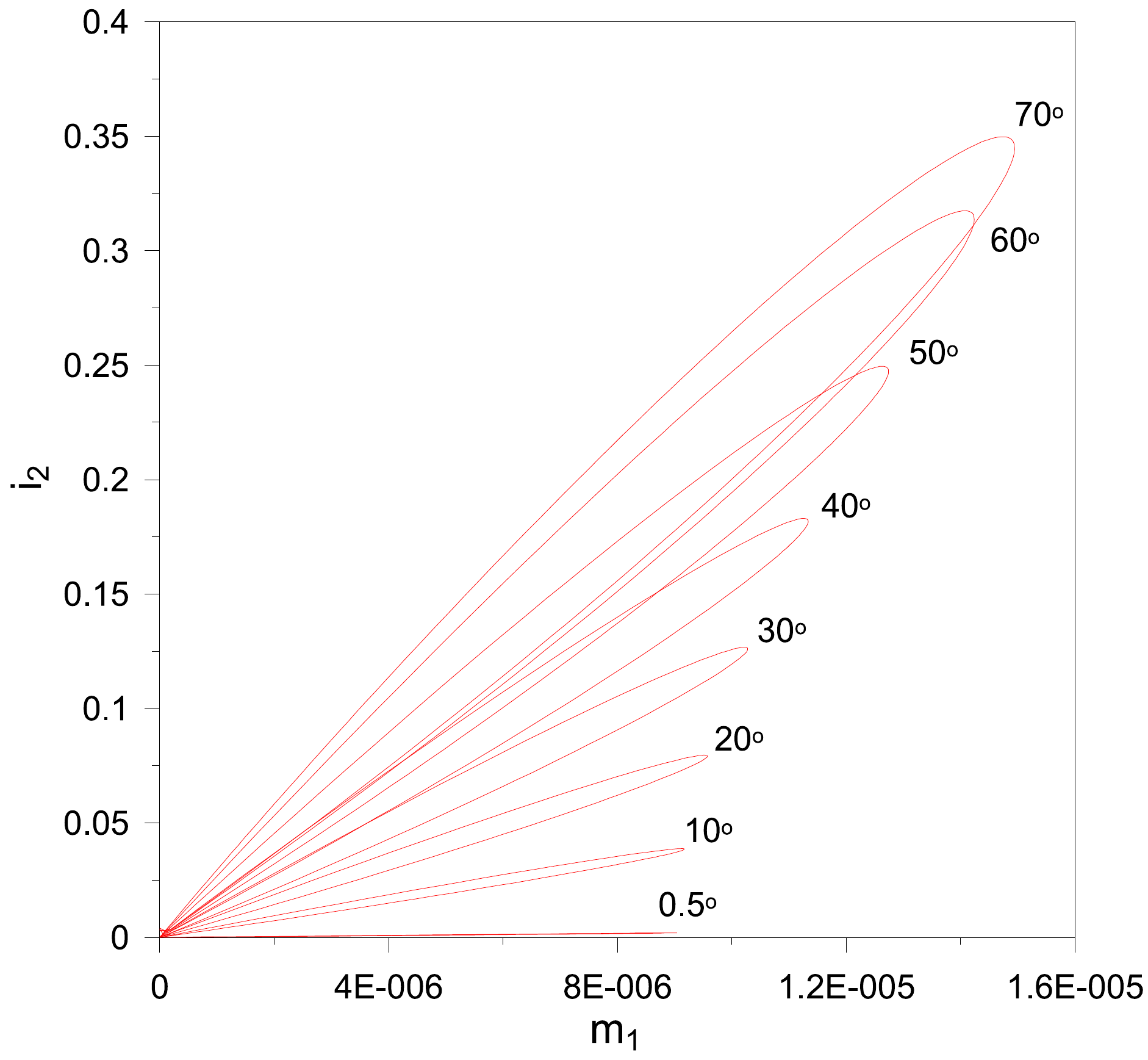}\\
\textnormal{(c)} &\textnormal{(d)}  
\end{array} $
\end{center}
\caption{Families $G_{c,m}^{2/1}$ of periodic orbits which are symmetric with respect to $x$-axis. The presentation is as in Fig. \ref{21XZm2}.}
\label{21Xm2}
\end{figure}

\begin{figure}
\begin{center}
$\begin{array}{cc}
\includegraphics[width=6cm]{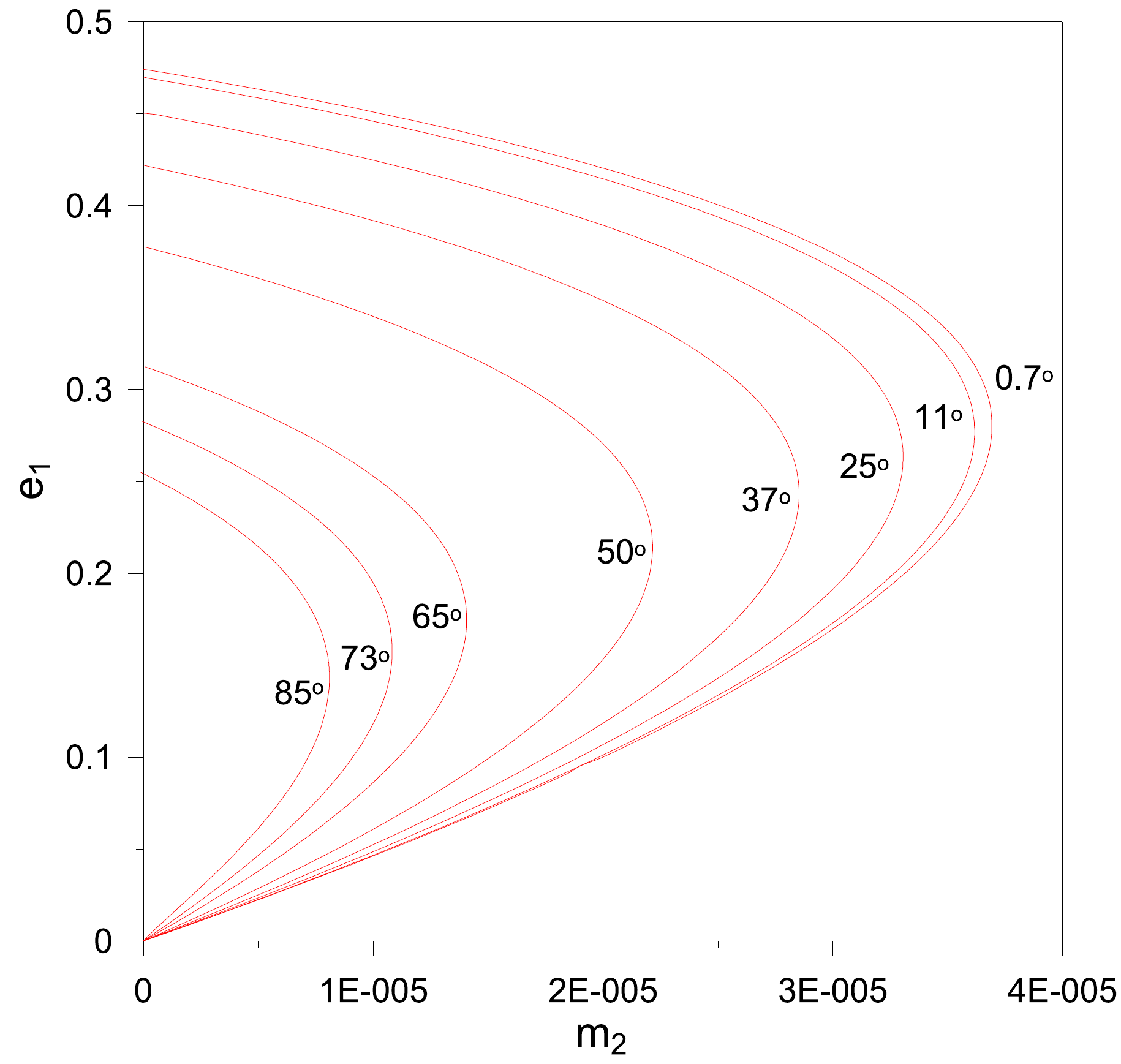} &\includegraphics[width=6cm]{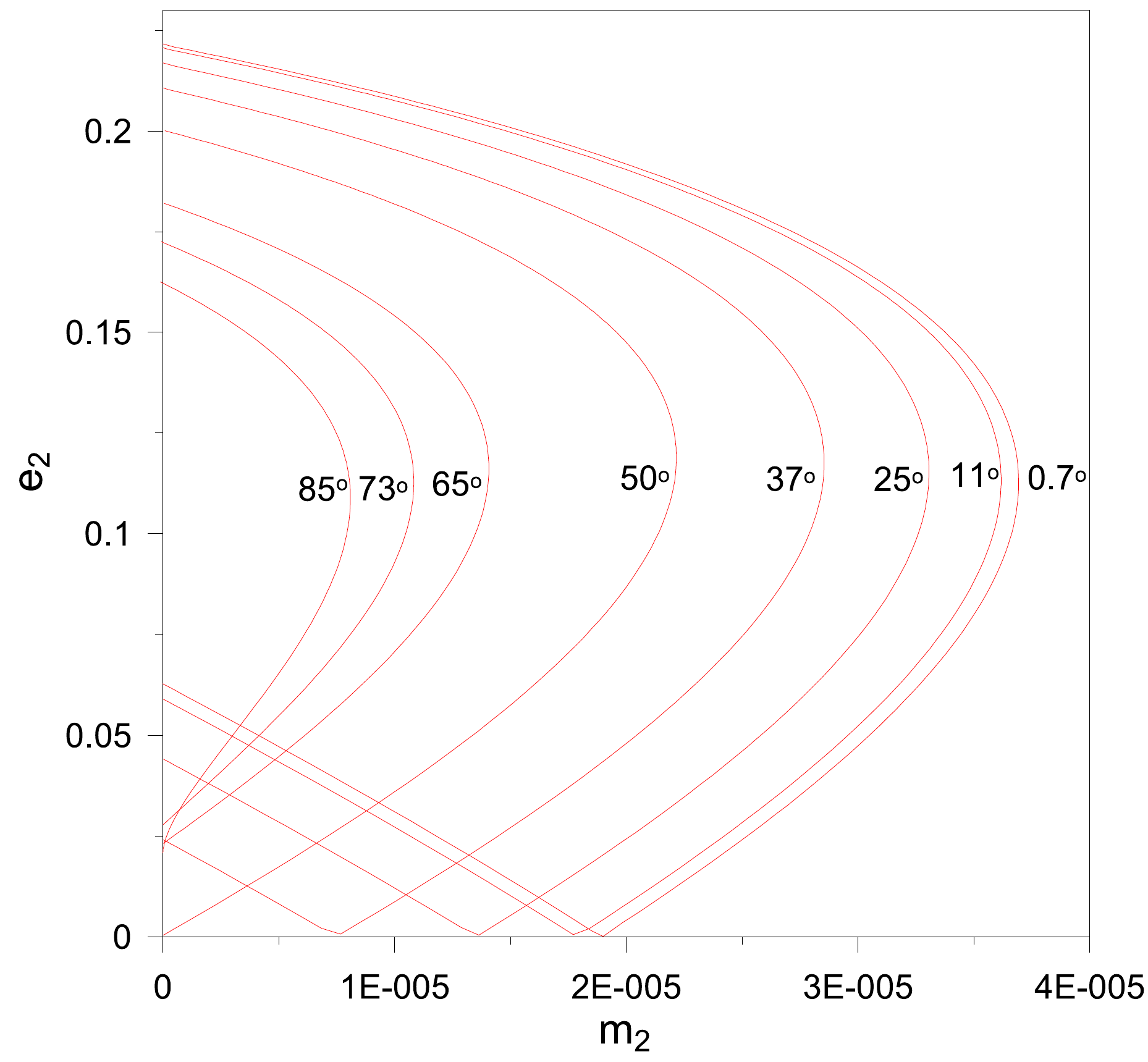}\\
\textnormal{(a)} &\textnormal{(b)}  
\end{array} $
\end{center}
\begin{center}
$\begin{array}{cc}
\includegraphics[width=6cm]{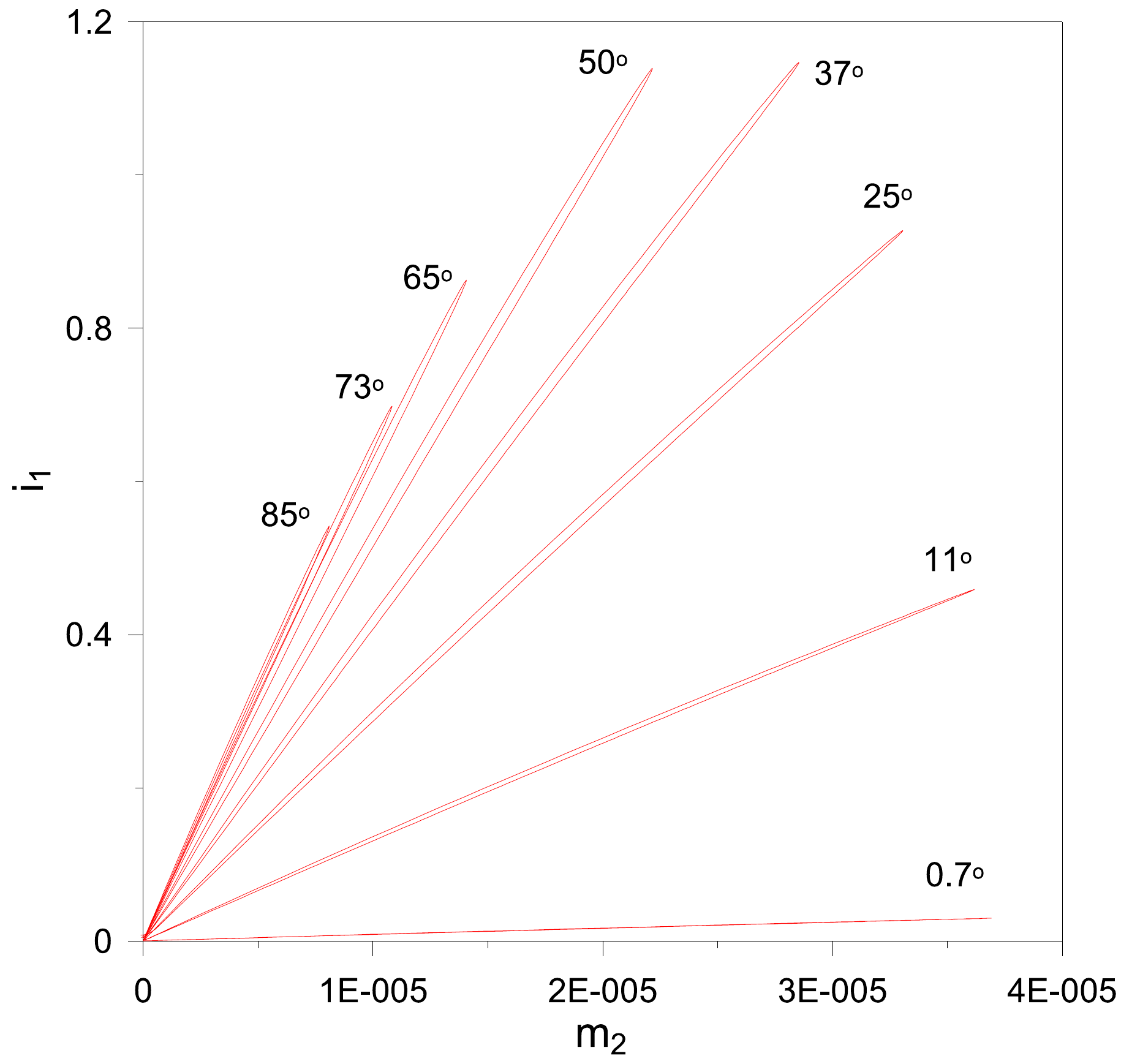} &\includegraphics[width=6cm]{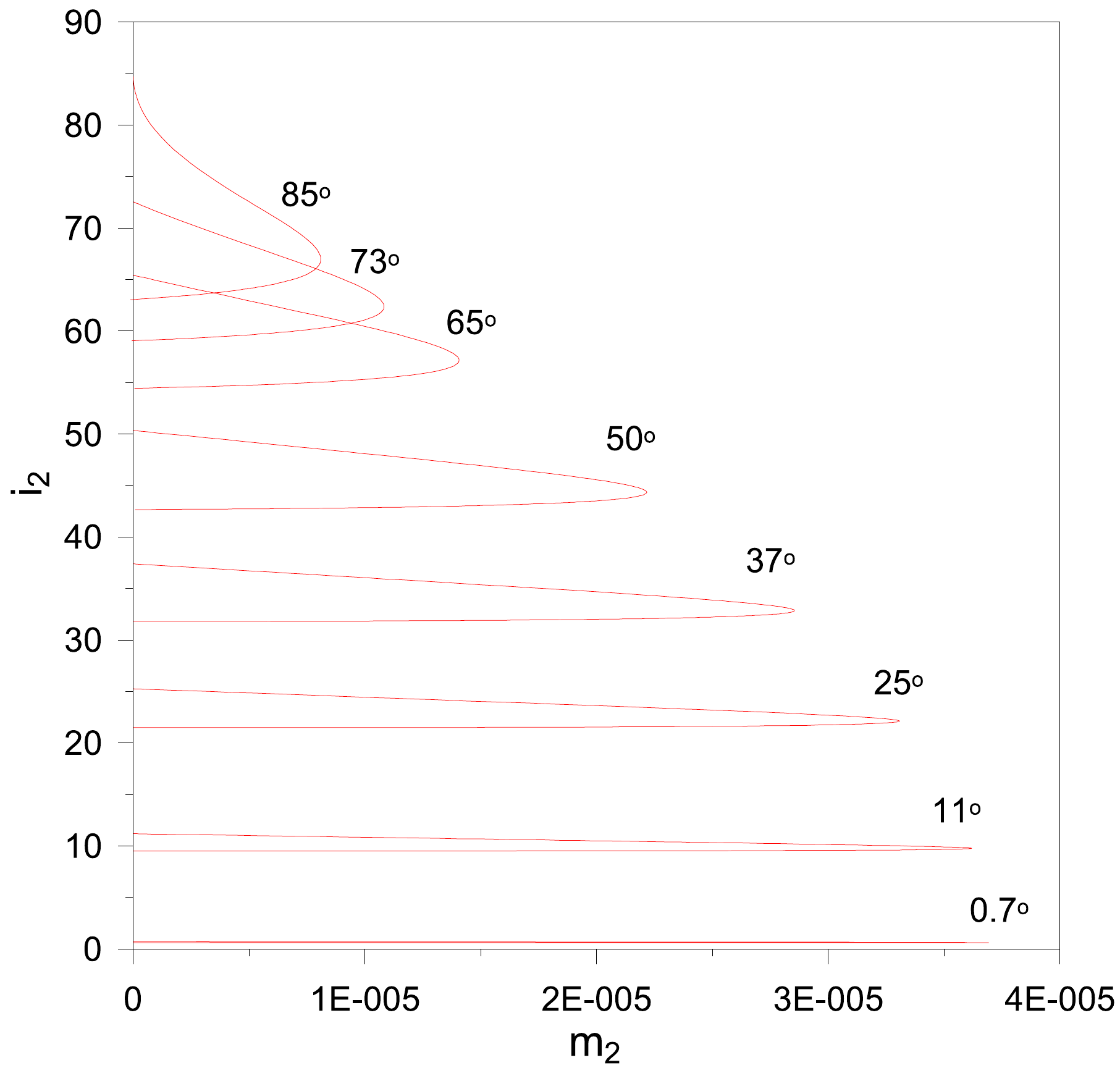}\\
\textnormal{(c)} &\textnormal{(d)}  
\end{array} $
\end{center}
\caption{Families $G_{c,m}^{1/2}$ of periodic orbits which are symmetric with respect to $x$-axis. The presentation is as in Fig. \ref{12XZm2}.}
\label{12Xm2}
\end{figure}

\subsubsection{Families  $G_{c,m}^{2/1}$: Continuation from orbits of family $G_{c,i}^{2/1}$ with respect to $m_1$}
In Fig. \ref{21Xm2}, we present the families $G_{c,m}^{2/1}$ of $x$-symmetric periodic orbits generated from orbits of the unstable family of the restricted problem. The continuation of these periodic orbits, which is performed by increasing the mass $m_1$, extends only up to a certain value $m_1^*$. Note that the value of $m_1^*$ for all families is relatively very small (of order $10^{-5}$) with respect to Jupiter's mass $m_2=10^{-3}$. Starting from $m_1=m_1^*$ and by decreasing $m_1$ we can obtain a different branch of the family down to $m_1=0$. However, at this point, as far as Jupiter is concerned, we have $e_2\neq 0$ and $i_2=0$, i.e. the families end at periodic orbits of the elliptic restricted problem (3D-ERTBP). Note, now, that Jupiter's orbit is almost planar in all cases.  

All orbits of families $G_{c,m}^{2/1}$ are unstable. Their resonant angles are 
$$\Delta \varpi=0^{\circ},\;\; \Delta \Omega=0^{\circ}, \;\;\sigma _1=180^{\circ}.$$

\subsubsection{Families  $G_{c,m}^{1/2}$ : Continuation from orbits of family $G_{c,i}^{1/2}$ with respect to $m_2$}
Similar behaviour as above (start and end) is obtained for the families $G_{c,m}^{1/2}$, where now Jupiter is the inner planet ($P_1$). All orbits are unstable (Fig. \ref{12Xm2}).

Concerning the resonant angles, the generating family $G_{c,i}^{1/2}$ of the restricted problem starts with orbits having  
$$\textnormal{(i)}\;\; \Delta \varpi=180^{\circ}, \;\; \Delta \Omega=0^{\circ}, \;\; \sigma _1=180^{\circ},$$
until $i_{20}<50^{\circ}$. For larger inclination values we have a change in the apsidal orientation and we get the values
$$\textnormal{(ii)}\;\; \Delta \varpi=0^{\circ},\;\; \Delta \Omega=0^{\circ}, \;\; \sigma _1=0^{\circ}$$ 

The generated families $G_{c,m}^{1/2}$ start with the values as above, (i) or (ii) for $i_{20}<50^\circ$ or $i_{20}>50^\circ$, respectively.  
For case (i), the eccentricity $e_2$ starts decreasing down to zero along the family. Then, the apsidal orientation of planet $P_2$ changes and the resonant angles take the values (ii).

\subsection{Scheme II: Continuation from 2D-GTBP to 3D-GTBP} 
\label{SII}
\begin{figure}
\begin{center}
$\begin{array}{cc}
\includegraphics[width=6cm]{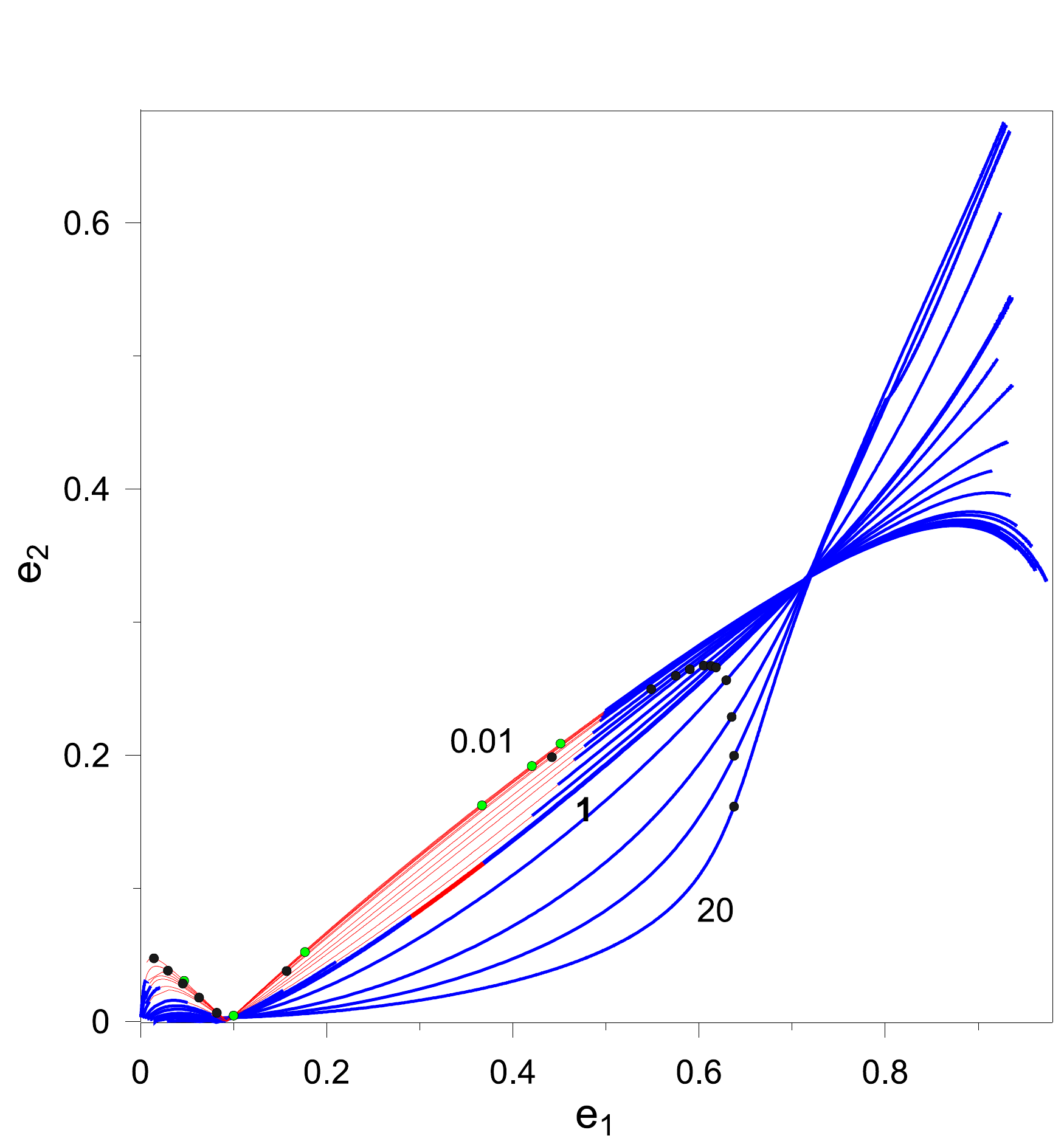} &
\includegraphics[width=6cm]{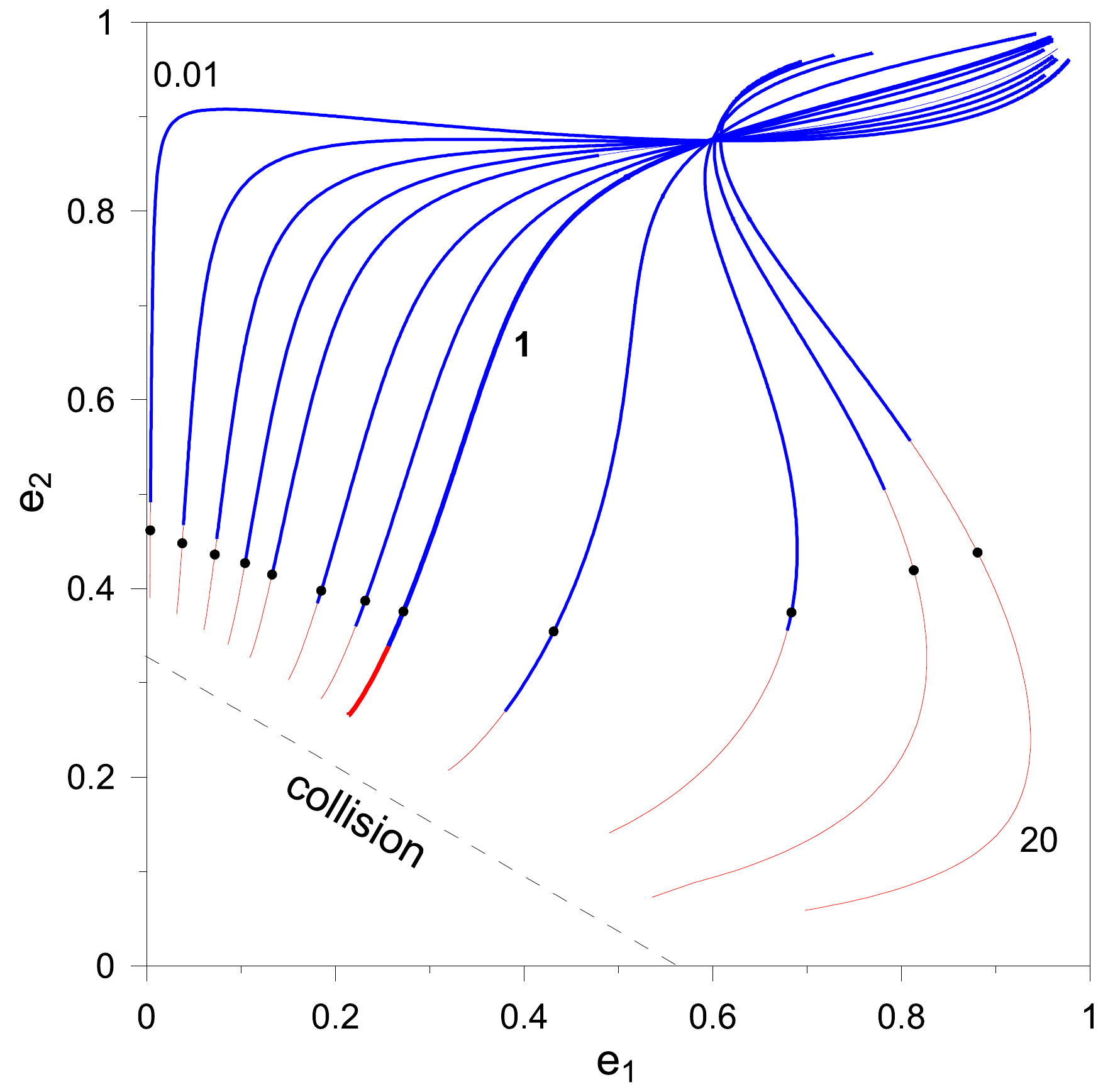}\\ 
\textnormal{(a)} &\textnormal{(b)}
\end{array} $
\end{center}
\caption{Families of symmetric periodic orbits of the planar general problem {\bf a} $f_1$ and {\bf b} $f_2$. Bold blue colour stands for stable periodic orbits and red for the unstable ones.}
\label{f12i}
\end{figure}

Resonant symmetric periodic orbits for the planar general problem (2D-GTBP) can be obtained by starting from either the circular family of orbits (e.g. \citet{hadj06}), or from resonant families of the planar restricted problem (e.g. \citet{avk11}). Particularly, for the 2/1 (or 1/2) resonance, planar periodic orbits have been computed by (\citealt{bmfm06,voyhadj05,vkh09}). Now, the distinction between the 2/1 and 1/2 resonance is not essential, since both planetary masses are nonzero and we refer to the mass ratio $\rho=m_2/m_1$, where $P_1$ is the inner planet. In the following computations, we always set $m_1=0.001$ (Jupiter's mass). In Fig. \ref{f12i}, we present the 1/2 resonant families $f_1$ and $f_2$ of the general planar problem. The families are given for various values of the mass ratio $\rho$ in the projection space defined by the planetary eccentricities.             

The variational equation that corresponds to the $z$-component of the motion of planet $P_2$ is written as\footnote{Note that the variational equation given in (\citealt{mich79}) holds only for the normalization $m_0+m_1=1$.}    
\begin{equation}
\ddot \zeta_2=A \zeta_2 +B \dot \zeta_2
\label{z22}
\end{equation}
where
\begin{equation}
\begin{array}{l}
A=-\frac{m m_0 [1-\frac{m_2 ( \dot \theta x_2+\dot y_2)}{m \dot \theta x_1}]}{(m_0+m_1)[(\frac{m_1 x_1}{m_0}+x_2)^2+y_2^2]^{3/2}} -\frac{m m_1 [1+\frac{m_0 m_2 (\dot \theta x_2+\dot y_2)}{m m_1 \dot \theta x_1}]}{(m_0+m_1)[(x_1 - x_2)^2+y_2^2]^{3/2}}\\[0.3cm]
B=\frac{m_0 m_2 y_2}{(m_0 + m_1)\dot \theta x_1 [(x_1-x_2)^2+y_2^2]^{3/2}}-\frac{m_0 m_2 y_2}{(m_0 + m_1)\dot \theta x_1 [(\frac{m_1 x_1}{m_0}+x_2)^2+y_2^2]^{3/2}}.
\label{AB}
\end{array}
\end{equation}
If $\Delta(T)=\{\xi_{ij}\}$, $i,j=1,2$, is the monodromy matrix of Eq. \ref{z22} for a planar periodic orbit of period $T$, then, this orbit has a {\em vertical critical index} (\citealt{mich79}) equal to
\begin{equation}
a_v=\frac{\xi_{11}\xi_{22}+\xi_{21}\xi_{12}}{\xi_{11}\xi_{22}-\xi_{21}\xi_{12}},
\label{av2}
\end{equation}
Any periodic orbit of the planar problem with $|a_v|=1$  is {\em vertical critical} (v.c.o.) and can be used as a starting orbit for analytical continuation in the spatial problem with the inclination of $P_2$ as a parameter (in computations we increase $z_2$ or $\dot z_2$). 

\begin{table}[h]
\caption{Eccentricity values for the v.c.o. of families $f_1$ from which $xz$-symmetric periodic orbits bifurcate in space.}  
\begin{tabular}{ccc}
				
    \hline
   $ \rho$  &  $e_1$ & $e_2$\\
	  \hline\hline

    $0.01$&$0.014402$&$0.047445$\\
    \hline
    $0.1$&$0.442008$&$0.198520$\\
    \hline
		$0.2$&$0.5486138$&$0.2496163$\\
		\hline
		$0.3$&$0.5750049$&$0.2596675$\\
		\hline
		$0.4$&$0.590191$&$0.264541$\\ \hline
		$0.6$&$0.6051763$&$0.2672597$\\
		\hline
		$0.8$&$0.6131065$&$0.2670630$\\
		\hline
		$1$&$0.6182749$&$0.2659223$\\
		\hline
		$2$&$0.6292770$&$0.2563407$\\
		\hline
		$5$&$0.6350957$&$0.2287521$\\
		\hline
		$10$&$0.6376757$&$0.1995789$\\
		\hline
		$20$&$0.6375954$&$0.1613843$\\
		\hline\end{tabular}
\label{vcp XZ f1i}
\end{table}

\begin{table}[h]
\caption{Eccentricity values for the v.c.o. of families $f_1$ from which $x$-symmetric periodic orbits bifurcate in space.} 
\begin{tabular}{ccc} 
\hline
$ \rho $ &  $e_1$ & $e_2$\\
\hline\hline
&  $0.046879$ & $0.030487$  \\[-1ex]
\raisebox{1.5ex}{$0.01$} & $0.451260$ & $0.208753$  \\
\hline
&  $0.099993$ & $0.004378$  \\[-1ex]
\raisebox{1.5ex}{$0.02$} & $0.420425$ & $0.191886$  \\ \hline
&  $0.176672$ & $0.052130$  \\[-1ex]
\raisebox{1.5ex}{$0.03$} & $0.366875$ & $0.162298$  \\
\hline 
\end{tabular}
\label{vcp X f1i}
\end{table}

\begin{table}[h]
\caption{Eccentricity values for the v.c.o. of families $f_2$ from which $xz$-symmetric periodic orbits bifurcate in space.}  
\begin{tabular}{ccc}
		
    \hline
   $ \rho $ &  $e_1$ & $e_2$\\
	  \hline\hline

    $0.01$&$0.003953$&$0.461514$\\
    \hline
    $0.1$&$0.037795$&$0.447835$\\
    \hline
		$0.2$&$0.072314$&$0.435873$\\
		\hline
		$0.3$&$0.104294$&$0.426776$\\
		\hline
		$0.4$&$0.133030$&$0.414625$\\ \hline
		$0.6$&$0.185104$&$0.397536$\\
		\hline
		$0.8$&$0.231733$&$0.386791$\\
		\hline
		$1$&$0.272251$&$0.375495$\\
		\hline
		$2$&$0.431450$&$0.354297$\\
		\hline
		$5$&$0.683522$&$0.374495$\\
		\hline
		$10$&$0.813132$&$0.419156$\\
		\hline
		$20$&$0.880779$&$0.437977$\\
		\hline
    
\end{tabular}
\label{vcp XZ f2i}
\end{table}

\begin{figure}
\begin{center}
$\begin{array}{cc}
\includegraphics[width=6cm]{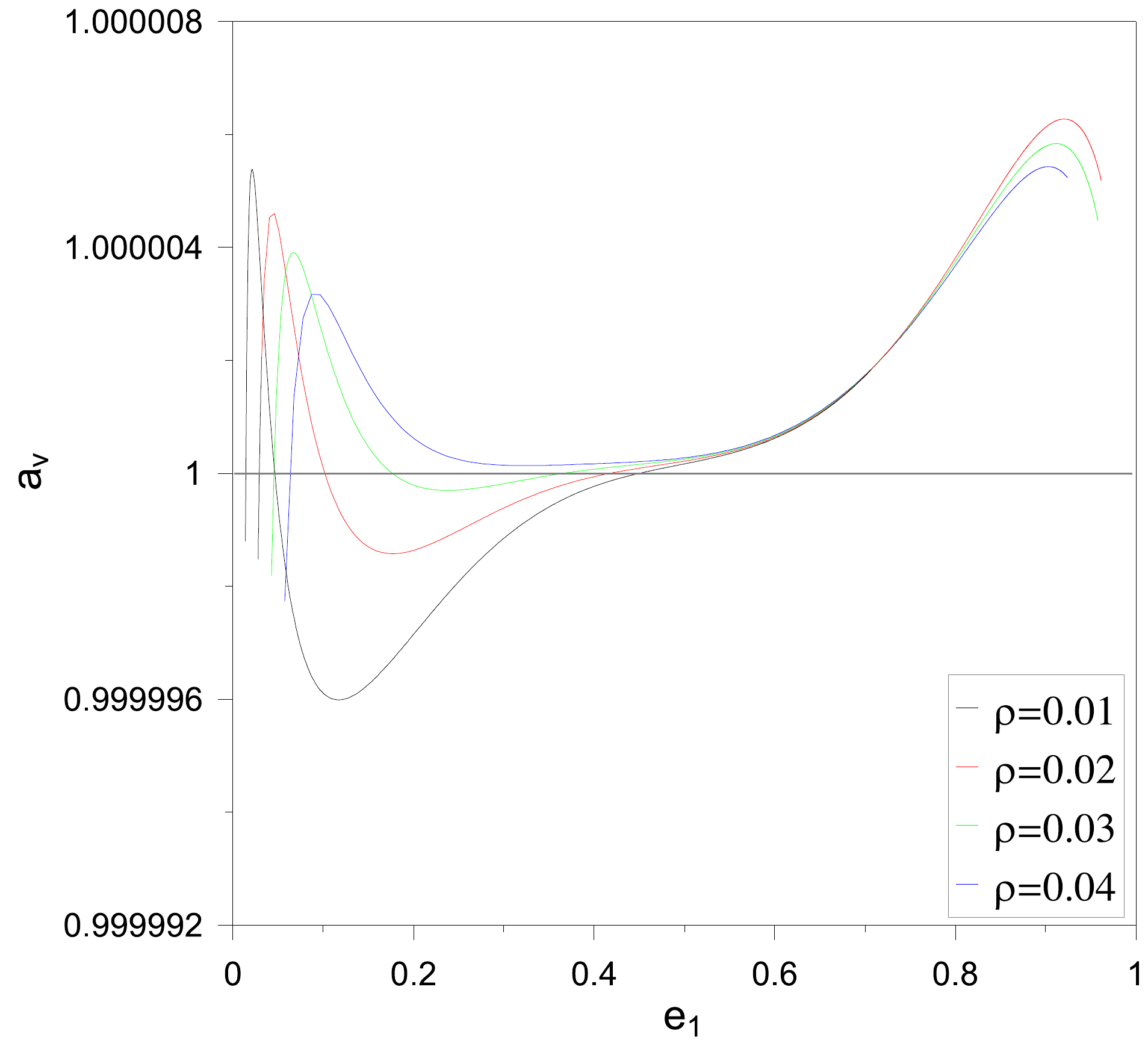} &
\includegraphics[width=6cm]{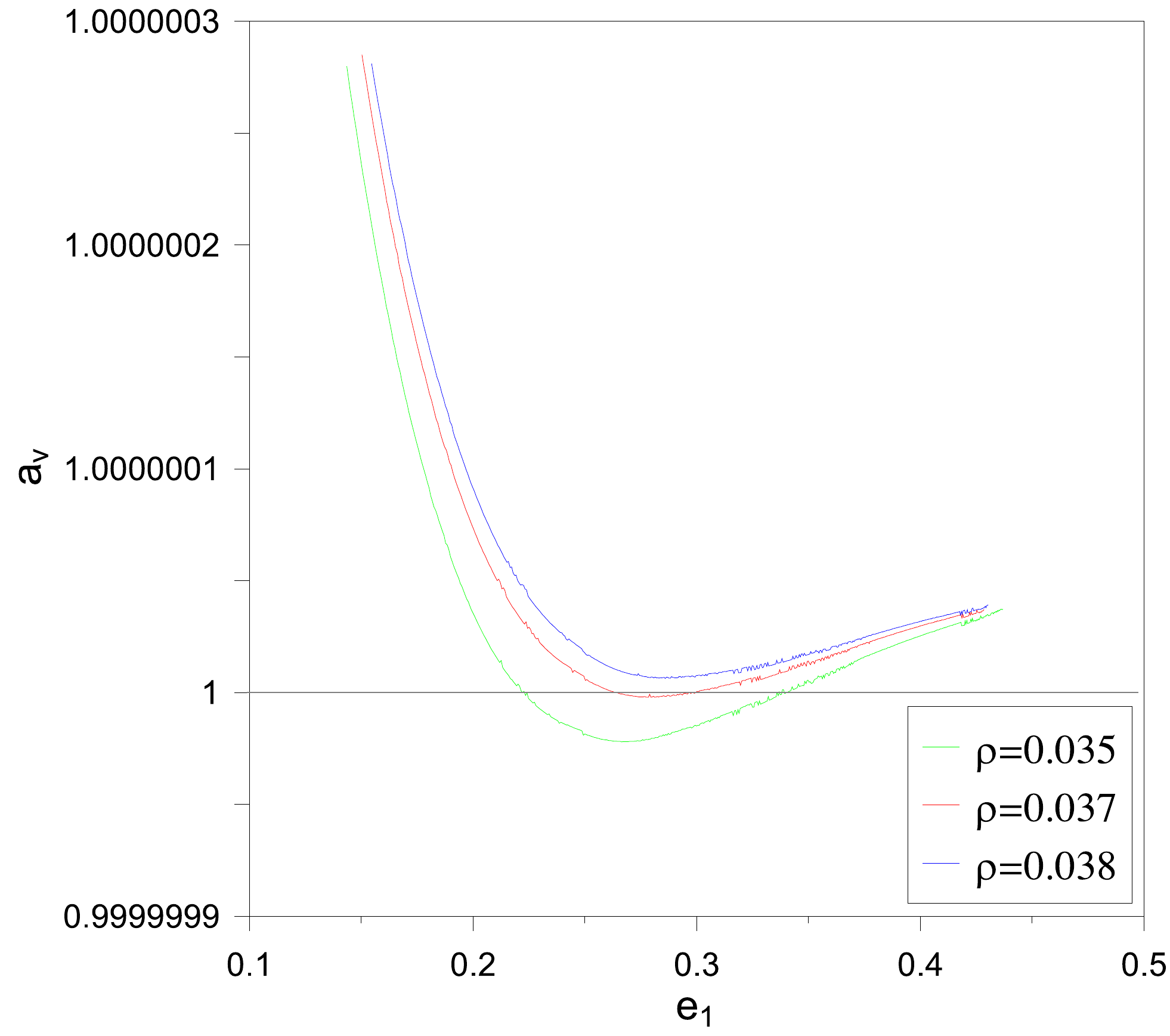}\\ 
\textnormal{(a)} &\textnormal{(b)}  
\end{array} $
\end{center}
\caption{{\bf a} The variation of of the index $a_v$ along the family $f_1$ ($e_1$ is used as the parameter of the family). For $\rho<0.0372$ the index becomes critical in three points, while for $\rho>0.0372$ we obtain only one vertical critical point. {\bf b} A magnification of panel (a) which illustrates the disappearance of the couple of critical orbits.}
\label{av}
\end{figure}

\begin{figure}
\begin{center}
$\begin{array}{cc}
\includegraphics[width=6cm]{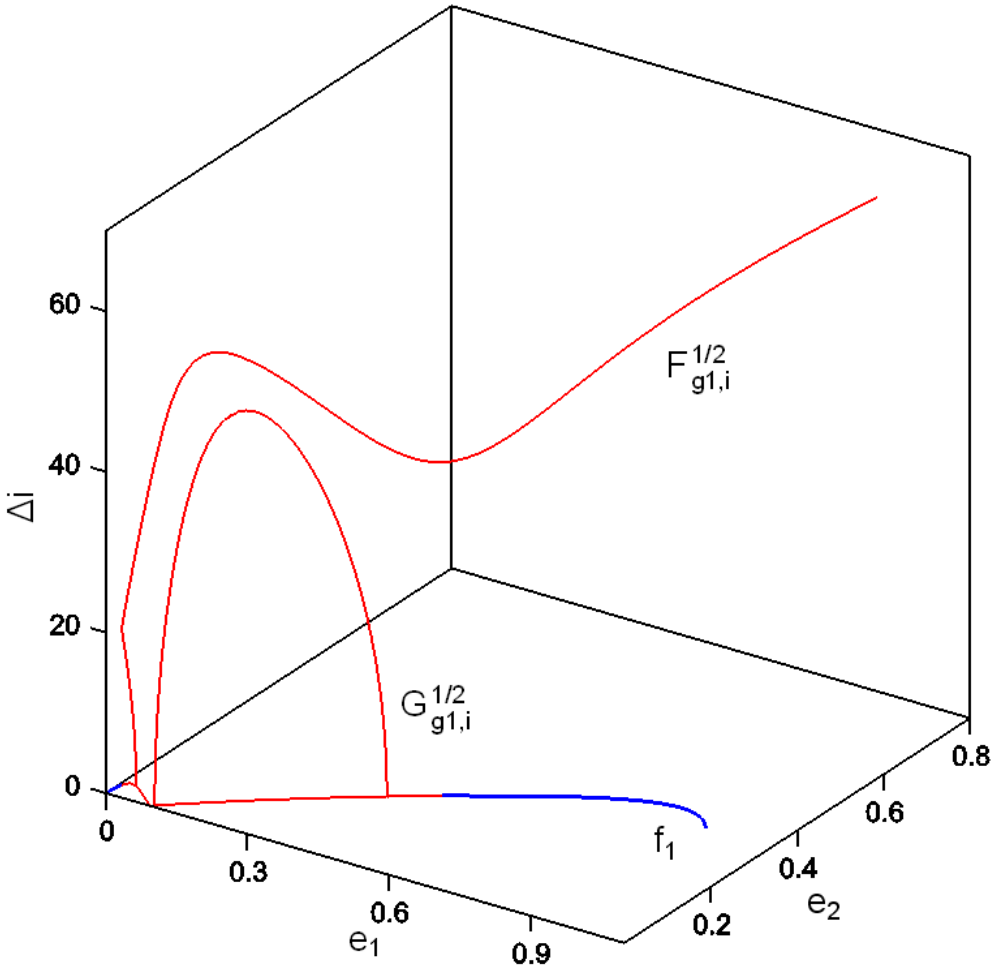} &\includegraphics[width=6cm]{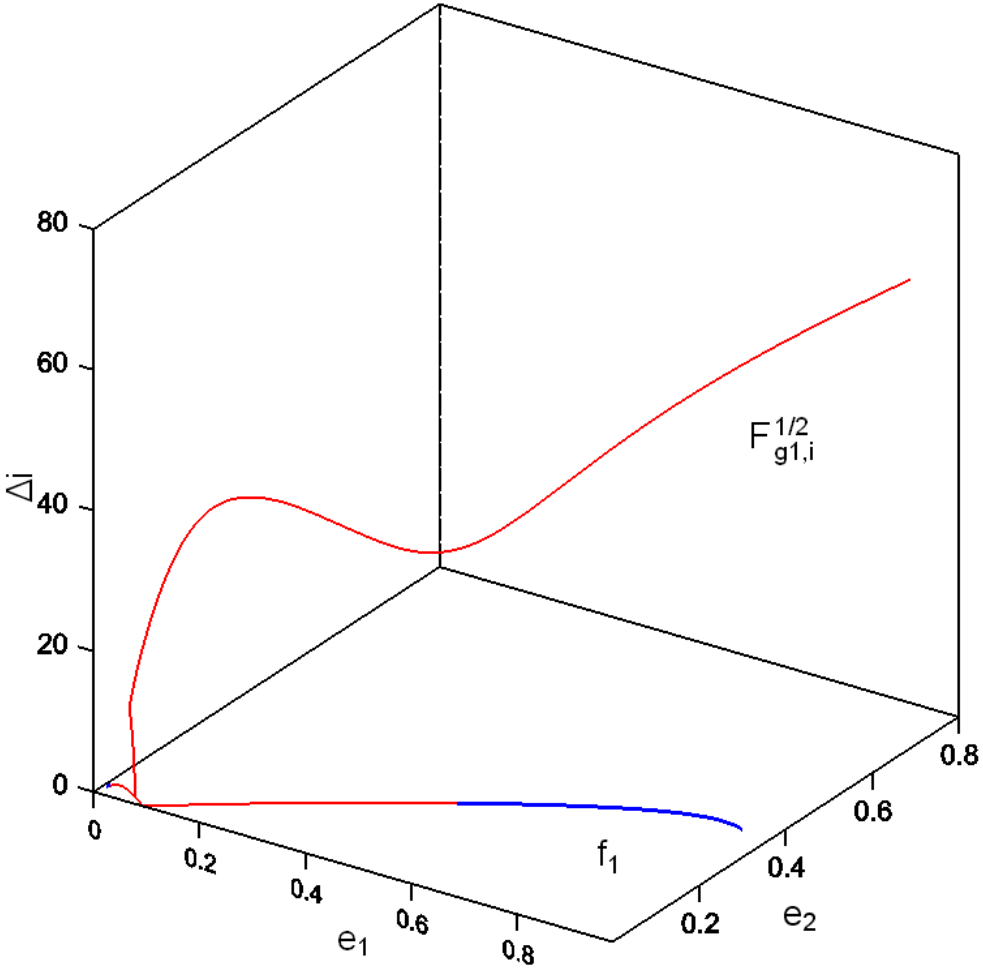}\\
\textnormal{$\rho$=0.02} &\textnormal{$\rho$=0.04}  
\end{array} $
\end{center}
\begin{center}
$\begin{array}{cc}
\includegraphics[width=6cm]{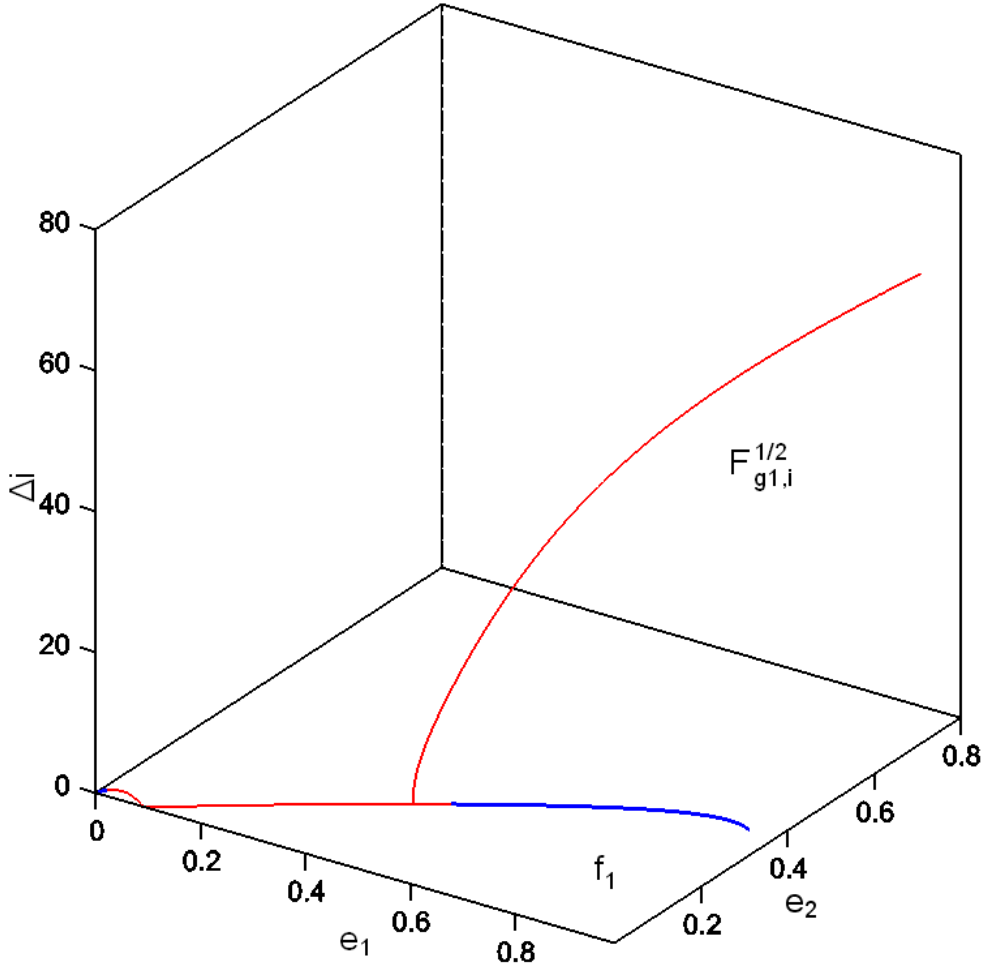} &\includegraphics[width=6cm]{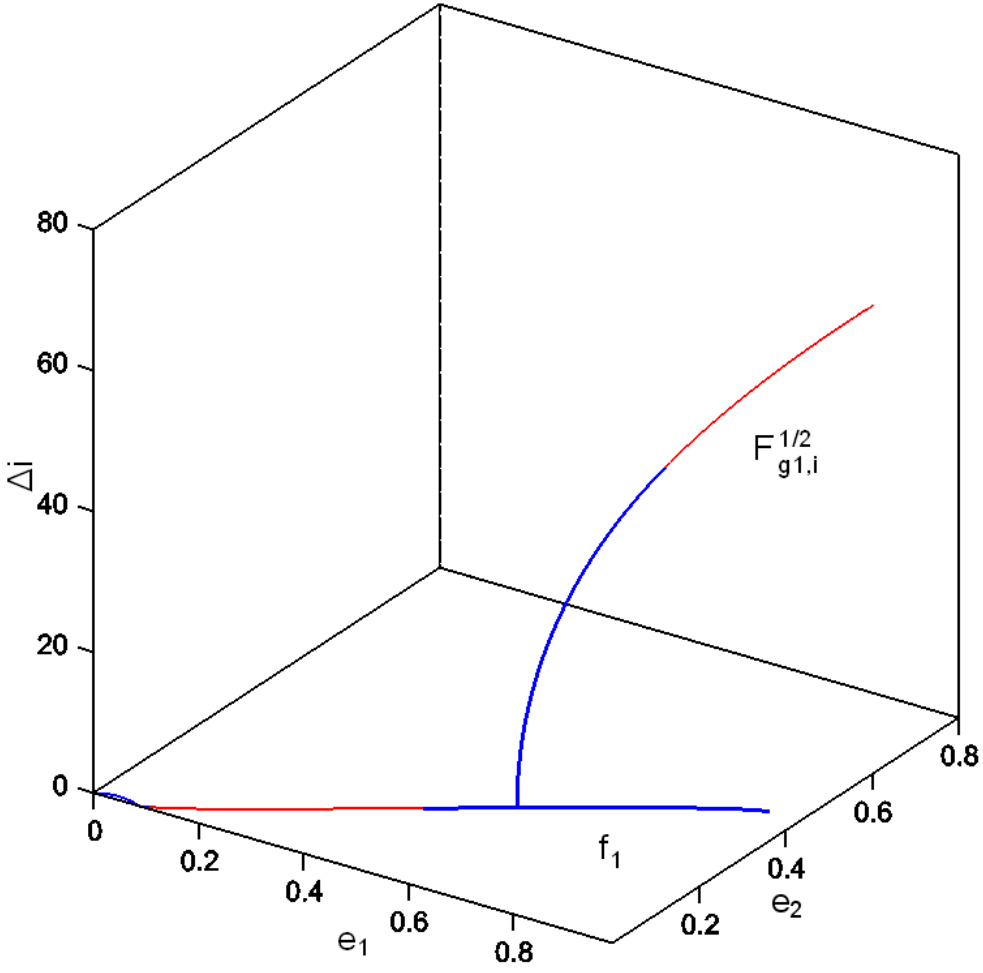}\\
\textnormal{$\rho$=0.1} &\textnormal{$\rho$=0.4}  
\end{array} $
\end{center}
\caption{Bifurcation of families $F^{1/2}_{g1,i}$ (and $G^{1/2}_{g1,i}$ only for $\rho<0.0372$) which are generated from vertical critical points of families $f_1$ given in Fig. \ref{f12i}a.  Bold blue colour stands for stable periodic orbits and red for the unstable ones.}
\label{3dpo}
\end{figure}

In Fig. \ref{f12i}, we present the v.c.o. by dots (see also Tables \ref{vcp XZ f1i}-\ref{vcp XZ f2i}). In Fig. \ref{av}a, we illustrate the variation of the index $a_v$ along the family $f_1$ for some values of $\rho$. For $\rho<0.0372$, we find three v.c.o. and starting from one of them (see Table \ref{vcp XZ f1i}) we can analytically continue $xz$-symmetric periodic orbits in space. We denote such families by $F^{1/2}_{g1,i}$ where the subscript $g$ indicates that the family bifurcates from the general planar problem (it is followed by a number that states the generating planar family) and $i$ indicates that the continuation takes place from plane to space. From the remaining couple of v.c.o. (see Table \ref{vcp X f1i}) $x$-symmetric periodic orbits in space are obtained forming the families $G^{1/2}_{g1,i}$. An example of the above families is given in Fig. \ref{3dpo}a in the projection space $e_1 e_2 \Delta i$, where $\Delta i$ is the mutual inclination. We obtain that the family $G^{1/2}_{g1,i}$ starts from and ends at v.c.o. of the same planar family $f_1$. For $\rho>0.0372$ the couple of v.c.o., from which we obtained the family $G^{1/2}_{g1,i}$, disappears. This transition is explained by the variation of the index $a_v$ along the family $f_1$ and is shown in Fig. \ref{av}b.   

Concerning the families $F^{1/2}_{g1,i}$, up to the mass ratio value $\rho^*\approx 0.12$ all the corresponding v.c.o. are unstable and, subsequently, the continuation generates families that start as unstable. The families continue to be unstable until the end of the computations. For $\rho>\rho^*$ the v.c.o. belong to the stable segment of family $f_1$. Now the generated families $F^{1/2}_{g1,i}$ start as stable and show a monotonical increase of the inclinations $i_1$ and $i_2$, and, subsequently, $\Delta i$, too (see Fig. \ref{3dpo}). The resonant angles that correspond to these periodic orbits are 
$$\Delta \varpi=0^\circ, \; \Delta \Omega=180^\circ, \; \sigma_1=0^\circ.$$ 

\begin{figure}
\begin{center}
\includegraphics[width=6cm]{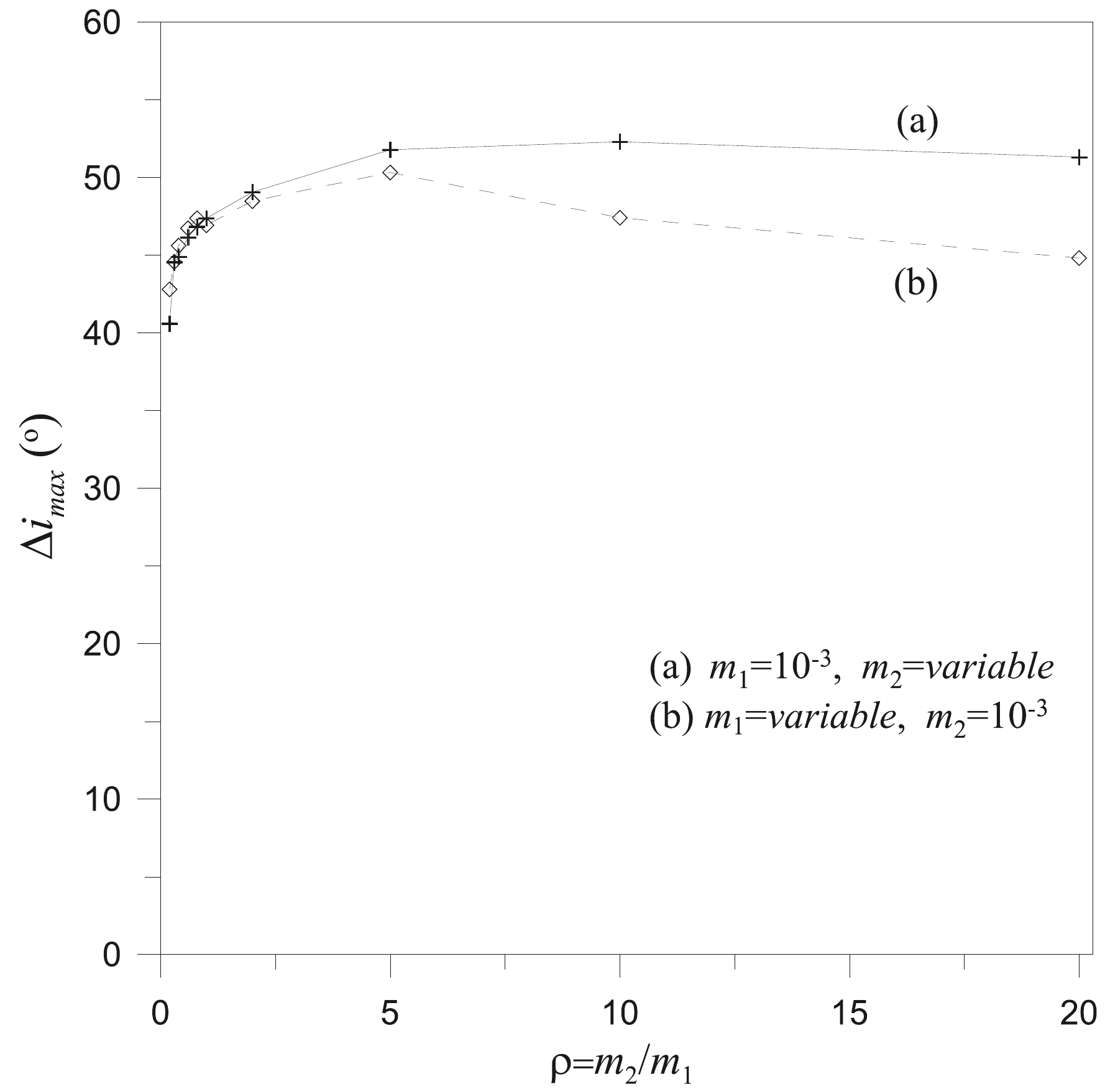} 
\end{center}
\caption{The maximum mutual inclination $\Delta i_{max}$ as a function of mass ratio $\rho=m_2/m_1$ up to which we obtain stable periodic orbits in families $F^{1/2}_{g1,i}$. Stability exists for $\rho>0.12$. In case (a) is $m_1=0.001$ (Jupiter is the inner planet, $P_1$) and in case (b) $m_2=0.001$ (Jupiter  is the outer planet, $P_2$)} 
\label{istab}
\end{figure}

The stability type along the families $F^{1/2}_{g1,i}$ changes above a maximum value for the mutual inclination, $\Delta i_{max}$ that depends on $\rho$, as it is shown in Fig. \ref{istab}. In case (a), Jupiter is the inner planet, $P_1$, and $P_2$ takes large mass values (up to 20 Jupiter masses), as $\rho$ increases.  In case (b), we present the same computations, but now Jupiter is the outer planet, $P_2$ and the other planet, $P_1$, takes smaller and smaller mass\rq{} values, as $\rho$ increases. We observe that we can have stable orbits up to mutual inclinations $40^\circ<\Delta i_{max}<50^\circ$ and there is no essential difference, whether Jupiter is the inner, or the outer planet in the system.      

\begin{figure}
\centering
\includegraphics[width=6cm]{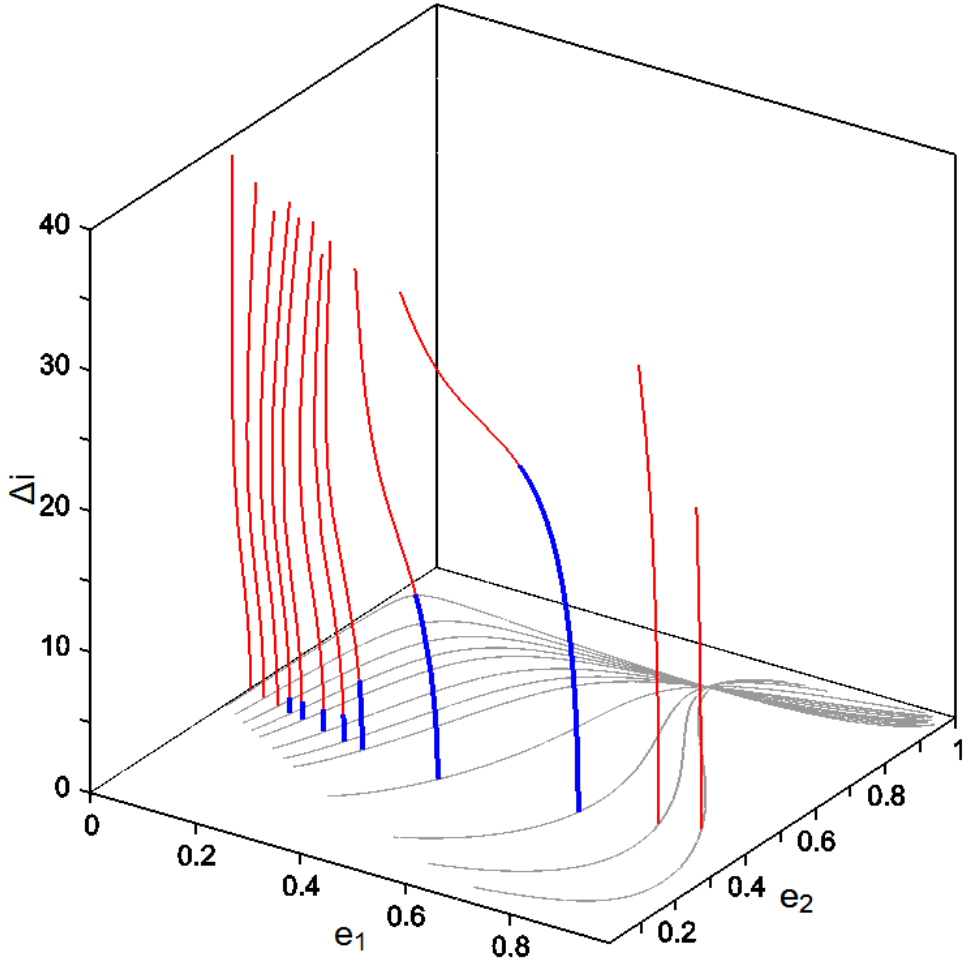}
\caption{The families $F^{1/2}_{g2,i}$ which bifurcate from the v.c.o. of families $f_2$ of the planar general problem.}
\label{ie}
\end{figure}

In Fig. \ref{ie}, we present the families $F^{1/2}_{g2,i}$, which bifurcate from the v.c.o. of families $f_2$ given in Fig. \ref{f12i}b. The corresponding resonant angles are
$$\Delta \varpi=180^\circ,\; \Delta \Omega=180^\circ,\; \sigma_1=180^\circ. $$ 
The families which start from stable v.c.o. (in the range $0.3<\rho<7$, approximately) show a segment of stable periodic orbits up to a critical mutual inclination value, $\Delta i_{max}$.  $\Delta i_{max}$ seems to increase, as $\rho$ increases in the above mentioned interval and reaches  $30^\circ$, approximately. The unstable segments show transitions from real to complex instability and vice versa (see e.g. Fig. \ref{eig}).  

\section{Miscellaneous cases}
\begin{figure}
\begin{center}
$\begin{array}{cc}
\includegraphics[width=6cm]{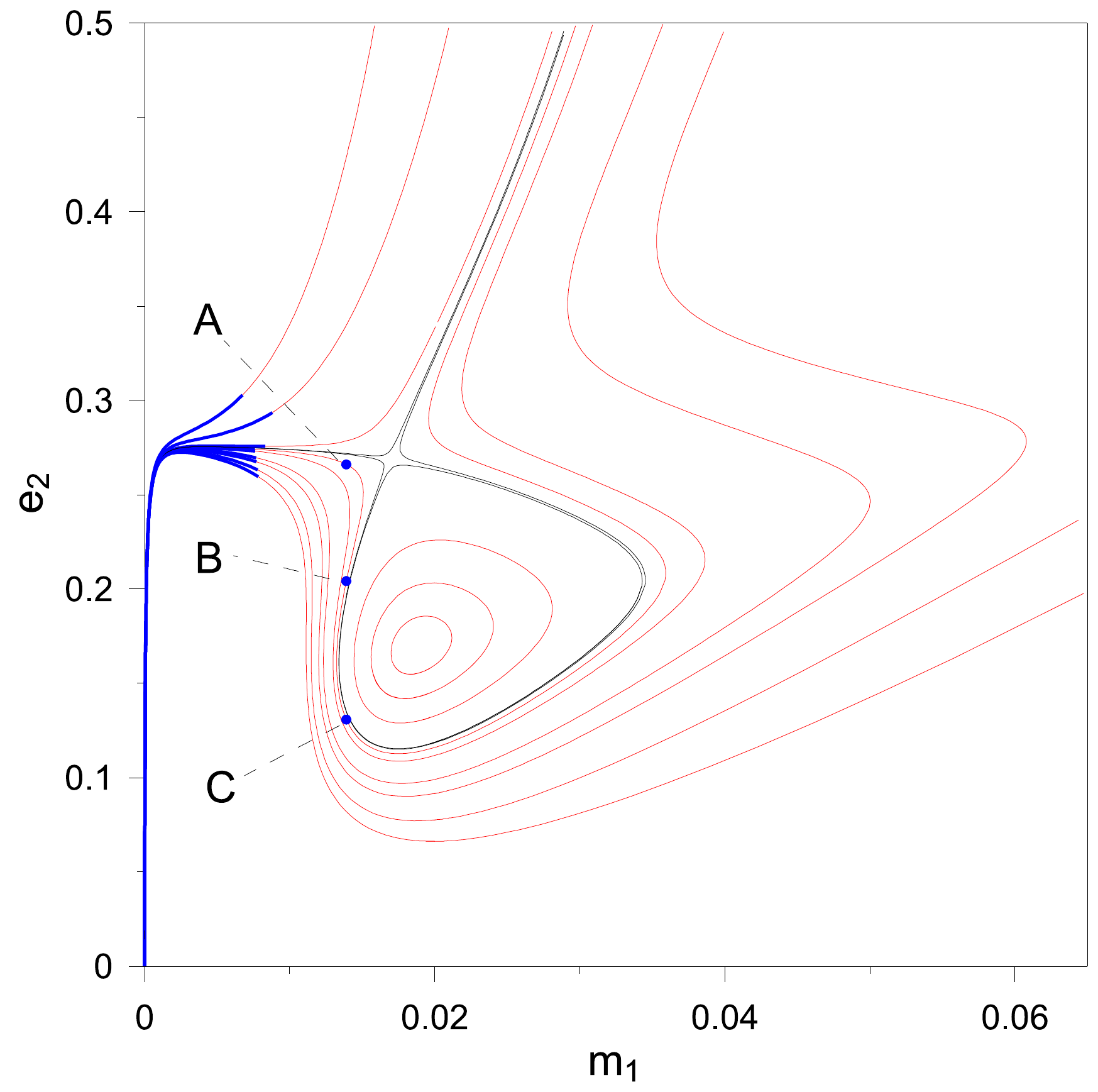} &\includegraphics[width=6.5cm]{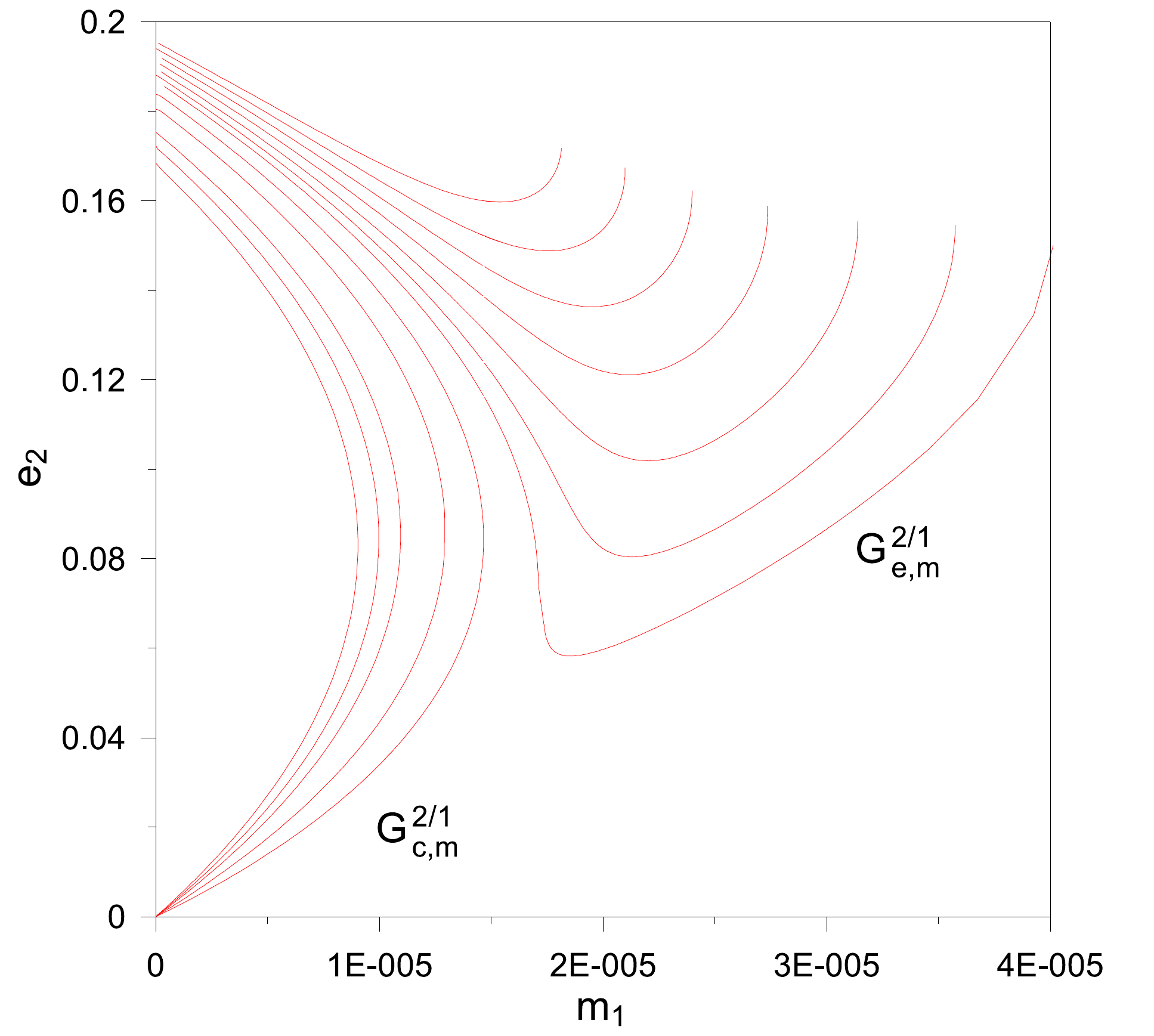}\\
\textnormal{(a)} &\textnormal{(b)}  
\end{array} $
\end{center}
\caption{ {\bf a} Families $F^{2/1}_{c,m}$ up to large mass values of $P_1$  {\bf b} Families $G^{2/1}_{c,m}$ and $G^{2/1}_{e,m}$ computed  by analytical continuation with respect to $m_1$.}
\label{combm2}
\end{figure}

The families $F^{2/1}_{c,m}$ are presented in Fig. \ref{21XZm2} only for $m_1<0.01$. Further continuation with respect to the mass, $m_1$, is possible and as a result, we obtain the characteristic curves of Fig. \ref{combm2}a. We observe that for $m_1>0.01$ foldings become apparent as in families $F^{1/2}_{c,m}$ (see Fig. \ref{12XZm2}). Also, we can obtain the formation of a loop, given by the {\em separatrix} characteristic curve, which includes closed families. Thus, along a family $F^{2/1}_{c,m}$ we may obtain more than one periodic orbits (e.g. the periodic orbits A, B and C) that correspond to the same planetary masses.

We have computed the families $G^{2/1}_{c,m}$ starting from generating orbits of the 3D-CRTBP. This explains the subscript $c$ in their name, although these families end at orbits of the 3D-ERTBP. All of these orbits extend up to a critical mass value,  $m_1^*$ and correspond to inclinations $i_{10}\precsim 70^\circ$. However, if we consider as starting points periodic orbits of the 3D-ERTBP with $i_{10}>70^\circ$ then, by continuation with respect to the mass, $m_1$, we obtain the families $G^{2/1}_{e,m}$ which do not end at the circular restricted problem. This is illustrated by the characteristic curves presented in Fig. \ref{combm2}b.  We found that a similar situation holds, also, for the 1/2 resonance, namely beside the families $G^{1/2}_{c,m}$ (see Fig. \ref{12Xm2}) there exist, also, families $G^{1/2}_{e,m}$. All of them are unstable. 

\begin{figure}
\centering
\includegraphics[width=8cm]{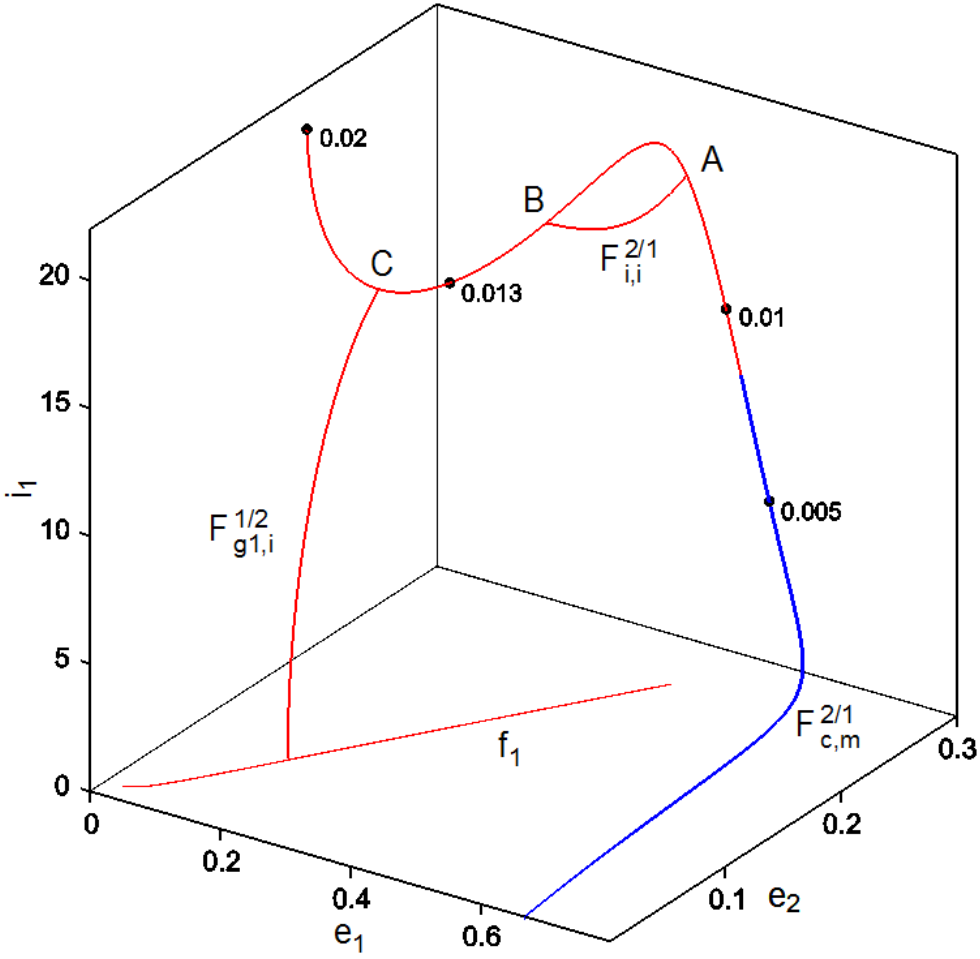}
\caption{ An example of a family $F_{g1,i}^{1/2}$ which appears as a bridge between a planar family $f_1$ and a family $F_{c,m}^{2/1}$. Also, a case of a family $F_{i,i}^{2/1}$, which starts and ends at orbits of family $F_{c,m}^{2/1}$, is shown. }   
\label{combex}
\end{figure}

The families obtained by the analytical continuation of Scheme I, consist of periodic orbits with $z_2\neq 0$ or $\dot z_2 \neq 0$, namely the planetary orbits are of nonzero inclination. Assuming any of the above mentioned periodic orbits we can apply the continuation of Scheme II, in order to decrease the inclination and terminate in the planar general problem. Then, this terminating point must be a v.c.o. of the planar problem. An example is shown in Fig. \ref{combex}. We start from the periodic orbit C that belongs to the family $F^{2/1}_{c,m}$ for $m_1=0.0139$. Along $F^{2/1}_{c,m}$ it is $z_2=0.01042$ (constant). By decreasing $z_2$, according to the Scheme II, we reach the planar general problem and a v.c.o that belongs to the family $f_1$ with $\rho=0.07$. Thus, we obtain a $F^{1/2}_{g1,i}$ family. However, this is not always the case. Considering, for example, as a starting periodic orbit, the orbit A, the continuation by varying $z_2$ and keeping $m_1$ constant results to a family that terminates at the orbit B without reaching the plane $z_2=0$. This family evolves exclusively in the 3D-GTBP and we denote it by $F^{2/1}_{i,i}$. The existence of a folding (namely two points corresponding to the same planetary masses) is a necessary condition for obtaining such a type of families.

\section{Conclusions}
In this paper, we presented 2:1 resonant families of periodic orbits in the three dimensional general TBP of planetary type considering a suitable rotating frame of reference, which reduces the system to four degrees of freedom. 

Symmetric periodic orbits are of two types: symmetric with respect to the $xz$-plane of the rotating frame ($\Delta\Omega=180^\circ$) and symmetric with respect to the $x$-axis ($\Delta\Omega=0^\circ$). For the computation of families of periodic orbits we can apply two continuation schemes:\\
i) we start from periodic orbits of the 3D circular or elliptic restricted TBP and continue them by varying the mass of the small planet (which is initially the massless body)\\
ii) we start from vertical critical periodic orbits that belong to the families of the general planar problem and continue them in space by varying the planetary inclination.

The computation and continuation procedures can be applied similarly to any other resonance. Moreover, similar procedures can be used for the computation of asymmetric periodic orbits too, but in this case, it is required to meet a larger number of periodic conditions. Also, we considered and presented orbits only for prograde planetary motion. 

Most of the periodic orbits found are linearly unstable. Especially, all periodic orbits symmetric with respect to the of $x$-axis are unstable.  In their neigbourhood in phase space, we have planetary orbits that evolve chaotically. We obtain relatively large excitations in eccentricity or inclination and after a relatively short-term evolution the planetary system is disrupted, due to close encounters. 

Nevertheless, families of $xz$-plane symmetry show segments which include stable periodic orbits. Such orbits can be obtained up to $50^\circ - 60^\circ$ of mutual planetary inclination. Particularly we found that all families which bifurcate from stable vertical orbits of the planar problem start having stable periodic orbits. For the case of the planar family $f_1$ ($\sigma_1=0^\circ$, $\Delta\varpi=0^\circ$),  stability of the bifurcated 3D periodic orbits is observed for a wide range of values of the planetary mass ratio, particularly for $\rho>0.12$ and at least up to $\rho=20$ (where we stopped our computations). For the families of 3D periodic orbits which bifurcate from the planar family $f_2$ ($\sigma_1=180^\circ$, $\Delta\varpi=180^\circ$) stable orbits exist for $0.3<\rho<7$. The maximum mutual inclination which is observed for stable orbits increases as $\rho$ increases in the above domain and reaches $30^\circ$, approximately. All stable periodic orbits found correspond to quite eccentric planetary orbits (for at least one of the planets). 

Starting with initial conditions sufficiently close to a stable periodic orbit the planetary evolution takes place regularly. The planetary system remains in the mean motion resonance and the eccentricities and inclinations show small oscillations. Also, all resonant angles librate around the values that correspond to the stable periodic orbit. Such regions near stable periodic orbits may be considered as strong candidates for hosting planetary systems. The relation between the dynamics of real planetary systems and 3D periodic orbits of the general TBP remains to be studied. 

\vspace{1cm}

{\bf Acknowledgments.} We thank Prof. J. Hadjidemetriou and Dr. K. Tsiganis for fruitful discussions about the presented paper. This research has been co-financed by the European Union (European Social Fund - ESF) and Greek national funds through the Operational Program "Education and Lifelong Learning" of the National Strategic Reference Framework (NSRF) - Research Funding Program: THALES. Investing in knowledge society through the European Social Fund.

\bibliographystyle{plainnat}
\bibliography{bib}

\begin{thebibliography}{42}
\providecommand{\natexlab}[1]{#1}
\providecommand{\url}[1]{\texttt{#1}}
\expandafter\ifx\csname urlstyle\endcsname\relax
  \providecommand{\doi}[1]{doi: #1}\else
  \providecommand{\doi}{doi: \begingroup \urlstyle{rm}\Url}\fi

\bibitem[Antoniadou et~al.(2011)Antoniadou, Voyatzis, and Kotoulas]{avk11}
K.~I. Antoniadou, G.~Voyatzis, and T.~Kotoulas.
\newblock On the bifurcation and continuation of periodic orbits in the three
  body problem.
\newblock \emph{International Journal of Bifurcation and Chaos}, 21:\penalty0
  2211, 2011.

\bibitem[Barnes et~al.(2011)Barnes, Greenberg, Quinn, McArthur, and
  Benedict]{bgqmb11}
R.~Barnes, R.~Greenberg, T.~R. Quinn, B.~E. McArthur, and G.~F. Benedict.
\newblock Origin and dynamics of the mutually inclined orbits of $\upsilon$
  andromedae c and d.
\newblock \emph{The Astrophysical Journal}, 726:\penalty0 71, 2011.

\bibitem[Beaug{\'e} et~al.(2006)Beaug{\'e}, Michtchenko, and
  Ferraz-Mello]{bmfm06}
C.~Beaug{\'e}, T.~A. Michtchenko, and S.~Ferraz-Mello.
\newblock Planetary migration and extrasolar planets in the 2/1 mean-motion
  resonance.
\newblock \emph{Monthly Notices of the Royal Astronomical Society},
  365:\penalty0 1160--1170, 2006.

\bibitem[Chatterjee et~al.(2008)Chatterjee, Ford, Matsumura, and
  Rasio]{chaford08}
S.~Chatterjee, E.~B. Ford, S.~Matsumura, and F.~A. Rasio.
\newblock Dynamical outcomes of planet-planet scattering.
\newblock \emph{The Astrophysical Journal}, 686:\penalty0 580--602, 2008.

\bibitem[Correia et~al.(2011)Correia, Laskar, Farago, and Bou{\'e}]{cor11}
A.~C.~M. Correia, J.~Laskar, F.~Farago, and G.~Bou{\'e}.
\newblock Tidal evolution of hierarchical and inclined systems.
\newblock \emph{Celestial Mechanics and Dynamical Astronomy}, 111:\penalty0
  105--130, 2011.

\bibitem[Ferraz-Mello et~al.(2003)Ferraz-Mello, Beaug{\'e}, and
  Michtchenko]{mebeaumich03}
S.~Ferraz-Mello, C.~Beaug{\'e}, and T.~A. Michtchenko.
\newblock Evolution of migrating planet pairs in resonance.
\newblock \emph{Celestial Mechanics and Dynamical Astronomy}, 87:\penalty0
  99--112, 2003.

\bibitem[Froeschl{\'e} et~al.(1997)Froeschl{\'e}, Lega, and Gonczi]{froe97}
C.~Froeschl{\'e}, E.~Lega, and R.~Gonczi.
\newblock Fast lyapunov indicators. application to asteroidal motion.
\newblock \emph{Celestial Mechanics and Dynamical Astronomy}, 67:\penalty0
  41--62, 1997.

\bibitem[Go{\'z}dziewski et~al.(2005)Go{\'z}dziewski, Konacki, and
  Wolszczan]{gozd05}
K.~Go{\'z}dziewski, M.~Konacki, and A.~Wolszczan.
\newblock Long-term stability and dynamical environment of the psr 1257+12
  planetary system.
\newblock \emph{The Astrophysical Journal}, 619:\penalty0 1084--1097, 2005.

\bibitem[Hadjidemetriou and Voyatzis(2000)]{hadjvoy00}
J.~Hadjidemetriou and G.~Voyatzis.
\newblock The 2/1 and 3/2 resonant asteroid motion: A symplectic mapping
  approach.
\newblock \emph{Celestial Mechanics and Dynamical Astronomy}, 78:\penalty0
  137--150, 2000.

\bibitem[Hadjidemetriou(2006{\natexlab{a}})]{hadj06}
J.~D. Hadjidemetriou.
\newblock Symmetric and asymmetric librations in extrasolar planetary systems:
  a global view.
\newblock \emph{Celestial Mechanics and Dynamical Astronomy}, 95:\penalty0
  225--244, 2006{\natexlab{a}}.

\bibitem[Hadjidemetriou(2006{\natexlab{b}})]{hadjbook06}
J.~D. Hadjidemetriou.
\newblock \emph{Periodic orbits in gravitational systems}, pages 43--79.
\newblock Springer Netherlands, 2006{\natexlab{b}}.

\bibitem[Hadjidemetriou and Voyatzis(2010)]{hadjvoy10}
J.~D. Hadjidemetriou and G.~Voyatzis.
\newblock On the dynamics of extrasolar planetary systems under dissipation:
  Migration of planets.
\newblock \emph{Celestial Mechanics and Dynamical Astronomy}, 107:\penalty0
  3--19, 2010.

\bibitem[Haghighipour et~al.(2003)Haghighipour, Couetdic, Varadi, and
  Moore]{haghi03}
N.~Haghighipour, J.~Couetdic, F.~Varadi, and W.~B. Moore.
\newblock Stable 1:2 resonant periodic orbits in elliptic three-body systems.
\newblock \emph{The Astrophysical Journal}, 596:\penalty0 1332--1340, 2003.

\bibitem[H\'enon(1973)]{hen}
M.~H\'enon.
\newblock Vertical stability of periodic orbits in the restricted problem. i.
  equal masses.
\newblock \emph{Astronomy and Astrophysics}, 28:\penalty0 415, 1973.

\bibitem[Ichtiaroglou and Michalodimitrakis(1980)]{ichmich80}
S.~Ichtiaroglou and M.~Michalodimitrakis.
\newblock Three-body problem - the existence of families of three-dimensional
  periodic orbits which bifurcate from planar periodic orbits.
\newblock \emph{Astronomy and Astrophysics}, 81:\penalty0 30--32, 1980.

\bibitem[Katopodis et~al.(1980)Katopodis, Ichtiaroglou, and
  Michalodimitrakis]{katichmich}
K.~Katopodis, S.~Ichtiaroglou, and M.~Michalodimitrakis.
\newblock The family i1v of the three-dimensional general three-body problem.
\newblock \emph{Astronomy and Astrophysics}, 90:\penalty0 102--105, 1980.

\bibitem[Kotoulas(2005)]{kot05}
T.~A. Kotoulas.
\newblock The dynamics of the 1:2 resonant motion with neptune in the 3d
  elliptic restricted three-body problem.
\newblock \emph{Astronomy and Astrophysics}, 429:\penalty0 1107--1115, 2005.

\bibitem[Kotoulas and Voyatzis(2005)]{kotvoy05}
T.~A. Kotoulas and G.~Voyatzis.
\newblock Three dimensional periodic orbits in exterior mean motion resonances
  with neptune.
\newblock \emph{Astronomy and Astrophysics}, 441:\penalty0 807--814, 2005.

\bibitem[Laughlin et~al.(2002)Laughlin, Chambers, and Fischer]{lauchfi02}
G.~Laughlin, J.~Chambers, and D.~Fischer.
\newblock A dynamical analysis of the 47 ursae majoris planetary system.
\newblock \emph{The Astrophysical Journal}, 579:\penalty0 455--467, 2002.

\bibitem[Lee and Peale(2002)]{leepeal02}
M.~H. Lee and S.~J. Peale.
\newblock Dynamics and origin of the 2:1 orbital resonances of the gj 876
  planets.
\newblock \emph{The Astrophysical Journal}, 567:\penalty0 596--609, 2002.

\bibitem[Lee and Thommes(2009)]{leetho09}
M.~H. Lee and E.~W. Thommes.
\newblock Planetary migration and eccentricity and inclination resonances in
  extrasolar planetary systems.
\newblock \emph{The Astrophysical Journal}, 702:\penalty0 1662--1672, 2009.

\bibitem[Libert and Tsiganis(2009{\natexlab{a}})]{litsi09}
A.-S. Libert and K.~Tsiganis.
\newblock Kozai resonance in extrasolar systems.
\newblock \emph{Astronomy and Astrophysics}, 493:\penalty0 677--686,
  2009{\natexlab{a}}.

\bibitem[Libert and Tsiganis(2009{\natexlab{b}})]{litsi09b}
A.-S. Libert and K.~Tsiganis.
\newblock Trapping in high-order orbital resonances and inclination excitation
  in extrasolar systems.
\newblock \emph{Monthly Notices of the Royal Astronomical Society},
  400:\penalty0 1373--1382, 2009{\natexlab{b}}.

\bibitem[Marchal(1990)]{marchal90}
C.~Marchal.
\newblock \emph{The three-body problem}.
\newblock Elsevier, Amsterdam, 1990.

\bibitem[Marzari and Weidenschilling(2002)]{mawei02}
F.~Marzari and S.~J. Weidenschilling.
\newblock Eccentric extrasolar planets: The jumping jupiter model.
\newblock \emph{Icarus}, 156:\penalty0 570--579, 2002.

\bibitem[Michalodimitrakis(1979{\natexlab{a}})]{mich79}
M.~Michalodimitrakis.
\newblock On the continuation of periodic orbits from the planar to the
  three-dimensional general three-body problem.
\newblock \emph{Celestial Mechanics}, 19:\penalty0 263--277, 1979/4/1
  1979{\natexlab{a}}.

\bibitem[Michalodimitrakis(1979{\natexlab{b}})]{mich79b}
M.~Michalodimitrakis.
\newblock General three-body problem - families of three-dimensional periodic
  orbits. ii.
\newblock \emph{Astrophysics and Space Science}, 65:\penalty0 459--475,
  1979{\natexlab{b}}.

\bibitem[Michalodimitrakis(1980)]{mich80}
M.~Michalodimitrakis.
\newblock General three-body problem - families of three-dimensional periodic
  orbits. i.
\newblock \emph{Astronomy and Astrophysics}, 81:\penalty0 113--120, 1980.

\bibitem[Michalodimitrakis(1981)]{mich81}
M.~Michalodimitrakis.
\newblock The families c 1 v and m 2 v of the three-dimensional general
  three-body problem.
\newblock \emph{Astronomy and Astrophysics}, 93:\penalty0 212--218, 1981.

\bibitem[Michtchenko and Ferraz-Mello(2011)]{mfpro}
T.~A. Michtchenko and S.~Ferraz-Mello.
\newblock The periodic and chaotic regimes of motion in the exoplanet 2/1
  mean-motion resonance.
\newblock \emph{Proceedings of Third La Plata International School on Astronomy
  and Geophysics: Chaos, diffusion and non-integrability in Hamiltonian Systems
  Applications to Astonomy}, 2011.

\bibitem[Michtchenko et~al.(2006{\natexlab{a}})Michtchenko, Beaug{\'e}, and
  Ferraz-Mello]{mbf06}
T.~A. Michtchenko, C.~Beaug{\'e}, and S.~Ferraz-Mello.
\newblock Stationary orbits in resonant extrasolar planetary systems.
\newblock \emph{Celestial Mechanics and Dynamical Astronomy}, 94:\penalty0
  411--432, 2006{\natexlab{a}}.

\bibitem[Michtchenko et~al.(2006{\natexlab{b}})Michtchenko, Ferraz-Mello, and
  Beaug{\'e}]{mfb06}
T.~A. Michtchenko, S.~Ferraz-Mello, and C.~Beaug{\'e}.
\newblock Modeling the 3-d secular planetary three-body problem. discussion on
  the outer $\upsilon$ andromedae planetary system.
\newblock \emph{Icarus}, 181:\penalty0 555--571, 2006{\natexlab{b}}.

\bibitem[Michtchenko et~al.(2008{\natexlab{a}})Michtchenko, Beaug{\'e}, and
  Ferraz-Mello]{mbf08a}
T.~A. Michtchenko, C.~Beaug{\'e}, and S.~Ferraz-Mello.
\newblock Dynamic portrait of the planetary 2/1 mean-motion resonance - i.
  systems with a more massive outer planet.
\newblock \emph{Monthly Notices of the Royal Astronomical Society},
  387:\penalty0 747--758, 2008{\natexlab{a}}.

\bibitem[Michtchenko et~al.(2008{\natexlab{b}})Michtchenko, Beaug{\'e}, and
  Ferraz-Mello]{mbf08b}
T.~A. Michtchenko, C.~Beaug{\'e}, and S.~Ferraz-Mello.
\newblock Dynamic portrait of the planetary 2/1 mean-motion resonance - ii.
  systems with a more massive inner planet.
\newblock \emph{Monthly Notices of the Royal Astronomical Society},
  391:\penalty0 215--227, 2008{\natexlab{b}}.

\bibitem[Psychoyos and Hadjidemetriou(2005)]{psyhadj05}
D.~Psychoyos and J.~D. Hadjidemetriou.
\newblock Dynamics of 2/1 resonant extrasolar systems application to hd82943
  and gliese876.
\newblock \emph{Celestial Mechanics and Dynamical Astronomy}, 92:\penalty0
  135--156, 2005.

\bibitem[Schwarz et~al.(2012)Schwarz, Bazso, {\'E}rdi, and Funk]{l4}
R.~Schwarz, {\'A}.~Bazso, B.~{\'E}rdi, and B.~Funk.
\newblock Stability of the lagrangian point $l_4$ in the spatial restricted
  three-body problem - application to exoplanetary systems.
\newblock \emph{Monthly Notices of the Royal Astronomical Society in press},
  2012.

\bibitem[Thommes and Lissauer(2003)]{thommes03}
E.~W. Thommes and J.~J. Lissauer.
\newblock Resonant inclination excitation of migrating giant planets.
\newblock \emph{The Astrophysical Journal}, 597:\penalty0 566--580, November
  2003.

\bibitem[Varadi(1999)]{varadi99}
F.~Varadi.
\newblock Periodic orbits in the 3:2 orbital resonance and their stability.
\newblock \emph{The Astronomical Journal}, 118:\penalty0 2526--2531, 1999.

\bibitem[Veras and Armitage(2004)]{verasa04}
D.~Veras and P.~J. Armitage.
\newblock The dynamics of two massive planets on inclined orbits.
\newblock \emph{Icarus}, 172:\penalty0 349--371, December 2004.

\bibitem[Voyatzis(2008)]{voyatzis08}
G.~Voyatzis.
\newblock Chaos, order, and periodic orbits in 3:1 resonant planetary dynamics.
\newblock \emph{The Astrophysical Journal}, 675:\penalty0 802--816, 2008.

\bibitem[Voyatzis and Hadjidemetriou(2005)]{voyhadj05}
G.~Voyatzis and J.~D. Hadjidemetriou.
\newblock Symmetric and asymmetric librations in planetary and satellite
  systems at the 2/1 resonance.
\newblock \emph{Celestial Mechanics and Dynamical Astronomy}, 93:\penalty0
  263--294, 2005.

\bibitem[Voyatzis et~al.(2009)Voyatzis, Kotoulas, and Hadjidemetriou]{vkh09}
G.~Voyatzis, T.~Kotoulas, and J.~D. Hadjidemetriou.
\newblock On the 2/1 resonant planetary dynamics - periodic orbits and
  dynamical stability.
\newblock \emph{Monthly Notices of the Royal Astronomical Society},
  395:\penalty0 2147--2156, 2009.

\end{thebibliography}

\end{document}